\documentclass[aps,prb,reprint,twocolumn,showpacs,preprintnumbers,superscriptaddress,amsmath,amssymb,floatfix]{revtex4-1}

\usepackage{graphicx}
\usepackage{bm}
\usepackage[mathlines]{lineno}
 
\begin{document}

\preprint{}

\title{Spin and orbital magnetic moments of size-selected iron, cobalt, and nickel clusters and their link to the bulk phase diagrams}

\author{A.~Langenberg}
\author{K.~Hirsch}
	\affiliation{Institut f\"{u}r Methoden und Instrumentierung der Forschung mit Synchrotronstrahlung, Helmholtz-Zentrum Berlin f\"{u}r Materialien und Energie GmbH, Albert-Einstein-Stra{\ss}e 15, 12489 Berlin, Germany}
	\affiliation{Institut f\"{u}r Optik und Atomare Physik, Technische Universit\"{a}t Berlin, Hardenbergstra{\ss}e 36, 10623 Berlin, Germany}

\author{A. {\L}awicki}
	\affiliation{Institut f\"{u}r Methoden und Instrumentierung der Forschung mit Synchrotronstrahlung, Helmholtz-Zentrum Berlin f\"{u}r Materialien und Energie GmbH, Albert-Einstein-Stra{\ss}e 15, 12489 Berlin, Germany}

\author{V.~Zamudio-Bayer}
	\affiliation{Institut f\"{u}r Methoden und Instrumentierung der Forschung mit Synchrotronstrahlung, Helmholtz-Zentrum Berlin f\"{u}r Materialien und Energie GmbH, Albert-Einstein-Stra{\ss}e 15, 12489 Berlin, Germany}
	\affiliation{Physikalisches Institut, Universit\"{a}t Freiburg, Stefan-Meier-Stra{\ss}e 21, 79104 Freiburg, Germany}

\author{M.~Niemeyer}
\author{P.~Chmiela}
\author{B.~Langbehn}
	\affiliation{Institut f\"{u}r Methoden und Instrumentierung der Forschung mit Synchrotronstrahlung, Helmholtz-Zentrum Berlin f\"{u}r Materialien und Energie GmbH, Albert-Einstein-Stra{\ss}e 15, 12489 Berlin, Germany}
	\affiliation{Institut f\"{u}r Optik und Atomare Physik, Technische Universit\"{a}t Berlin, Hardenbergstra{\ss}e 36, 10623 Berlin, Germany}

\author{A.~Terasaki}
	\affiliation{Cluster Research Laboratory, Toyota Technological Institute, 717-86 Futamata, Ichikawa, Chiba 272-0001, Japan}
	\affiliation{Department of Chemistry, Kyushu University, 6-10-1 Hakozaki, Higashi-ku, Fukuoka 812-8581, Japan}

\author{B.~v.~Issendorff}
	\affiliation{Physikalisches Institut, Universit\"{a}t Freiburg, Stefan-Meier-Stra{\ss}e 21, 79104 Freiburg, Germany}

\author{J.~T.~Lau}
	\email{tobias.lau@helmholtz-berlin.de}
	\affiliation{Institut f\"{u}r Methoden und Instrumentierung der Forschung mit Synchrotronstrahlung, Helmholtz-Zentrum Berlin f\"{u}r Materialien und Energie GmbH, Albert-Einstein-Stra{\ss}e 15, 12489 Berlin, Germany}

\date{\today}

\begin{abstract}
Spin and orbital magnetic moments of cationic iron, cobalt, and nickel clusters have been determined from x-ray magnetic circular dichroism spectroscopy. In the size regime of $n = 10 - 15$ atoms, these clusters show strong ferromagnetism with maximized spin magnetic moments of 1~$\mu_B$ per empty $3d$ state because of completely filled $3d$ majority spin bands. The only exception is $\mathrm{Fe}_{13}^+$ where an unusually low average spin magnetic moment of $0.73 \pm 0.12$~$\mu_B$ per unoccupied $3d$ state is detected; an effect, which is neither observed for $\mathrm{Co}_{13}^+$ nor $\mathrm{Ni}_{13}^+$.\@ This distinct behavior can be linked to the existence and accessibility of antiferromagnetic, paramagnetic, or nonmagnetic phases in the respective bulk phase diagrams of iron, cobalt, and nickel. Compared to the experimental data, available density functional theory calculations generally seem to underestimate the spin magnetic moments significantly. In all clusters investigated, the orbital magnetic moment is quenched to $5 - 25$\,\% of the atomic value by the reduced symmetry of the crystal field. The magnetic anisotropy energy is well below 65 $\mu$eV per atom.
\end{abstract}


\maketitle

\section{Introduction}
Only very few elements in the periodic table show magnetic order at room temperature. In bulk iron, cobalt, and nickel, this magnetic order arises from the balance of electron localization that gives rise to magnetic moments, and of electron delocalization as a prerequisite for direct exchange interaction \cite{Stoehr06}.\@ 
Even in these ferromagnetic $3d$ transition elements, the magnetic moments are smaller than in the corresponding atoms because electron delocalization reduces the spin magnetic moment to $30 - 50$\,\% of their atomic values, while crystal field effects quench the orbital magnetic moment to less than 5\,\%.\@
The evolution of spin and orbital magnetic moments from atoms via nanoscale matter to the bulk has therefore been studied intensively by theory and experiment \cite{Kodama99, Alonso00, Binns01a, Batlle02, Bansmann05}.\@
In neutral clusters of the ferromagnetic $3d$ transition elements iron, cobalt, and nickel \cite{deHeer90, Bucher91, Billas93, Louderback93, Billas94, Cox94, Apsel96, Billas97}, enhanced magnetic moments and superparamagnetic behavior \cite{Bean59} were found because of narrow $3d$ bands that are created by the low coordination of surface atoms in clusters. Furthermore, transition elements that show antiferromagnetic order or even no magnetic order in the bulk can exhibit finite magnetic moments in small clusters \cite{Cox93a, Knickelbein01, Knickelbein04a, Payne06}.\@ Metastable magnetic species of iron and cobalt clusters have also been observed \cite{Xu11}.\@
When deposited on surfaces, adatoms \cite{Gambardella03, Meier08, Brune09}, clusters \cite{Lau02a, Lau02b, Ballentine07, Glaser12}, and nanoparticles \cite{Edmonds99, Pietzsch04, Kleibert09, Bansmann10} were investigated by means of x-ray magnetic circular dichroism spectroscopy \cite{Koide01, Lau02a, Lau02b, Fauth04a, Boyen05a, Gambardella05a, Bansmann10, Glaser12}, spin polarized scanning tunneling microscopy \cite{Wiesendanger90, Wulfhekel07} or scanning tunneling spectroscopy \cite{Brune06, Hirjibehedin07, Balashov09, Donati13}, giving access to spin and orbital magnetic moments \cite{}, magnetic anisotropy energies \cite{}, or exchange coupling constants \cite{} of these systems.
Large magnetic anisotropy energies up to 9 meV per atom of iron and cobalt impurities deposited on platinum surfaces \cite{Gambardella03, Balashov09} as well as ferromagnetic behavior at room temperature of iron-cobalt nanoparticles on a silicon surface \cite{Getzlaff06} have been reported. However, many properties of deposited clusters and nanoparticles are caused by the strong interaction with the support that can enhance or even quench magnetic moments \cite{Fauth04a}, and that also governs the anisotropy energy. 
To gain insight into the intrinsic unperturbed properties of these systems without coupling to a support or matrix, gas phase studies of free particles are mandatory.
Stern-Gerlach deflection experiments \cite{Cox85, deHeer90, Apsel96, Knickelbein06, Payne07, Xu11, Rohrmann13}, which yield total magnetic moments with high precision \cite{Xu11}, have been the method of choice for the determination of magnetic properties of free clusters for a long time.
In recent experimental studies, x-ray magnetic circular dichroism (XMCD) spectroscopy \cite{Erskine75, Schuetz87, Thole92, Carra93, Chen95} has been applied successfully to clusters that are stored in an ion trap \cite{Peredkov11b, Niemeyer12, Hirsch13a, ZamudioBayer13}, giving access to the intrinsic spin and orbital magnetic moments of isolated clusters. 
These studies showed a peculiar reduction of the average spin magnetic moment \cite{Niemeyer12} in $\mathrm{Fe}_{13}^+$ as well as the strong quenching of the orbital magnetic moment already for very small iron clusters of only three atoms.
\newline
Here, we show that the reduced spin magnetic moment that is observed for $\mathrm{Fe}_{13}^+$ is neither present in $\mathrm{Co}_{13}^+$ nor $\mathrm{Ni}_{13}^+$ clusters but is unique among the ferromagnetic $3d$ elements in this size range. In fact, with the exception of $\mathrm{Fe}_{13}^+$, all  Fe$_n^+$, Co$_n^+$, and Ni$_n^+$ clusters with $n = 10 - 15$ atoms per cluster are strong ferromagnets with magnetic moments of $\approx 1\, \mu_B$ per empty $3d$ state. This particular behavior of iron clusters might be related to the complex magnetic properties of bulk iron and to the existence of antiferromagnetic or nonmagnetic phases in the bulk phase diagram \cite{Bancroft56, Jamieson62, Taylor91, Wang98b, Rueff99, Mathon04, Ishimatsu07}.\@ 

\section{Experimental Setup}
Cluster ions are produced in a liquid nitrogen cooled magnetron gas aggregation source, size selected in a quadrupole mass filter, and guided into a radio-frequency quadrupole ion trap where x-ray absorption and x-ray magnetic circular dichroism (XMCD) spectroscopy \cite{Erskine75, Schuetz87, Thole92, Carra93, Chen95} of size-selected cluster ions \cite{Peredkov11b, Niemeyer12, ZamudioBayer13, Hirsch13a} is performed in ion yield mode \cite{Lau08, Lau09a, Lau09b, Hirsch09, Lau11, Vogel12, Hirsch12b}at the $L_{2,3}$ edges of iron, cobalt, and nickel.
\newline
The total magnetic moments of size-selected gas phase clusters are aligned by a homogeneous magnetic field $\mu_{0} H \le 5$ T of a superconducting solenoid that is placed around the ion trap. The ion trap housing and electrodes are cooled down to a temperature of $4 - 6$ K by a flow of liquid helium, and the cluster ions are thermalized by a constant flow of purified and pre-cooled helium buffer gas in the presence of the applied magnetic field. The ion temperature $T_{\,\text{ion}}$ is typically 10 to15 K.\@ The helium atom density in the ion trap corresponds to $10^{-4}$ to $10^{-3}$ mbar at room temperature. 
\newline
Along the ion trap axis a monochromatic and elliptically polarized soft-x-ray synchrotron radiation beam from an undulator beamline (UE52-SGM and UE52-PGM) at the Berlin synchrotron radiation facility BESSY II is coupled in for photoexcitation of the clusters in the vicinity of the transition metal $L_{2,3}$ edges. 
This interaction leads to dipole-allowed transitions from atomic $2p$ core levels into unoccupied $d$ and $s$ valence states as well as to direct valence and core-level photoionization. In this process, $3d$ states are predominantly probed because of the large transition matrix element \cite{vanderLaan91, Hirsch12a}, which leads to x-ray absorption cross sections that are larger by one order of magnitude than transitions into higher $ns$ and $nd$ $(n > 3)$ states. 
This excitation scheme allows to probe the magnetic moments of iron, cobalt, and nickel clusters, which are carried by the $3d$ electrons while magnetic contributions from $4s$ and $4p$ states are negligible for clusters with more than 10 atoms \cite{AguileraGranja98} as well as in the bulk \cite{Soederlind92, AlvaradoLeyva13, Wu12b, Yuan13}.\@ 
\newline
The $2p$ core hole that is created in the x-ray absorption process relaxes via Auger decay cascades \cite{}. This leads to highly charged clusters, which disintegrate predominantly into monomer cations \cite{Lau08, Hirsch09} that are also stored in the ion trap. Bunches of parent ions and product ions are extracted from the ion trap by a pulsed exit aperture potential, and are detected by a reflectron time-of-flight mass spectrometer with a mass resolution of $m / \Delta m \approx 3000$ that operates in the inhomogeneous stray field of the superconducting solenoid and therefore is mounted in-line with the ion trap for maximum transmission. 
\newline
To obtain ion yield spectra as a measure of x-ray absorption, time-of-flight mass spectra were recorded for parallel ($\sigma^+$) and antiparallel ($\sigma^-$) orientation of the magnetic field $\mu_{0} H$ and photon helicity $\sigma$ with a total data acquisition time of $8 - 24$ s per photon energy step.
These x-ray absorption and XMCD spectra were recorded with a target density of $\approx 5 \times 10^{7}~\text{ions cm}^{-3}$, a typical photon flux of $1 - 5 \times 10^{12}$ photons per second, and a typical photon energy resolution of 250 meV.\@ All spectra were normalized to the incident photon flux, monitored with a GaAsP photodiode mounted on-axis behind the ion trap. The XMCD asymmetry was corrected for the circular polarization degree of 90\% ($P_3 = 0.9$) of the elliptically polarized soft x-ray beam. The XMCD asymmetry was typically recorded by inversion of the photon helicity, but also cross-checked for systematic errors by inverting the direction of the applied magnetic field, the results of which are identical within the error bars. 
\newline
The incident soft x-ray beam also generates parasitic helium ions by direct photoionization of the helium buffer gas. These helium ions where continuously removed from the ion trap by radio-frequency ion cyclotron resonance excitation of $\mathrm{He}^+$ in the applied magnetic field in order to prevent the parent ions from being pushed out of the trap by the high space charge on the ion trap axis.

\section{Experimental Methods}

\subsection{X-ray absorption and XMCD spectroscopy of size-selected free cluster ions}
\begin{figure}
	\includegraphics{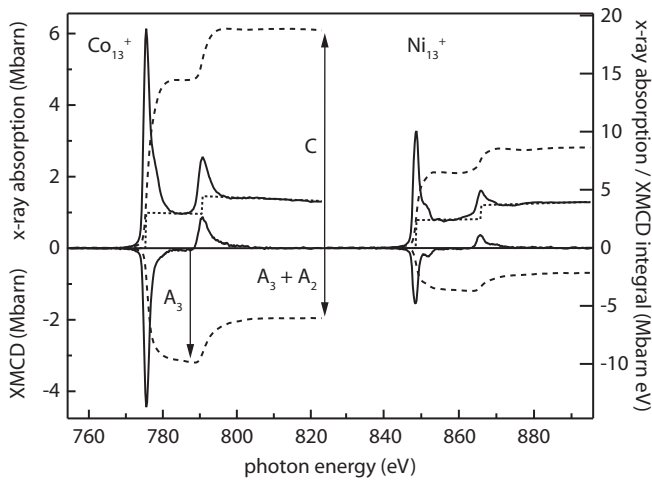}
	\caption{$L_{2,3}$-edge x-ray absorption and XMCD spectra (solid lines) of $\mathrm{Co}_{13}^+$ and $\mathrm{Ni}_{13}^+$; integrated $2p \rightarrow 3d$ x-ray absorption cross sections (dashed lines); and the two step functions (dotted lines) that approximate direct $2p$ photoionization and $2p \rightarrow ns, nd\ (n > 3)$ contributions.}
	\label{fig:Co13XMCD}
\end{figure}
Typical x-ray absorption spectra, 
\begin{equation}
	\label{eq:XAS}
	I_{\text{XAS}}= 1/2 \left( I(\sigma^+) + I(\sigma^-) \right), 
\end{equation}
XMCD spectra, 
\begin{equation}
	\label{eq:XMCD}
	I_{\text{XMCD}} = I(\sigma^+) - I(\sigma^-), 
\end{equation}
and the integrated $2p \rightarrow 3d$ x-ray absorption cross sections with respect to the photon energy are shown in Fig.\ \ref{fig:Co13XMCD} for $\mathrm{Co}_{13}^+$ and $\mathrm{Ni}_{13}^+$ clusters. The signal-to-noise ratio in this data is comparable to x-ray absorption spectra of bulk metals \cite{Chen95} even though the target density in the ion trap corresponds to only $10^{-6} - 10^{-5}$ atomic monolayers. This signal-to-noise ratio is achieved because x-ray absorption spectra of free clusters in ion yield mode are not obscured by substrate background absorption. 
\newline
For a quantitative comparison to bulk data, the continuum absorption of the x-ray absorption spectra shown in Fig.\ \ref{fig:Co13XMCD} has been scaled to the calculated cross section of the direct $2p$ photoionization $\approx 40$ eV above the $L_3$ absorption band of $1.34$ Mbarn at 815 eV for cobalt, and $1.28$ Mbarn at 890 eV for nickel \cite{Yeh85, Henke93}.\@ 
Good agreement with the absolute absorption cross section in the corresponding bulk spectra at the $L_3$ line of $\approx 6.5$ Mbarn for cobalt and $\approx 4.6$ Mbarn for nickel \cite{Chen91, Chen95, Regan01, Stoehr06} is found. Even though the line shape is almost bulk-like, the line width \cite{Lau08, Niemeyer12, Hirsch12b} is narrower for the clusters investigated here. 
\newline
In addition to the x-ray absorption spectra, two step functions are shown in Fig.\ \ref{fig:Co13XMCD}.\@ These step functions approximate the intensity $I_c$ of Rydberg ($2p \rightarrow ns, nd; n \ge 4$) and continuum ($2p \rightarrow \epsilon s, \epsilon d$) transitions that lead to the absorption edge. The intensity of these transitions is subtracted from the experimental spectra to extract only the intensity $I_{3d}$ of $2p \rightarrow 3d$ transitions, as these contain information about the magnetic properties that are carried mainly by the $3d$ electrons.
As shown by \citeauthor{Hirsch12a} \cite{Hirsch12a}, the positioning of these two step functions at the peak of the $L_{2,3}$ transitions, as it is done for the bulk \cite{Chen95}, is also justified for finite ionic systems, although for the latter case the direct $2p$ photoionization threshold that marks the onset of continuum excitations is shifted by several eV to higher photon energy. \cite{Hirsch12a, Vogel12, Bahn12} 
This approach is valid, because the x-ray absorption cross section of transitions into higher $ns, nd\ (n \ge 4)$ states is similar to that of direct photoionization from the $2p$ core levels, but these $2p \rightarrow ns, nd$ transitions do not shift relative to the $2p \rightarrow 3d$ excitation and therefore still create the step edges underneath the $L_3$ and $L_2$ lines. \cite{Hirsch12a}
\newline
In order to obtain quantitative magnetic moments, well known XMCD sum rules \cite{Thole92, Carra93, Chen95} are used to link measured x-ray absorption and XMCD spectra to the projection $m_S$ of the spin magnetic moment $\mu_S$
\begin{equation}
	\label{eq:mu_s}
	m_S = -2 \mu_B n_h \frac{A_3 - 2 A_2}{C}- 7 \frac{\mu_B}{\hbar} \left \langle T_z \right \rangle
\end{equation}
and to the projection $m_L$ of the orbital magnetic moment $\mu_L$ 
\begin{equation}
	\label{eq:mu_l}
	m_L = -\frac{4}{3}\mu_B n_h \frac{A_3 + A_2}{C}
\end{equation}
onto the direction of the magnetic field $\mu_{0} H$ at a given temperature. Here, $n_h$ is the number of unoccupied $3d$ states and $\left \langle T_z \right \rangle$ is the spin magnetic dipole term. The quantities 
\begin{equation}
	\label{eq:A3}
	A_3 = \int \limits_{L_3} I_{3d}(\sigma^+) - I_{3d}(\sigma^-) \,dE
\end{equation}
and 
\begin{equation}
	\label{eq:A2}
	A_2 = \int \limits_{L_2} I_{3d}(\sigma^+) - I_{3d}(\sigma^-) \,dE 
\end{equation}
correspond to the integrated XMCD asymmetry at the $L_3$ and $L_2$ absorption edges, respectively, and 
\begin{equation}
	\label{eq:C}
	C = 1 / 2 \int\limits_{L_3 + L_2}I_{3d}(\sigma^+) + I_{3d}(\sigma^-) \,dE 
\end{equation}
to the integrated $2p \rightarrow 3d$ x-ray absorption spectrum after step-edge correction. These integrals $A_3$, $A_2$, and $C$ are marked in Fig.\ \ref{fig:Co13XMCD} with arrows.
\begin{figure}
	\includegraphics[width=0.50\textwidth]{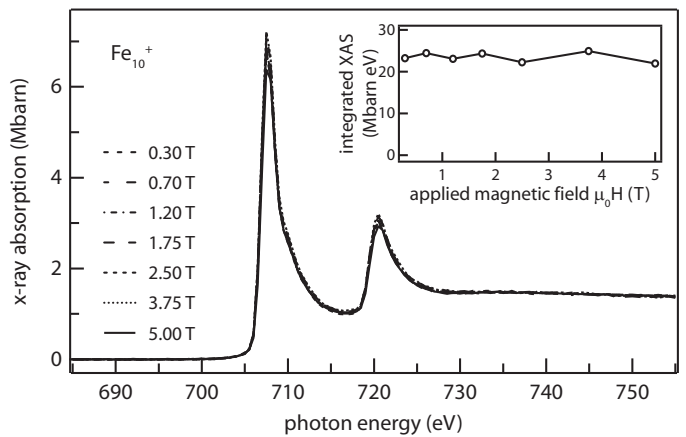}
	\caption{L$_{2,3}$-edge x-ray absorption of $\mathrm{Fe}_{10}^+$ at different applied magnetic fields. The identical line shape for $0 - 5$ T shows the absence of natural linear dichroism or spatial alignment. Integrated spectra (inset) are constant within an error of $\pm$ 6\%.}
	\label{fig:Fe10XAS}
\end{figure}
\newline
Even though $\left\langle T_z \right\rangle$ can be a significant contribution to the spin sum rule at single crystal surfaces \cite{Wu94}, there is no contribution of $\left\langle T_z \right\rangle$ to the spin sum rule for randomly oriented samples \cite{Stoehr95a}.\@ 
To illustrate the situation in free transition metal clusters, Fig.\ \ref{fig:Fe10XAS} shows a series of x-ray absorption spectra of $\mathrm{Fe}_{10}^{+}$ for applied magnetic fields ranging from $\mu_{0} H = 0.3$ up to $5.0$ T, corresponding to an alignment of the magnetic moment of 16\,\% to 91\,\% at 15 K ion temperature if a total magnetic moment of $36\ \mu_B$ is assumed.
All $\mathrm{Fe}_{10}^{+}$ x-ray absorption spectra in this series exhibit an identical line shape, independent of the applied magnetic field. 
Small differences in the peak intensities at the $L_{2,3}$ edges are caused by statistical variations but are not correlated to the magnetic field strength. The approximate absolute x-ray absorption cross sections that were obtained by scaling these spectra to the calculated $2p$ direct photoionization cross section of iron, $1.45$ Mbarn \cite{Yeh85, Henke93} at 748 eV, are in good agreement with the absolute cross section of $\approx 7$ Mbarn at the $L_3$ line of bulk iron spectra \cite{Chen95, Regan01, Stoehr06}.\@
The integrated x-ray absorption spectrum of $\mathrm{Fe}_{10}^{+}$ is constant within 6\,\%, as can be seen in the inset of Fig.\ \ref{fig:Fe10XAS}.\@ 
If $\left\langle T_z \right\rangle$ were large, i.e., if bonding and charge distribution in the clusters were strongly anisotropic, then clusters that would be spatially aligned by the magnetic field should show a strong linear dichroism in their x-ray absorption spectra. This is not the case as can be seen from Fig.\ \ref{fig:Fe10XAS}.\@ Hence, the clusters are randomly oriented in the ion trap and $\left\langle T_z \right\rangle$ averages to zero.

\subsection{Constant number $\bm{n_h}$ of unoccupied $\bm{3d}$ states in small clusters}
\begin{figure}[t]
	\includegraphics[width=0.5\textwidth]{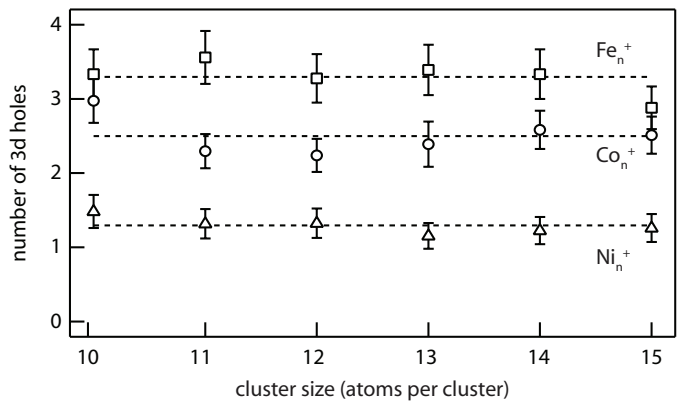}
	\caption{Experimentally determined variation in the number $n_h$ of unoccupied $3d$ states for $\mathrm{Fe}_{n}^{+}$, $\mathrm{Co}_{n}^{+}$, and $\mathrm{Ni}_{n}^{+}$ clusters. Within the error bars, $n_h$ is constant for clusters of a given element.}
	\label{fig:3dholes}
\end{figure}

\begin{table}[tb]
	\begin{ruledtabular}	
	\begin{tabular}{l c c c c}
	 & bulk & clusters & dimers & $n_h$ 
	\tabularnewline 
	 Fe 	& $3.34 - 3.44$ 	& $2.9 - 3.5$ 	& $3.25$ 	& $3.3 \pm 0.2$
	\tabularnewline
	 Co 	& $ 2.5 $ 	& $2.3 - 2.5$ 	& $2.5$ 	& $2.5 \pm 0.2$ 
	\tabularnewline
	 Ni 	& $ 1.5 $ 	& $0.7 - 1.2$ 	& $1.2$ 	& $1.3 \pm 0.2$ 
	\tabularnewline
	\end{tabular}
	\end{ruledtabular}
	\caption{\label{tab:nh}Calculated number of unoccupied $3d$ states for bulk iron, cobalt, and nickel \cite{Wu93, Wu94, Guo94}; iron ($n = 2 - 89$), cobalt ($n = 1 - 5$), and nickel ($n = 2 - 6$) clusters \cite{Basch80, Pacchioni87, Sipr05, Minar06, Gutsev12, Wu12b, AlvaradoLeyva13}; and cationic diatomic molecules \cite{Gutsev03a}.\@ In this work, we assume $n_h$ as given in the last column.}
\end{table}
The number $n_h$ of empty $3d$ states is a parameter in the XMCD sum rules \cite{Thole92, Carra93, Chen95} (Eqs.\ \ref{eq:mu_s} and \ref{eq:mu_l}) that is required to obtain the magnetic moment per atom. For the cluster ions investigated here, $n_h$ for each element has been taken from the available theoretical values for the number of unoccupied $3d$ states in iron, cobalt, and nickel bulk metals \cite{Wu93, Wu94, Guo94, Soederlind92, Daalderop94, Srivastava97, Dhesi00}, clusters \cite{Basch80, Pacchioni87, Sipr05, Minar06, Wu12b, AlvaradoLeyva13}, and diatomic molecular cations \cite{Gutsev03a}.\@ 
Calculations for iron clusters \cite{Sipr05} have shown that $n_h$ is close to the bulk value and does not undergo significant variations over a large range of cluster sizes. This also applies to calculated values of $n_h$ for cobalt and nickel clusters, which are again close to the corresponding bulk values as can be seen from Table \ref{tab:nh}.\@ We thus assume $n_h = 3.3 \pm 0.2$ for iron, $n_h = 2.5 \pm 0.2$ for cobalt, and $n_h = 1.3 \pm 0.2$ for nickel clusters. 
\newline
Experimentally, $n_h$ can be determined from the charge sum rule \cite{vanderLaan91, Pearson93, Ankudinov01, Graetz04, Stoehr95b, Stoehr06} of x-ray absorption
\begin{equation}
\label{eq:nh}
	C \propto n_h 
\end{equation}
that relates $n_h$ to the integrated x-ray absorption cross section $C$ (cf.\ Eq.\ \ref{eq:C}) of $2p \rightarrow 3d$ transitions. 
This proportionality can also be seen in Fig.\ \ref{fig:Co13XMCD}, where the integrated x-ray absorption cross section $C$ of $2p \rightarrow 3d$ transitions is 17.7 Mbarn eV for $\mathrm{Co}_{13}^{+}$, but 8.4 Mbarn eV for $\mathrm{Ni}_{13}^{+}$, reflecting the larger number of unoccupied $3d$ states $n_h = 2.5$ for cobalt than $n_h = 1.3$ for nickel clusters. 
\newline
For $\mathrm{Fe}_{n}^{+}$, $\mathrm{Co}_{n}^{+}$, and $\mathrm{Ni}_{n}^{+}$ with $n = 10 - 15$, the experimentally determined variation in the number of empty $3d$ states is shown in Fig.\ \ref{fig:3dholes}.\@ 
The values given here were determined from the charge sum rule \cite{vanderLaan91, Pearson93, Ankudinov01, Graetz04, Stoehr06}, cf.\ Eq.\ \ref{eq:nh}, by integrating the x-ray absorption spectrum, and calibrating the average integrated intensity $C$ of $2p \rightarrow 3d$ transitions (cf.\ Eq.\ \ref{eq:C}) with the $n_h$ value \cite{Basch80, Pacchioni87, Wu93, Wu94, Guo94, Gutsev03a, Minar06, Wu12b} from Table \ref{tab:nh} for a given element.
The scatter in the experimentally determined $n_h$ for different cluster sizes of the same element in Fig.\ \ref{fig:3dholes} is of the same order of magnitude as the scatter for an individual cluster at different magnetic fields as shown in the inset of Fig.\ \ref{fig:Fe10XAS} for the case of $\mathrm{Fe}_{10}^{+}$.\@ 
This implies that the observed scatter in the number of unoccupied $3d$ states is due to experimental uncertainties rather than to a real variation of $n_h$, and the experimental data thus indeed reveal a nearly constant $n_h$ within the error bars for a given element in the cluster size range considered here. 
\newline
The experimental proportionality constant of $6.4 - 7.3$ Mbarn eV per empty $3d$ state is similar for iron, cobalt, and nickel clusters, but is only $\approx 2/3$ of the reported bulk value \cite{Stoehr06} in our data. 
This might be related to the partial ion yield detection technique, which is a good approximation of, but not identical to, x-ray absorption. Since $2p \rightarrow 3d$ transitions and direct $2p$ photoionization lead to charge states of the core-excited cluster that differ by 1, the product ion distribution after multiple Auger decay and fragmentation will not be identical for both excitation channels. Direct $2p$ photoionization typically leads to product ions with a larger charge-to-mass ratio $q / m$ than $2p \rightarrow 3d$ transitions. Since all spectra shown here were recorded on the dominant singly charged metal ion, the continuum step edge might be slightly overestimated in the x-ray absorption spectra shown here. When scaling the direct $2p$ photoionization to calculated subshell photoionization cross sections \cite{Yeh85, Henke93}, the absolute absorption cross sections of $2p \rightarrow 3d$ transitions, given in Figs.\ \ref{fig:Co13XMCD} and \ref{fig:Fe10XAS}, would thus be underestimated. However, the partial ion yield detection scheme only changes the ratio of $2p \rightarrow 3d$ transitions, $I_{3d}(\sigma^+) + I_{3d}(\sigma^-)$ to direct photoionization, $I_c$, but does not alter the relative intensities of $I_{3d}(\sigma^+)$ and $I_{3d}(\sigma^-)$ which enter into the XMCD sum rules via Eqs.\ \ref{eq:A3} to \ref{eq:C} after the step edge $I_c$ has been subtracted. Hence, the magnetic moments derived from the XMCD sum rules are not affected by partial ion yield detection.

\subsection{Magnetization curves, magnetic moments, and electronic temperature}
\begin{figure}
	\includegraphics[width=0.50\textwidth]{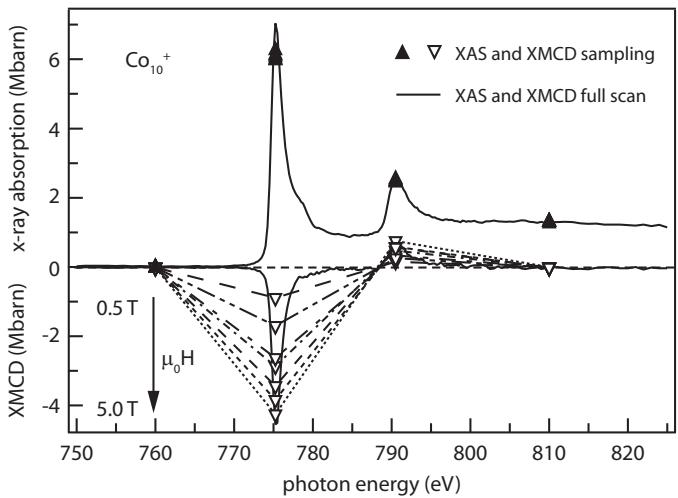}
	\caption{X-ray absorption and XMCD spectra taken at $\mu_{0} H = 5$ T (solid lines) and low resolution peak intensity scans (triangles) of the XMCD and x-ray absorption signal at $\mu_{0} H = 0.5 - 5.0$ T, from which magnetization curves are derived.}
	\label{fig:Co10_4point}
\end{figure}
It is important to recall that XMCD spectroscopy, just like Stern-Gerlach experiments \cite{}, does not determine absolute magnetic moments $\mu$ of size selected clusters, but measures the magnetization $m$ as a response to an applied magnetic field at a given temperature. The magnetic moment is then deduced from this magnetization.
For a precise determination of magnetic moments and temperature of the clusters, magnetization curves have been measured for $\mathrm{Fe}_{13}^+$, $\mathrm{Co}_{10-15}^+$, and $\mathrm{Ni}_{13}^+$ by recording the magnetization as a function of the applied magnetic field $\mu_{0} H$ at constant temperature, where the magnetic field, applied along the $z$ direction, acts on the total magnetic moment $\mu_J = g\, \mu_B / \hbar\, J_z$ of the cluster, with $\left< J_z \right> = \left< L_z \right> + \left< S_z \right>$ in $LS$ coupling, where $\left< L_z \right>$ and $\left< S_z \right>$ are derived from XMCD sum rules for a given applied magnetic field. 
\newline
In practice, magnetization curves have been derived by sampling and reconstructing the full XMCD and x-ray absorption spectra at the $L_{2,3}$ peak intensities with low photon energy resolution of 1250 meV, as illustrated in Fig.\ \ref{fig:Co10_4point}.\@ 
Since natural dichroism has not been observed in these spectra at different magnetic fields, i.e., the line shape does not change as shown in Fig.\ \ref{fig:Fe10XAS}, the integrated XMCD and x-ray absorption cross sections ($A_3$, $A_2$, and $C$ in Fig.\ \ref{fig:Co13XMCD}) are proportional to the peak intensities of the $L_{2,3}$-edges of the XMCD (open triangles in Fig.\ \ref{fig:Co10_4point}) and x-ray absorption (solid triangles in Fig.\ \ref{fig:Co10_4point}) curves that are repeatedly sampled at four different photon energies for 60 s each. 
As expected, the XMCD signal diminishes with lower magnetic fields while the x-ray absorption intensities remain constant, confirming stable and reproducible experimental conditions along the acquisition of a complete magnetization curve. The sum and difference of the sampled $L_3$ and $L_2$ peak intensities at $\mu_{0} H = 5$ T were scaled to the corresponding integrals $A_3$, $A_2$, and $C$ of full spectra measured at $\mu_{0} H = 5$ T, shown as a solid line in Fig.\ \ref{fig:Co10_4point}, for the application of XMCD sum rules. The same scaling factors were then used to convert peak intensities into integrals $A_{1}'$, $A_{2}'$, and $C'$ for magnetic fields $\mu_{0} H < 5$ T.\@
\begin{figure}
	\includegraphics[width=0.50\textwidth]{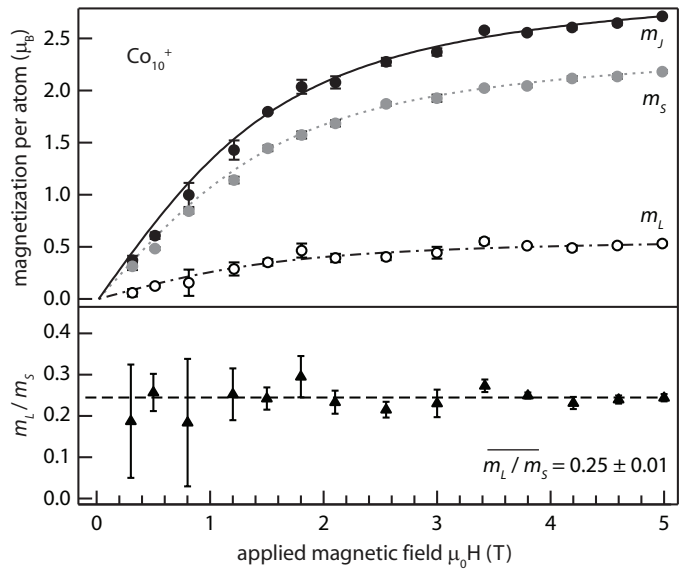}
	\caption{Upper panel: Total ($m_J$, filled circles), spin ($m_S$, light gray circles), and orbital ($m_L$, open circles) magnetization of $\mathrm{Co}_{10}^+$ as a function of the applied magnetic field. A Brillouin fit to the total magnetization $m_J$ (solid line) yields the cluster ion temperature $T_{\,\text{ion}}=14.6 \pm 0.3$ K and the total magnetic moment $\mu_J = 3.15 \pm 0.09$ $\mu_B$ per atom. The same Brillouin fit is scaled to $m_S$ and $m_L$ as a guide to the eye. 
\newline
Lower panel: The average ratio of $m_L / m_S = \mu_L / \mu_S = 0.25 \pm 0.01$ (filled triangles and dashed line) is constant within the error bars and is used to decompose $\mu_J$ into $\mu_S$ and $\mu_L$. The error bars increase for smaller applied magnetic fields because of decreasing magnetization.}
	\label{fig:Co10MC}
\end{figure}
\newline
As an example, the magnetization curve $m_J \left(\mu_{0} H\right)$ of $\mathrm{Co}_{10}^+$ is shown in Fig.\ \ref{fig:Co10MC} in the 0 to 5 T range. A fit to the experimental data with a Brillouin function, shown as a solid line in the upper panel of Fig.\ \ref{fig:Co10MC}, yields the electronic cluster ion temperature $T_{\,\text{ion}}$ and the total magnetic moment $\mu_J$ per atom as independent fit parameters. 
The total magnetization $m_J$ in Fig.\ \ref{fig:Co10MC} is also decomposed into spin and orbital contributions $m_S$ and $m_J $ for illustration, although $\mu_{0} H$ only acts on $J$ but neither on $L$ nor on $S$ separately. The corresponding spin and orbital magnetic moments $\mu_S$ and $\mu_L$ per atom are derived from a decomposition of the total magnetic moment $m_J$ according to the mean value of the experimentally determined ratio $m_L / m_S$ of orbital to spin magnetization that is recorded for a given cluster at different applied magnetic fields, as is shown for $\mathrm{Co}_{10}^+$ in the lower panel of Fig.\ \ref{fig:Co10MC}.\@ 
\newline
The analysis of the magnetization curves does not only yield the magnetic moments $\mu_J$, $\mu_S$, and $\mu_L$ but also the electronic temperature $T_{\,\text{ion}}$ of the cluster ions. To the best of our knowledge, this is the only method so far to determine the electronic temperature of isolated systems in an ion trap \cite{Niemeyer12}.\@ The typical electronic cluster temperature under our experimental conditions is $T_{\,\text{ion}} = 12 \pm 3$ K, and is increased by $7 \pm 3$ K over the ion trap electrode temperature $T_{\text{trap}} = 5 \pm 1$ K because of radio frequency heating in the presence of a helium buffer gas \cite{Gerlich92}.\@ 
The electronic temperature corresponds to the vibrational or rotational cluster ion temperature \cite{Moriwaki92, Gronert98, Gerlich09, Asvany09, Otto13} because these degrees of freedom are in thermal equilibrium by multiple collisions with the helium buffer gas. 

\section{Results}

\subsection{Spin and orbital magnetic moments}
\begin{figure}[t]
	\includegraphics[width=0.50\textwidth]{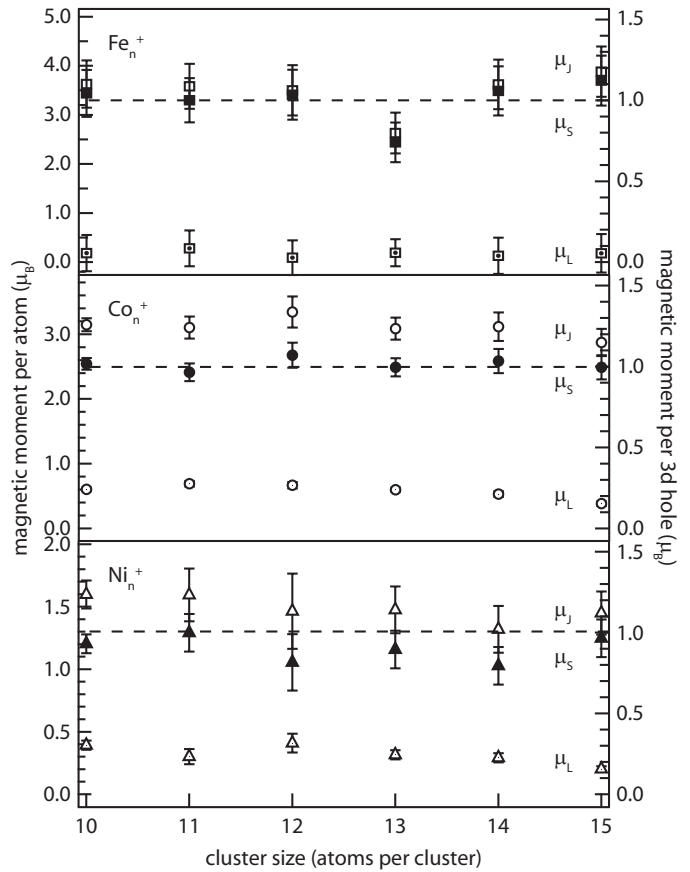}
	\caption{Total ($\mu_J$, open symbols), spin ($\mu_S$, filled symbols), and orbital ($\mu_L$, open symbols with dot) magnetic moments per atom and per unoccupied $3d$ state of iron, cobalt, and nickel clusters. With the exception of $\mathrm{Fe}_{13}^+$, all clusters carry $\approx 1$ $\mu_B$ per $3d$ hole.}
 	\label{fig:FeCoNi}
\end{figure}

\begin{table*}[bt]
	\begin{ruledtabular}
	\begin{tabular}{l c c c c c c c c c}
 & $n_e$ & $\mu_S$ & $\mu_L$ & $\mu_S^{\text{cl}}$ & $\mu_L^{\text{cl}}$ & $2S$ & $L$ & $2S^{\text{\,theo}}$ & $L^{\text{\,theo}}$
\tabularnewline
 & & $(\mu_B)$ & $(\mu_B)$ & $(\mu_B)$ & $(\mu_B)$ & ($\hbar$) & ($\hbar$) & ($\hbar$) & ($\hbar$)
\vspace{2mm}
\tabularnewline
 $\mathrm{Co}_{10}^+$ & 89 & $2.54 \pm 0.09$ & $0.61 \pm 0.03$ & $25.4 \pm 0.9$ & $6.1 \pm 0.3$ & {\bf{25}} & 6
\tabularnewline
 Co$_{11}^+$ & 98 & $2.41 \pm 0.14$ & $0.69 \pm 0.04$ & $26.5 \pm 1.5$ & $7.6 \pm 0.5$ & 26; {\bf{28}} & 8
\tabularnewline
 Co$_{12}^+$ & 107 & $2.68 \pm 0.19$ & $0.67 \pm 0.05$ & $32.1 \pm 2.3$ & $8.0 \pm 0.6$ & {\bf{31}}; 33 & 8 & 21 \footnotemark[1]
\tabularnewline
 $\mathrm{Co}_{13}^+$ & 116 & $2.49 \pm 0.14$ & $0.60 \pm 0.03$ & $32.4 \pm 1.8$ & $7.8 \pm 0.4$ & {\bf{32}}; 34 & 8 & 26 \footnotemark[1]
\tabularnewline
 Co$_{14}^+$ & 125 & $2.58 \pm 0.19$ & $0.53\ \pm 0.04$ & $36.2 \pm 2.6$ & $7.4 \pm 0.6$ & {\bf{35}}; 37 & 7; 8
\tabularnewline
 Co$_{15}^+$ & 134 & $2.49 \pm 0.19$ & $0.38 \pm 0.04$ & $37.3 \pm 2.8$ & $5.7 \pm 0.5$ & 36; {\bf{38}}; 40 & 6
\vspace{2mm}
\tabularnewline
 $\mathrm{Fe}_{10}^+$ & 79 & $3.19 \pm 0.19$ & $0.26 \pm 0.04$ & $31.9 \pm 1.9$ & $2.6 \pm 0.4$ & 31; {\bf{33}} & 3 & 29 \footnotemark[1] \footnotemark[2] & $1.4 - 1.8$ \footnotemark[2] 
\vspace{2mm}
\tabularnewline
 $\mathrm{Ni}_{13}^+$ & 129 & $1.20 \pm 0.14$ & $0.32 \pm 0.06$ & $15.6 \pm 1.9$ & $4.2 \pm 0.8$ & 15; {\bf{17}} & 4; 5 & 9 \footnotemark[1] & $3.4 - 4.2$ \footnotemark[3] 
\tabularnewline
	\end{tabular}
	\end{ruledtabular}
	\footnotetext[1]{from Ref.~\onlinecite{Gutsev13}.} 
	\footnotetext[2]{from Ref.~\onlinecite{Yuan13}.}
	\footnotetext[3]{from Ref.~\onlinecite{GuiradoLopez03}.}
	\caption{Total number $n_e$ of valence electrons per cluster; experimentally determined spin and orbital magnetic moments per atom, $\mu_S$ and $\mu_L$, and per cluster, $\mu_S^{\text{cl}}$ and $\mu_L^{\text{cl}}$; spin multiplicities $2S$ and orbital angular momenta $L$ per cluster that are compatible with the experimental error bars, compared to available theoretical values $2S^{\text{\,theo}}$ and $L^{\text{\,theo}}$ of the total spin \cite{Gutsev13, Yuan13} and orbital \cite{GuiradoLopez03, Yuan13} angular momentum. Potential contributions of $4s$ and $4p$ electrons to the experimental spin magnetic moment $m_S$ are neglected. Values of $2S \approx n \cdot n_h$ are marked in bold face.}
	\label{tab:ms}
\end{table*}
In Fig.\ \ref{fig:FeCoNi}, the total, spin, and orbital magnetic moments are presented for iron \cite{Niemeyer12}, cobalt, and nickel clusters in the size range of $n = 10 - 15$.\@ 
For cobalt, magnetic moments have been determined from magnetization curves for all cluster sizes, which leads to the smallest error. 
Except for $\mathrm{Fe_{10}^+}$ and $\mathrm{Ni_{13}^+}$, the data for iron and nickel clusters were taken at a fixed ion trap temperature and fixed magnetic field of 5 T.\@ To obtain magnetic moments, the total magnetization that was obtained from the XMCD analysis under these conditions was Brillouin corrected. 
The error bars in Fig.\ \ref{fig:FeCoNi} correspond to statistical intensity variations of the XMCD spectra or, as for cobalt clusters, from Brillouin fits to magnetization curves. Systematic errors, e.g.\ uncertainties in the number of unoccupied $3d$ states ($\Delta \mu_J \approx 0.2$ $\mu_B$) and in the position of the two step functions ($\Delta \mu_J \approx 0.1 - 0.2$ $\mu_B$) sum up to $0.3$ $\mu_B$ per atom for cobalt and $0.4$ $\mu_B$ per atom for iron and nickel clusters but are omitted from Fig.\ \ref{fig:FeCoNi}.\@
\newline 
Table \ref{tab:ms} lists the average experimental $3d$ spin and orbital magnetic moments per atom and per cluster. Even though XMCD spectroscopy at the $L_{2, 3}$ edges of $3d$ transition elements is only sensitive to the $3d$ magnetic moments, these values will be close to the total magnetic moments because $4s$ and $4p$ electron contributions to the spin magnetic moment are expected to be negligible \cite{Soederlind92, AguileraGranja98, AlvaradoLeyva13, Yuan13, Wu12b}.\@ 
\newline
Within the range defined by the error bars of these experimental magnetic moments, only integral values for the spin multiplicity $2S$ and the orbital angular momentum $L$ per cluster are allowed because the angular momentum that leads to the magnetic moment is quantized. 
A further constraint on the spin magnetic moment is imposed by the total number of valence electrons $n_e$ per cluster, which defines whether the spin multiplicity of teh cluster is odd or even. For $\mathrm{Fe}_{n}^+$, $\mathrm{Co}_{n}^+$, and $\mathrm{Ni}_{n}^+$ with $3d^6\,4s^2$, $3d^7\,4s^2$, and $3d^8\,4s^2$ atomic configurations, the total number of electrons per cluster is listed in Table \ref{tab:ms} as $n_e = 8 n - 1$, $n_e = 9 n - 1$, and $n_e = 10 n - 1$, respectively. 
The values of $2S$ and $L$ of iron, cobalt, and nickel clusters that are compatible with these constraints are also listed in Table \ref{tab:ms}.\@
Note that for the determination of $\mu_S$ and $\mu_L$, the number of empty $3d$ states, given in Table \ref{tab:nh}, is assumed to be constant, even though $n_h$ will have to vary slightly with cluster size under the constraints of an integral number of atoms per cluster $n$ and integer total spin multiplicity $2S$ for a strong ferromagnet without $sp$ contributions to the total magnetic moment. This variation is below the error bars of the experimentally determined $n_h$ data that is shown in Fig.\ \ref{fig:3dholes}.\@ The values of $2S \approx n \cdot n_h$, which are close to a pattern where each atom contributes on average $1/2\, n_h$ $\hbar$ to the total spin, are marked in bold face in Table \ref{tab:ms}.\@ 
\newline
From these numbers, the exact spin state can be given experimentally for $\mathrm{Co}_{10}^+$ as $2 S = 25$, while two different spin states are possible within the experimental error bars for $\mathrm{Co}_{11}^+$ to $\mathrm{Co}_{14}^+$; $\mathrm{Fe}_{10}^+$, and $\mathrm{Ni}_{13}^+$.\@ For $\mathrm{Co}_{15}^+$, three different spin states are compatible with the experimental data. 
As can be seen from the available theoretical values, also included in Table \ref{tab:ms}, the spin magnetic moments seem to be underestimated by DFT in comparison to the experimental data for iron, cobalt, and nickel, with a difference of $0.4 - 0.8$ $\mu_B$ per atom. In this respect, the experimental data presented here can serve as a benchmark for further development of theoretical methods to predict spin and orbital magnetic moments.

\subsection{Magnetic anisotropy energy of transition metal clusters}
The magnetic anisotropy energy $E_{\mathrm{MAE}}$ is a consequence of spin--orbit and orbit--lattice coupling, and is responsible for the preferred alignment of the magnetic moment along easy crystallographic axes. It is therefore an important property of magnetic materials, in particular in view of technological applications. For transition metal dimers, large magnetic anisotropy energies in the $1-10$ meV range have been predicted \cite{Strandberg07, Fritsch08}, while transition metal clusters of $\approx 1000$ atoms were found experimentally \cite{Jamet04} to have magnetic anisotropy energies that are comparable to or even lower than the bulk values. 
Even though we do not determine $E_{\mathrm{MAE}}$ directly, we can give an estimate of the upper limit of $E_{\mathrm{MAE}}$ in small iron, cobalt, and nickel clusters from energy considerations \cite{Niemeyer12} and from the absence of natural linear dichroism in the x-ray absorption spectra taken at different magnetic fields $\mu_{0} H$, as presented in Fig.\ \ref{fig:Fe10XAS} for the case of $\mathrm{Fe}_{10}^+$.\@ The absence of linear dichroism is due to a random orientation of the clusters inside the ion trap. This implies that the magnetic anisotropy energy $E_{\text{MAE}}$ is smaller than the thermal (rotational) energy $E_{\text{rot}} = 1/2 \,k_B T$ per degree of freedom \cite{Niemeyer12}.\@ Under these conditions, the total magnetic moment of the cluster can overcome the anisotropy barrier to rotate freely and align with the applied magnetic field, independent of the spatial cluster orientation. This leads to the observed superparamagnetic \cite{Bean59, Billas97} behavior. 
\newline
With the knowledge of the cluster ion temperature from magnetization curve thermometry, we can deduce an upper limit of the magnetic anisotropy energy, 
\begin{equation}
	\label{eq:MAE}
	E_{\text{MAE}} < 1/2\, k_B T_{\,\text{ion}}.
\end{equation}
Since this upper limit on $E_{\mathrm{MAE}}$ is mainly determined by the cluster ion temperature, $E_{\mathrm{MAE}}$ is most likely overestimated. Therefore, Eq.\ \ref{eq:MAE} yields $E_{\text{MAE}} \ll 52$ $\mu$eV per atom for a fixed rotation axis of $\mathrm{Co}_{10}^+$ and $\mathrm{Ni}_{10}^+$ clusters. 
For $\mathrm{Fe}_{10}^+$, a slightly increased upper limit of $E_{\text{MAE}} \ll 65$ $\mu$eV per atom is estimated \cite{Niemeyer12} from Eq.\ \ref{eq:MAE} because of different experimental conditions with higher cluster ion temperature. 
These experimental upper limits on $E_{\mathrm{MAE}}$ are close to the magnetic anisotropy energy of 60 $\mu$eV for hexagonal close-packed cobalt \cite{Stiles01, Yang01} with a strongly preferred $c$ axis, which leads to an anisotropy energy that is much larger than $E_{\mathrm{MAE}} = 1.4$ $\mu$eV in body-centered cubic iron and $E_{\mathrm{MAE}} = 2.8$ $\mu$eV in face-centered cubic nickel\cite{Stiles01, Yang01}.\@ 
For nearly spherical and compact clusters in the size range considered here, we expect $E_{\mathrm{MAE}}$ to be smaller than in crystalline bulk because of the absence of strongly preferred axes. For the same reason, we also expect magnetic anisotropy energies in iron, cobalt, and nickel clusters to be closer to each other than in bulk, because the difference in $E_{\mathrm{MAE}}$ should mainly be due to the amount of orbital magnetism. This would lead to $E_{\mathrm{MAE}} \left( \text{Co} \right) > E_{\mathrm{MAE}} \left( \text{Ni} \right) \ge E_{\mathrm{MAE}} \left( \text{Fe} \right)$ from the data presented in Table \ref{tab:ms}.\@ DFT calculations by \citeauthor{AlvaradoLeyva13} for the cationic $\mathrm{Fe}_{13}^+$ cluster obtain a magnetic anisotropy energy of $E_{\mathrm{MAE}} = 66$ $\mu$eV per cluster or 5 $\mu$eV per atom \cite{AlvaradoLeyva13}, which is close to the bulk value and consistent with our interpretation of the experimental data. 
\newline
In contrast to deposited clusters or adatoms, where the magnetic anisotropy energy is in the meV range \cite{Gambardella03, Lehnert10} and is strongly enhanced by the two-dimensional geometry or the interaction with the support that induces large orbital magnetic moments, the intrinsic magnetic anisotropy energy is about two to three orders of magnitude lower in free clusters. 

\section{Discussion}

\subsection{Strong ferromagnetism in $\bm{\mathrm{Fe}_{n}^{+}}$, $\bm{\mathrm{Co}_{n}^{+}}$, and $\bm{\mathrm{Ni}_{n}^{+}}$ clusters}
Fig.\ \ref{fig:FeCoNi} displays the measured total, spin, and orbital magnetic moments per atom and per unoccupied $3d$ state. 
As can be seen, all clusters except for $\mathrm{Fe}_{13}^+$ exhibit a completely filled $3d$ majority spin band that, within the error bars, leads to a spin magnetic moment of $\mu_S = 1$ $\mu_B$ per unoccupied $3d$ state, marked by the dashed lines in Fig.\ \ref{fig:FeCoNi}.\@ 
These maximized spin magnetic moments reveal a strong ferromagnetism in iron, cobalt, and nickel clusters, which is in contrast to the bulk metals. For comparison, the majority spin bands of bulk iron, cobalt, and nickel are not completely filled but the bulk spin magnetic moments correspond to approximately 0.62, 0.63, and 0.38 $\mu_B$ per $3d$ hole \cite{Soederlind92, Wu94, Stoehr06}, respectively. 
Interestingly, the total number of empty $3d$ states (cf. Table \ref{tab:nh} and Fig.\ \ref{fig:3dholes}) is predicted \cite{Sipr05} to be nearly identical in small clusters and in the bulk, which indicates a similar amount of $spd$ hybridization, but only the distribution of $3d$ holes over the majority and minority states is different.
The enhancement of the spin magnetic moments of clusters over the bulk values would thus indeed be caused by a redistribution of electrons from minority into majority $3d$ states because of a narrower $3d$ band as a consequence of the lower average coordination in clusters \cite{Apsel96, Eriksson92, Liu89}, where most of the atoms reside at the surface in the size range considered here.

\subsection{Reduced spin magnetic moment of $\bm{\mathrm{Fe}_{13}^{+}}$}
\begin{figure}[t]
	\includegraphics[width=0.50\textwidth]{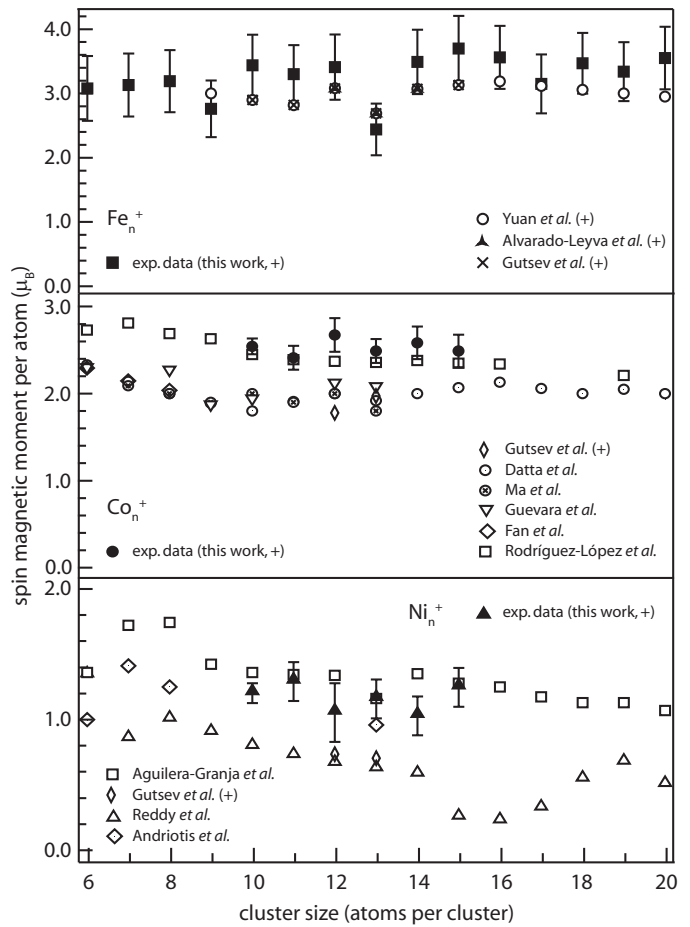}
	\caption{Calculated spin magnetic moments (open symbols) of iron, cobalt, and nickel clusters \cite{Wu12b, Dieguez01, Dunlap90, RodriguezLopez03, Gutsev13, Datta07, Ma06a, Guevara99, Fan97, AguileraGranja98, Reddy98, Andriotis98} compared to spin magnetic moments, derived in this XMCD study (filled symbols). The $+$ symbol denotes results for cationic clusters. Systematic errors are not shown.}
	\label{fig:Co_calc_moments}
\end{figure}

$\mathrm{Fe}_{13}^{+}$ is the only exception from this general finding of strong ferromagnetism. As reported previously \cite{Niemeyer12}, $\mathrm{Fe}_{13}^+$ has an average spin magnetic moment of $\mu_S = 2.4 \pm 0.4$ $\mu_B$ per atom, much closer to the bulk value than any other cluster in this size range. 
Since the number $n_h$ of $3d$ holes is nearly constant for $\mathrm{Fe}_{n}^{+}\ (n = 10 - 15)$, as has been shown experimentally in Fig.\ \ref{fig:3dholes}, the average spin magnetic moment is only $0.73 \pm 0.12$ $\mu_B$ per $3d$ hole in $\mathrm{Fe}_{13}^{+}$.\@ 
This value is significantly reduced by $\Delta \mu_S = 1.0 \pm 0.5$ $\mu_B$ per atom in comparison to adjacent cluster sizes.
In contrast to iron, the situation is different for cobalt and nickel clusters. Here, our data give spin magnetic moments of $\mathrm{Co}_{13}^+$ and $\mathrm{Ni}_{13}^+$ that do not show a significant reduction when compared to neighboring cluster sizes. 
\newline
There is a controversy about the origin of the reduced spin magnetic moment of $\mathrm{Fe}_{13}^{+}$ in the literature. DFT models of cationic $\mathrm{Fe}_{13}^{+}$ find either an antiferromagnetically coupled spin magnetic moment of the central atom \cite{AlvaradoLeyva13, Yuan13} or a ferromagnetic alignment of the $3d$ spins at all atoms but a nearly quenched spin magnetic moment of the central atom \cite{Wu12b, Gutsev12}.\@ In spite of these differences, all these studies predict the same average magnetic moment of $\mu_S = 2.69$ $\mu_B$ per atom \cite{Wu12b, Gutsev12, AlvaradoLeyva13, Yuan13} for $\mathrm{Fe}_{13}^{+}$ and reproduce our experimental value of $\mu_S = 2.4 \pm 0.4$ $\mu_B$ within the error bars, cf.\ Fig.\ \ref{fig:Co_calc_moments}, even though the modeled reduction of $\Delta \mu_S = 0.4$ $\mu_B$ per atom for $\mathrm{Fe}_{13}^{+}$ in comparison to $\mathrm{Fe}_{12}^{+}$ and $\mathrm{Fe}_{14}^{+}$ is less pronounced in the DFT calculations \cite{Wu12b, Gutsev12, Yuan13} than in the experiment. 
Not only for $\mathrm{Fe}_{13}^{+}$ but also for other $\mathrm{Fe}_{n}^{+}$ clusters, all existing calculations \cite{Wu12b, Gutsev12, AlvaradoLeyva13, Yuan13} determine identical spin magnetic moments as can be seen from Fig.\ \ref{fig:Co_calc_moments}, although the geometric and electronic ground state structures of \citeauthor{Wu12b} \cite{Wu12b} and \citeauthor{Gutsev12} \cite{Gutsev12} differ from those found by \citeauthor{AlvaradoLeyva13} \cite{AlvaradoLeyva13} and \citeauthor{Yuan13} \cite{Yuan13}\@ These distinctions in the calculated geometric and electronic structures lead to different explanations about the origin of the reduced spin moment of $\mathrm{Fe}_{13}^+$: While \citeauthor{Wu12b} conclude a symmetry-driven spin quenching in a highly-symmetric icosahedral structure, \citeauthor{AlvaradoLeyva13} report a $T_h$ distorted icosahedral structure of $\mathrm{Fe}_{13}^+$ in combination with an electronic shell closure that leads to antiferromagnetic spin coupling \cite{AlvaradoLeyva13}.\@
\newline
It should be noted that the results of our XMCD study yield average spin and orbital magnetic moments but do not give direct access to the relative orientation of spins. Therefore, both possible configurations, i.e., antiferromagnetic \cite{AlvaradoLeyva13, Yuan13} or ferromagnetic with nearly quenched spin of the central atom \cite{Wu12b, Gutsev12}, are compatible with our experimental results within the error bars of the spin magnetic moment. 
Indirect evidence for antiferromagnetic spin coupling in $\mathrm{Fe}_{13}^+$ might be obtained from the reduced number $n_h = 2.92$ of unoccupied $3d$ states that is predicted for $\mathrm{Fe}_{13}^+$ in the studies of \citeauthor{Wu12b} and \citeauthor{Gutsev12} in contrast to $n_h = 3.34$ predicted by \citeauthor{AlvaradoLeyva13} in their study. Even though a slight reduction of $n_h$ in $\mathrm{Fe}_{13}^+$ cannot be excluded within the experimental error bars given in Fig.\ \ref{fig:3dholes}, we do not see experimental evidence for a strong deviation of $\mathrm{Fe}_{13}^+$ from the constant number $n_h \approx 3.3$ of unoccupied $3d$ states in iron clusters \cite{Minar06}.\@
\newline
Also for neutral iron clusters, DFT modeling \cite{Dieguez01, BobadovaParvanova02a} in some cases has assigned a reduction in the average magnetic moment of neutral $\mathrm{Fe}_{13}$ to an antiferromagnetic spin coupling of the central atom to the shell atoms, and has demonstrated that the antiferromagnetic state becomes more stable as the interatomic distance decreases \cite{BobadovaParvanova02a, Singh08}.\@ Interestingly, more recent calculations \cite{Rollmann06, Wu12b, Gutsev12, Yuan13} rather find a larger spin magnetic moment in neutral $\mathrm{Fe}_{13}$ than in $\mathrm{Fe}_{12}$ or $\mathrm{Fe}_{14}$ by $0.23 -0.24$ $\mu_B$ per atom, which is increased because of larger $4s$ contributions to the magnetic moment \cite{Wu12b, Gutsev12, Yuan13} in the calculations. 
\newline
Similar to the antiferromagnetic alignment of the central atom in several studies \cite{Dieguez01, BobadovaParvanova02a} of neutral $\mathrm{Fe}_{13}$, the reduction of the average spin magnetic moment in cationic $\mathrm{Fe}_{13}^+$ is correlated with a compression of the interatomic distances by $2 - 3$\,\% when compared to the modeled ferromagnetic ground state of neutral $\mathrm{Fe}_{13}$ as well as to cationic $\mathrm{Fe}_{12}^+$ and $\mathrm{Fe}_{14}^+$ in DFT modeling \cite{Dunlap90, Singh08, BobadovaParvanova02a, AlvaradoLeyva13, Yuan13, Gutsev12, Wu12b}.\@ This reduction in bond length corresponds to a spherical volume compression of $V/V_0 = 0.91 - 0.94$ and might be correlated to the magnetic properties via the bulk phase diagram of iron, where a transition from the ferromagnetic body-centered cubic (bcc) $\alpha$ phase to the nonmagnetic hexagonal close packed (hcp) $\epsilon$ phase of iron occurs around $12-16$ GPa at room temperature \cite{Bancroft56, Jamieson62, Taylor91, Wang98b, Rueff99, Mathon04}.\@ 
This transition occurs around a volume compression of $V/V_0 = 0.94$ \cite{Wilburn78, Iota07, Dewaele08}, which corresponds to the calculated compression ratio in $\mathrm{Fe}_{13}^+$ \cite{Dunlap90, Singh08, BobadovaParvanova02a, AlvaradoLeyva13, Yuan13, Gutsev12, Wu12b}, even though the modeled $\mathrm{T_h}$ symmetry of $\mathrm{Fe}_{13}^+$ \cite{AlvaradoLeyva13, Wu12b, Yuan13, Gutsev12}can be interpreted as a distorted fcc \cite{AlvaradoLeyva13} rather than hcp structure. 
The face-centered cubic (fcc) $\gamma$ phase of bulk iron is stable only above the Curie temperature of $\alpha$ iron and is paramagnetic, but an antiferromagnetic fcc phase \cite{Abrahams62} of iron with a N\'{e}el temperature of 67 K can be stabilized in iron precipitates \cite {Gonser63, Johanson70, Onodera94}.\@ Complex antiferromagnetic order in thin films of fcc iron grown on copper (001) has also been observed experimentally \cite{Keune77, Himpsel91a, Li94, Meyerheim09}.\@
\newline
In bulk cobalt, a comparable transition from the ferromagnetic hcp $\epsilon$ phase to the nonmagnetic fcc $\beta$ phase occurs at much higher pressure of $80 - 130$ GPa and higher volume compression of $V/V_0 \approx 0.75$ \cite{Yoo98, Yoo00, Goncharov04, Iota07, Ishimatsu07, Antonangeli08, Ishimatsu11, Torchio11a}.\@ When approximating $\mathrm{Co}_{13}^+$ as a sphere, this volume compression would correspond to a large radial bond compression of $\approx 10$\% that is not expected to occur in $\mathrm{Co}_{13}^+$ and might be a hint at why no reduction of the spin magnetic moment is observed in this case. 
Similarly, no phase transition has been observed in bulk nickel up to a pressure of 200 GPa and a volume compression of $V/V_0 \approx 0.65$ \cite{Iota07, Ishimatsu07, Torchio11b}, where ferromagnetic fcc nickel is still stable. Therefore, neither a change of magnetic order nor a reduction of the magnetic moment of the central atom by bond compression is likely to occur in $\mathrm{Ni}_{13}^+$.\@
This way, the bulk phase diagrams elucidate why the peculiar behavior of antiferromagnetic alignment or spin quenching at the central atom of the thirteen-atom cluster is observed only for iron but does not show up in cobalt or nickel clusters, although the bulk phase diagrams do not explain what mechanism would be responsible for such a spontaneous bond contraction and change of magnetic order. 
\newline
Even for iron, this unexpected magnetic behavior is only observed for $\mathrm{Fe}_{13}^+$.\@ This finding might either be related to electronic shell closure in the spin-up and spin-down states, or to the geometric structure with one central atom and 12 shell atoms. 
The modeled electronic structure and geometry of $\mathrm{Fe}_{13}^+$ should, however, be taken with caution: Even though many likely geometric structures have been explored in electronic structure calculations \cite{Wu12b, Gutsev12, AlvaradoLeyva13, Yuan13}, no unbiased global geometry optimization has been performed for cationic iron clusters, to the best of our knowledge.
Apart from adsorption or titration studies \cite{Parks88, Parks92, Parks93a} that cannot give detailed information on interatomic distances, no experimental structure determination has been reported for free iron, cobalt, and nickel clusters in this size range so far. 
The experimental bond energies of cationic iron clusters show a local maximum \cite{Lian92a} at $\mathrm{Fe}_{13}^+$, which is in agreement with its enhanced abundance in mass spectrometry \cite{Sakurai99} studies. The ionization potentials of neutral iron clusters also show a pronounced maximum around $\mathrm{Fe}_{13}$ and $\mathrm{Fe}_{14}$ \cite{Rohlfing84, Yang90, Parks90}.\@
To test the calculated bond length compression in $\mathrm{Fe}_{13}^+$ experimentally, trapped ion electron diffraction \cite{Rapps13}, infrared multiphoton dissociation \cite{Gehrke09}, or extended x-ray absorption fine structure studies \cite{Kakar97} would be required with their sensitivity to geometric structure and interatomic distances. 

\subsection{Comparison of experimental and theoretical spin magnetic moments of $\bm{\mathrm{Fe}_{n}^{+}}$, $\bm{\mathrm{Co}_{n}^{+}}$, and $\bm{\mathrm{Ni}_{n}^{+}}$ clusters}
For iron, cobalt, and nickel clusters, nearly all theoretical studies performed on neutral \cite{Dunlap90, Reuse95b, Fan97, Andriotis98, Reddy98, AguileraGranja98, Guevara99, Dieguez01, RodriguezLopez03, Ma06a, Datta07, Wu12b} as well as on cationic \cite{AlvaradoLeyva13, Wu12b, Yuan13, Gutsev12, Gutsev13} clusters seem to underestimate the average spin magnetic moments by $0.4 - 0.8$ $\mu_B$ per atom, as can be seen in Fig.\ \ref{fig:Co_calc_moments} and Table \ref{tab:ms}.\@ Such an underestimation may originate from the known inability of standard approximations to the exchange-correlation functional to describe highly correlated electrons as in the $3d$ transition metals. 
This problem is avoided in studies of \citeauthor{RodriguezLopez03} and \citeauthor{AguileraGranja98} \cite{RodriguezLopez03, AguileraGranja98} by using semi-empirical many-body potentials and a self-consistent tight-binding method, where elastic constants or hopping and exchange integrals are fitted to reproduce bulk properties, at the expense of an \emph{ab initio} approach. As can be seen, the calculated magnetic moments of \citeauthor{RodriguezLopez03} and \citeauthor{AguileraGranja98} (open squares in Fig.\ \ref{fig:FeCoNi}) agree better with the experimental magnetic moments that are determined in our study. However, the results of semi-empirical calculations depend strongly on the parametrization and might not be able to reproduce the significant change in the geometric and electronic structure that is apparently induced by a single elementary charge difference in neutral and cationic $\mathrm{Fe}_{13}^+$ as found by \textit{ab initio} calculations \cite{AlvaradoLeyva13, Wu12b, Yuan13, Gutsev12, Gutsev13}.\@ 

\subsection{Reduction of atomic spin and orbital magnetic moments in small iron, cobalt, and nickel clusters}
\begin{figure}
	\includegraphics[width=0.50\textwidth]{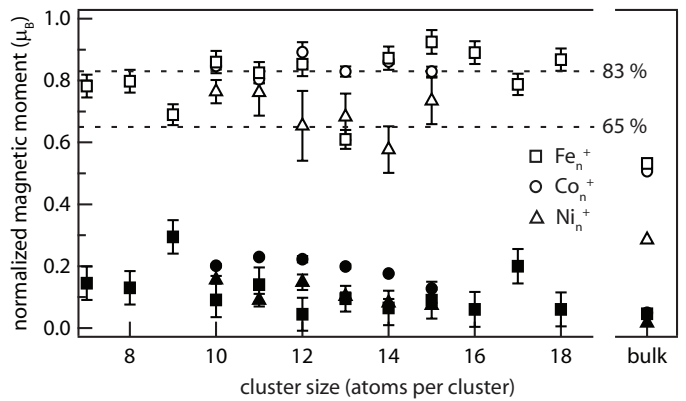}
	\caption{Spin (open symbols) and orbital magnetic moments (filled symbols) scaled to atomic values of the Fe ($3d^6$), Co ($3d^7$), and Ni ($3d^8$) configurations. Dashed lines (Fe, Co: 83\,\%; Ni: 65\,\%) mark the ratio of unoccupied $3d$ states in atoms and clusters.}
	\label{fig:FeCoNi_atom}
\end{figure}
Fig.\ \ref{fig:FeCoNi_atom} illustrates the reduction of spin magnetic moments and the quenching of orbital magnetic moments for iron, cobalt, and nickel clusters in comparison to the free atom values. Here, the experimentally determined spin and orbital magnetic moments have been normalized to the corresponding $3d$ spin magnetic moments of the atomic $3d^6$, $3d^7$, and $3d^8$ configuration of iron, cobalt, and nickel. Normalization to the atomic instead of the cationic configuration was chosen, even though for Co/$\mathrm{Co}^+$ and Ni/$\mathrm{Ni}^+$ the configuration changes from $3d^7$ to $3d^8$ and from $3d^8$ to $3d^9$ (Ref.\ \onlinecite{Shenston70, Litzen93, Pickering96, Pickering98, Hirsch12a}) but for clusters larger than $n =10$ atoms with $\ge 60$ $3d$ valence electrons, the effect of a single $s-d$ promotion on the total number of $3d$ electrons is below 2\,\% and the error that is introduced by this normalization can be neglected. For Fe/$\mathrm{Fe}^+$, the $3d^6$ configuration is identical for atoms and cations \cite{Johansson78, Nave94, Hirsch12a}.\@
\newline
As can be seen in Fig.\ \ref{fig:FeCoNi_atom}, the spin magnetic moments of iron, cobalt, and nickel clusters are reduced to a similar amount of $60 - 90$\,\% of their atomic spin magnetic moments of 4, 3, and 2 $\mu_B$ per atom, respectively. Since clusters in this size regime as well as atoms carry the maximum spin magnetic moment of $\mu_S = 1.0$ $\mu_B$ per $3d$ hole because of filled majority bands for clusters or Hund's rules for atoms, the observed reduction in the spin magnetic moment is simply caused by a different number $n_h$ of unoccupied $3d$ states. 
Therefore, the average reduction of the spin magnetic moments that is shown in Fig.\ \ref{fig:FeCoNi_atom} is close to the ratio of $3d$ holes in atoms and clusters, which is 0.83 for iron, 0.83 for cobalt, and 0.65 for nickel \cite{Sipr05, Minar06, Pacchioni87, Basch80}.\@ In the bulk metals, spin magnetic moments of iron, cobalt, and nickel are further reduced by the formation of wider $3d$ bands, which creates empty states in the $3d$ majority spin bands with average spin magnetic moments $\mu_S < 1$ $\mu_B$ per unoccupied $3d$ state. 
Even though $n_h$ is already bulk-like in small clusters, the distribution of unoccupied states over the majority and minority states is still different from the bulk but changes with cluster size. 
Consequently, the magnetic moments of neutral clusters reach the bulk value when the formation of bulk-like $spd$ bands leads to the loss of strong ferromagnetism around 400 to 600 atoms per cluster \cite{Billas93, Billas94, Apsel96, Billas97, Payne07}.\@
\newline
In contrast to the spin moment, orbital magnetic moments of bulk iron, cobalt, and nickel are strongly quenched by the crystal field and are only restored by the spin--orbit interaction to 2-3\,\% of their atomic values \cite{Eriksson90}.\@ This quenching is already very pronounced in small clusters \cite{Niemeyer12} with $n \ge 3$ because of the reduced rotational symmetry. 
Different from bulk bcc iron, hcp cobalt and fcc nickel, the modeled structures of iron, cobalt, and nickel clusters are quite similar \cite{Gutsev12, Ma07a, Dieguez01, Datta07, Ma06a}.\@ 
Since the strength of the crystal field is nearly constant along the $3d$ transition metal series \cite{Eriksson90}, the orbital magnetic moments of iron, cobalt, and nickel clusters are reduced to a similar amount of $5 - 25$\,\% of their atomic value.
Even though the orbital magnetic moments are strongly quenched when compared to the atom, where $\mu_L = 2$ $\mu_B$ for iron and $\mu_L = 3$ $\mu_B$, for cobalt and nickel, they are significantly enhanced when compared to the bulk values. For clusters, we generally find $\mu_L (\text{Co}) > \mu_L (\text{Ni}) \ge \mu_L (\text{Fe})$.\@
\newline
The enhancement of the orbital magnetic moments in clusters compared to bulk is most likely due to a reduced crystal field interaction that is caused by a lower coordination of cluster surface atoms. In a similar way, surface enhanced orbital magnetic moments of $\mu_L = 0.06 - 0.3$ $\mu_B$ have been calculated for iron, cobalt, and nickel surfaces and clusters \cite{Eriksson92, GuiradoLopez03, Yuan13}.\@
\begin{figure}
	\includegraphics[width=0.50\textwidth]{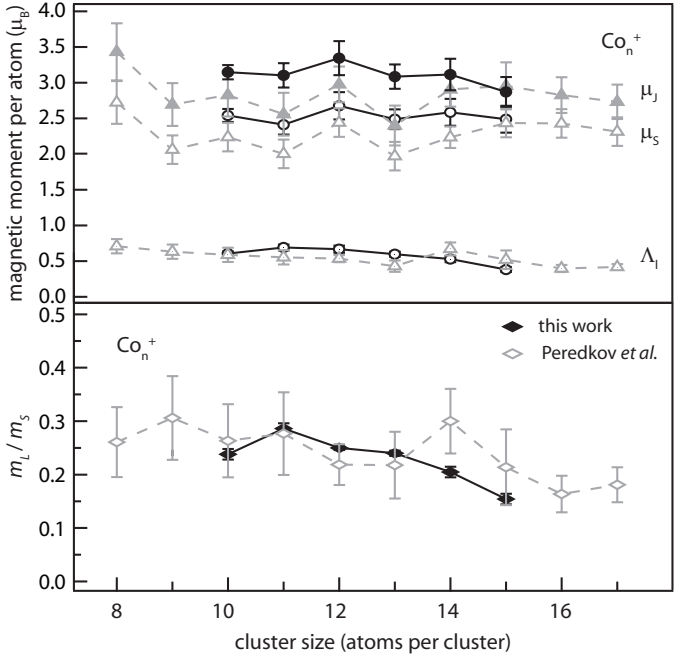}
	\caption{Upper panel: Comparison of spin, orbital, and total magnetic moments of the present (black symbols) and a former (light gray symbols, Ref.\ \onlinecite{Peredkov11b} ) XMCD study on free cobalt clusters. Lower panel: Good agreement of the ratio of spin-to-orbital magnetization of the present study and that of Ref.\ \onlinecite{Peredkov11b}.}
	\label{fig:Co_SternGerlach}
\end{figure}

\subsection{Coupling of spin and orbital angular momenta}
In a previous XMCD study of size-selected cobalt cluster ions by \citeauthor{Peredkov11b} \cite{Peredkov11b}, decoupling of the spin and orbital angular momenta has been postulated, because the authors found different Langevin scaling for orbital and spin magnetic moments in their experimental data \cite{Peredkov11b}.\@ 
Decoupled spin and orbital angular momenta would be surprising as these would not only require significant interatomic orbit--orbit coupling for a total $\mu_L$ to interact with the magnetic field, but this decoupling would also imply that the interaction of the spin magnetic moment with the applied magnetic field were much stronger than the $3d$ spin--orbit interaction in the $3d$ transition elements. However, the energy gain by alignment of a $2.5$ $\mu_B$ spin magnetic moment in a field of 7 T is of the order of 1 meV, which is small compared to the $3d$ spin--orbit coupling energy of $50 - 100$ meV per atom \cite{Popescu01, SanchezBarriga10} in iron, cobalt, and nickel. 
\newline
Experimentally, the expected coupling of spin and orbital angular momenta, $S$ and $L$, to a total angular momentum $J$can be shown from the field dependence of the ratio of orbital to spin magnetization. This $m_L / m_S$ ratio is shown in the lower panel of Fig.\ \ref{fig:Co10MC} for $\mathrm{Co}_{10}^+$ versus the applied $0 - 5$ T magnetic field. As can be seen, $m_L / m_S = 0.25 \pm 0.01$ is constant and independent of the applied magnetic field. This finding is the same for all clusters investigated and confirms the well-known Russel-Saunders coupling but contradicts the unexpected finding of \citeauthor{Peredkov11b}\@
\newline
After rescaling for $LS$ coupling, the spin and orbital magnetic moments of $\mathrm{Co}_{n}^+$ in the study of \citeauthor{Peredkov11b} agree with the data presented here within the error bars, even though the magnetic moments of \citeauthor{Peredkov11b} seem to be systematically lower than in our study, as shown in the upper panel of Fig.\ \ref{fig:Co_SternGerlach}.\@ 
Most important, the ratio $m_L / m_S$ of orbital to spin magnetization of $\mathrm{Co}_{n}^+$ in the lower panel of Fig.\ \ref{fig:Co_SternGerlach} agrees very well, as this ratio can be determined from the XMCD sum rules to a higher precision than spin and orbital magnetic moments separately because of the cancellation of potential errors in the degree of circular polarization, in the number of unoccupied $3d$ states, or in the normalization to the isotropic x-ray absorption spectrum. This quantitative agreement of two independent studies confirms the reliability of XMCD studies of size-selected cluster ions.

\subsection{Comparison of XMCD results to Stern-Gerlach experiments}
In a size range that is similar to the one considered here, Stern-Gerlach deflection experiments yield total magnetic moments of $\mu_J \approx 3.0 - 5.5$ $\mu_B$ for neutral iron clusters of 10 to 50 atoms \cite{Cox85, deHeer90, Billas93, Billas94, Billas97, Knickelbein02, Xu11}; $\mu_J \approx 2.25 - 3.9$ $\mu_B$ for neutral cobalt clusters of 10 to 50 atoms \cite{Bucher91, Douglass93, Billas94, Billas97, Xu05, Knickelbein06, Payne07, Datta07, Xu08, Xu11}; and $\mu_J \approx 0.8 - 1.3$ $\mu_B$ for neutral nickel clusters of 10 to 15 atoms \cite{Billas94, Billas97, Louderback93, Apsel96, Knickelbein02b}.\@
Even though there is a large scatter in the various Stern-Gerlach data, our average XMCD results for iron ($\mu_J \approx 3.5$ $\mu_B$) and cobalt ($\mu_J \approx 3.0$ $\mu_B$) cluster ions fall well within the range spanned for neutral clusters. For nickel clusters, we find a larger total magnetic moment ($\mu_J \approx 1.5$ $\mu_B$) than Stern-Gerlach results.
\newline
In addition to total magnetic moments, our XMCD results also yield an orbital magnetic moment that is typically $\le 25$\,\% of the spin magnetic moment for clusters in the size range considered here. This implies that an upper limit for the total magnetic moment per atom can be estimated as the full atomic spin magnetic moment plus the reduced orbital contribution. Therefore, total magnetic moments $> 5.0$ $\mu_B$ per atom that have been reported for small iron clusters \cite{Knickelbein02}, and $> 3.75$ $\mu_B$ per atom that have been reported for cobalt clusters \cite{Payne07} are in disagreement with our study and seem very unlikely. 
\newline
Consistent with our results and with the observed absence of spatial alignment of the clusters in the ion trap, superparamagnetic behavior \cite{Bean59} of $3d$ transition metal clusters is also found in Stern-Gerlach experiments on neutral $3d$ transition metal cluster beams \cite{Jensen91, Ballone91, Maiti93, Visuthikraisee96, Billas97, Hamamoto00, Knickelbein04b, Das05, Xu05, Rohrmann13}.\@ 

\section{Conclusion}
In conclusion, spin and orbital magnetic moments of iron, cobalt and nickel clusters were determined and discussed in the size regime of $n = 10 - 15$ atoms per cluster. In this size range $\mathrm{Fe}_{13}^+$ has a significantly reduced spin magnetic moment that cannot be observed for $\mathrm{Co}_{13}^+$ and $\mathrm{Ni}_{13}^+$, as can be rationalized from the bulk phase diagrams. In other aspects iron, cobalt and nickel clusters behave quite similar: Except for $\mathrm{Fe}_{13}^+$, all these clusters are strong ferromagnets with completely filled majority states and 1 $\mu_B$ spin magnetic moment per unoccupied $3d$ state. They are characterized by low magnetic anisotropy energies of $E_{\text{MAE}} \ll 65$ $\mu$eV per atom for iron and $E_{\text{MAE}} \ll 52$ $\mu$eV for cobalt and nickel clusters, which leads to superparamagnetic behavior. The average number of unoccupied $3d$ states in these clusters is nearly constant for a given element and is close to the respective bulk value, even though the distribution of $3d$ holes over minority and majority states is different from the bulk. The orbital magnetic moments of iron, cobalt, and nickel clusters are quenched to $\le 25$\,\% while the spin magnetic moments remain at 60 to 90\,\% of the atomic values. 
In comparison with existing DFT studies, we note a discrepancy between measured and calculated magnetic moments for iron, cobalt, and nickel clusters. The spin magnetic moments that are calculated within DFT seem to be underestimated. In this context, the experimental data of this work can act as a benchmark for theoretical studies on small and medium sized transition metal clusters. 
Element specific XMCD of gas phase ions is a versatile technique that is not limited to pure transition metal clusters but will also allow the study of alloys, compounds, oxides, or complexes.

\begin{acknowledgements}
Beam time for this project was provided at BESSY II beamlines UE52-SGM and UE52-PGM, operated by Helmholtz-Zentrum Berlin. We thank T. Blume, S. Krause, P. Hoffmann, and E. Suljoti for technical assistance with the apparatus and during beam times. This work was supported by the Special Cluster Research Project of Genesis Research Institute, Inc.\ and was partially funded by the German Ministry for Education and Research (BMBF) under grant No.\ BMBF-05K13VF2.\@ BvI acknowledges travel support by HZB.\@
\end{acknowledgements}


\begin{thebibliography}{178}%
\makeatletter
\providecommand \@ifxundefined [1]{%
 \@ifx{#1\undefined}
}%
\providecommand \@ifnum [1]{%
 \ifnum #1\expandafter \@firstoftwo
 \else \expandafter \@secondoftwo
 \fi
}%
\providecommand \@ifx [1]{%
 \ifx #1\expandafter \@firstoftwo
 \else \expandafter \@secondoftwo
 \fi
}%
\providecommand \natexlab [1]{#1}%
\providecommand \enquote  [1]{``#1''}%
\providecommand \bibnamefont  [1]{#1}%
\providecommand \bibfnamefont [1]{#1}%
\providecommand \citenamefont [1]{#1}%
\providecommand \href@noop [0]{\@secondoftwo}%
\providecommand \href [0]{\begingroup \@sanitize@url \@href}%
\providecommand \@href[1]{\@@startlink{#1}\@@href}%
\providecommand \@@href[1]{\endgroup#1\@@endlink}%
\providecommand \@sanitize@url [0]{\catcode `\\12\catcode `\$12\catcode
  `\&12\catcode `\#12\catcode `\^12\catcode `\_12\catcode `\%12\relax}%
\providecommand \@@startlink[1]{}%
\providecommand \@@endlink[0]{}%
\providecommand \url  [0]{\begingroup\@sanitize@url \@url }%
\providecommand \@url [1]{\endgroup\@href {#1}{\urlprefix }}%
\providecommand \urlprefix  [0]{URL }%
\providecommand \Eprint [0]{\href }%
\@ifxundefined \urlstyle {%
  \providecommand \doi  [0]{\begingroup \@sanitize@url \@doi}%
  \providecommand \@doi [1]{\endgroup \@@startlink {\doibase
  #1}doi:\discretionary {}{}{}#1\@@endlink }%
}{%
  \providecommand \doi  [0]{doi:\discretionary{}{}{}\begingroup
  \urlstyle{rm}\Url }%
}%
\providecommand \doibase [0]{http://dx.doi.org/}%
\providecommand \Doi [0]{\begingroup \@sanitize@url \@Doi }%
\providecommand \@Doi  [1]{\endgroup\@@startlink{\doibase#1}\@@Doi}%
\providecommand \@@Doi [1]{#1\@@endlink}%
\providecommand \selectlanguage [0]{\@gobble}%
\providecommand \bibinfo  [0]{\@secondoftwo}%
\providecommand \bibfield  [0]{\@secondoftwo}%
\providecommand \translation [1]{[#1]}%
\providecommand \BibitemOpen [0]{}%
\providecommand \bibitemStop [0]{}%
\providecommand \bibitemNoStop [0]{.\EOS\space}%
\providecommand \EOS [0]{\spacefactor3000\relax}%
\providecommand \BibitemShut  [1]{\csname bibitem#1\endcsname}%
\bibitem [{\citenamefont {St\"ohr}\ and\ \citenamefont
  {Siegmann}(2006)}]{Stoehr06}%
  \BibitemOpen
  \bibfield  {author} {\bibinfo {author} {\bibfnamefont {J.}~\bibnamefont
  {St\"ohr}}\ and\ \bibinfo {author} {\bibfnamefont {H.}~\bibnamefont
  {Siegmann}},\ }\href@noop {} {\emph {\bibinfo {title} {Magnetism from
  Fundamentals to Nanoscale Dynamics}}},\ \bibinfo {series} {Springer series in
  solid-state sciences}, Vol.\ \bibinfo {volume} {152}\ (\bibinfo  {publisher}
  {Springer-Verlag},\ \bibinfo {address} {Berlin},\ \bibinfo {year}
  {2006})\BibitemShut {NoStop}%
\bibitem [{\citenamefont {Kodama}(1999)}]{Kodama99}%
  \BibitemOpen
  \bibfield  {author} {\bibinfo {author} {\bibfnamefont {R.~H.}\ \bibnamefont
  {Kodama}},\ }\Doi {10.1016/S0304-8853(99)00347-9} {\bibfield  {journal}
  {\bibinfo  {journal} {J. Magn. Magn. Mater.},\ }\textbf {\bibinfo {volume}
  {200}},\ \bibinfo {pages} {359} (\bibinfo {year} {1999})}\BibitemShut
  {NoStop}%
\bibitem [{\citenamefont {Alonso}(2000)}]{Alonso00}%
  \BibitemOpen
  \bibfield  {author} {\bibinfo {author} {\bibfnamefont {J.~A.}\ \bibnamefont
  {Alonso}},\ }\Doi {10.1021/cr980391o} {\bibfield  {journal} {\bibinfo
  {journal} {Chem. Rev.},\ }\textbf {\bibinfo {volume} {100}},\ \bibinfo
  {pages} {637} (\bibinfo {year} {2000})}\BibitemShut {NoStop}%
\bibitem [{\citenamefont {Binns}(2001)}]{Binns01a}%
  \BibitemOpen
  \bibfield  {author} {\bibinfo {author} {\bibfnamefont {C.}~\bibnamefont
  {Binns}},\ }\Doi {10.1016/S0167-5729(01)00015-2} {\bibfield  {journal}
  {\bibinfo  {journal} {Surf. Sci. Rep.},\ }\textbf {\bibinfo {volume} {44}},\
  \bibinfo {pages} {1} (\bibinfo {year} {2001})}\BibitemShut {NoStop}%
\bibitem [{\citenamefont {Batlle}\ and\ \citenamefont
  {Labarta}(2002)}]{Batlle02}%
  \BibitemOpen
  \bibfield  {author} {\bibinfo {author} {\bibfnamefont {X.}~\bibnamefont
  {Batlle}}\ and\ \bibinfo {author} {\bibfnamefont {A.}~\bibnamefont
  {Labarta}},\ }\Doi {10.1088/0022-3727/35/6/201} {\bibfield  {journal}
  {\bibinfo  {journal} {J. Phys. D: Appl. Phys.},\ }\textbf {\bibinfo {volume}
  {35}},\ \bibinfo {pages} {R15} (\bibinfo {year} {2002})}\BibitemShut
  {NoStop}%
\bibitem [{\citenamefont {Bansmann}\ \emph {et~al.}(2005)\citenamefont
  {Bansmann}, \citenamefont {Baker}, \citenamefont {Binns}, \citenamefont
  {Blackman}, \citenamefont {Bucher}, \citenamefont {Dorantes-D\'avila},
  \citenamefont {Dupuis}, \citenamefont {Favre}, \citenamefont {Kechrakos},
  \citenamefont {Kleibert}, \citenamefont {Meiwes-Broer}, \citenamefont
  {Pastor}, \citenamefont {Perez}, \citenamefont {Toulemonde}, \citenamefont
  {Trohidou}, \citenamefont {Tuaillon},\ and\ \citenamefont
  {Xie}}]{Bansmann05}%
  \BibitemOpen
  \bibfield  {author} {\bibinfo {author} {\bibfnamefont {J.}~\bibnamefont
  {Bansmann}}, \bibinfo {author} {\bibfnamefont {S.~H.}\ \bibnamefont {Baker}},
  \bibinfo {author} {\bibfnamefont {C.}~\bibnamefont {Binns}}, \bibinfo
  {author} {\bibfnamefont {J.~A.}\ \bibnamefont {Blackman}}, \bibinfo {author}
  {\bibfnamefont {J.~P.}\ \bibnamefont {Bucher}}, \bibinfo {author}
  {\bibfnamefont {J.}~\bibnamefont {Dorantes-D\'avila}}, \bibinfo {author}
  {\bibfnamefont {V.}~\bibnamefont {Dupuis}}, \bibinfo {author} {\bibfnamefont
  {L.}~\bibnamefont {Favre}}, \bibinfo {author} {\bibfnamefont
  {D.}~\bibnamefont {Kechrakos}}, \bibinfo {author} {\bibfnamefont
  {A.}~\bibnamefont {Kleibert}}, \bibinfo {author} {\bibfnamefont {K.~H.}\
  \bibnamefont {Meiwes-Broer}}, \bibinfo {author} {\bibfnamefont {G.~M.}\
  \bibnamefont {Pastor}}, \bibinfo {author} {\bibfnamefont {A.}~\bibnamefont
  {Perez}}, \bibinfo {author} {\bibfnamefont {O.}~\bibnamefont {Toulemonde}},
  \bibinfo {author} {\bibfnamefont {K.~N.}\ \bibnamefont {Trohidou}}, \bibinfo
  {author} {\bibfnamefont {J.}~\bibnamefont {Tuaillon}}, \ and\ \bibinfo
  {author} {\bibfnamefont {Y.}~\bibnamefont {Xie}},\ }\Doi
  {10.1016/j.surfrep.2004.10.001} {\bibfield  {journal} {\bibinfo  {journal}
  {Surf. Sci. Rep.},\ }\textbf {\bibinfo {volume} {56}},\ \bibinfo {pages}
  {189} (\bibinfo {year} {2005})}\BibitemShut {NoStop}%
\bibitem [{\citenamefont {de~Heer}\ \emph {et~al.}(1990)\citenamefont
  {de~Heer}, \citenamefont {Milani},\ and\ \citenamefont
  {Ch\^atelain}}]{deHeer90}%
  \BibitemOpen
  \bibfield  {author} {\bibinfo {author} {\bibfnamefont {W.~A.}\ \bibnamefont
  {de~Heer}}, \bibinfo {author} {\bibfnamefont {P.}~\bibnamefont {Milani}}, \
  and\ \bibinfo {author} {\bibfnamefont {A.}~\bibnamefont {Ch\^atelain}},\
  }\Doi {10.1103/PhysRevLett.65.488} {\bibfield  {journal} {\bibinfo  {journal}
  {Phys. Rev. Lett.},\ }\textbf {\bibinfo {volume} {65}},\ \bibinfo {pages}
  {488} (\bibinfo {year} {1990})}\BibitemShut {NoStop}%
\bibitem [{\citenamefont {Bucher}\ \emph {et~al.}(1991)\citenamefont {Bucher},
  \citenamefont {Douglass},\ and\ \citenamefont {Bloomfield}}]{Bucher91}%
  \BibitemOpen
  \bibfield  {author} {\bibinfo {author} {\bibfnamefont {J.~P.}\ \bibnamefont
  {Bucher}}, \bibinfo {author} {\bibfnamefont {D.~C.}\ \bibnamefont
  {Douglass}}, \ and\ \bibinfo {author} {\bibfnamefont {L.~A.}\ \bibnamefont
  {Bloomfield}},\ }\Doi {10.1103/PhysRevLett.66.3052} {\bibfield  {journal}
  {\bibinfo  {journal} {Phys. Rev. Lett.},\ }\textbf {\bibinfo {volume} {66}},\
  \bibinfo {pages} {3052} (\bibinfo {year} {1991})}\BibitemShut {NoStop}%
\bibitem [{\citenamefont {Billas}\ \emph {et~al.}(1993)\citenamefont {Billas},
  \citenamefont {Becker}, \citenamefont {Ch\^atelain},\ and\ \citenamefont
  {de~Heer}}]{Billas93}%
  \BibitemOpen
  \bibfield  {author} {\bibinfo {author} {\bibfnamefont {I.~M.~L.}\
  \bibnamefont {Billas}}, \bibinfo {author} {\bibfnamefont {J.~A.}\
  \bibnamefont {Becker}}, \bibinfo {author} {\bibfnamefont {A.}~\bibnamefont
  {Ch\^atelain}}, \ and\ \bibinfo {author} {\bibfnamefont {W.~A.}\ \bibnamefont
  {de~Heer}},\ }\Doi {10.1103/PhysRevLett.71.4067} {\bibfield  {journal}
  {\bibinfo  {journal} {Phys. Rev. Lett.},\ }\textbf {\bibinfo {volume} {71}},\
  \bibinfo {pages} {4067} (\bibinfo {year} {1993})}\BibitemShut {NoStop}%
\bibitem [{\citenamefont {Louderback}\ \emph {et~al.}(1993)\citenamefont
  {Louderback}, \citenamefont {Cox}, \citenamefont {Lising}, \citenamefont
  {Douglass},\ and\ \citenamefont {Bloomfield}}]{Louderback93}%
  \BibitemOpen
  \bibfield  {author} {\bibinfo {author} {\bibfnamefont {J.~G.}\ \bibnamefont
  {Louderback}}, \bibinfo {author} {\bibfnamefont {A.~J.}\ \bibnamefont {Cox}},
  \bibinfo {author} {\bibfnamefont {L.~J.}\ \bibnamefont {Lising}}, \bibinfo
  {author} {\bibfnamefont {D.~C.}\ \bibnamefont {Douglass}}, \ and\ \bibinfo
  {author} {\bibfnamefont {L.~A.}\ \bibnamefont {Bloomfield}},\ }\href@noop {}
  {\bibfield  {journal} {\bibinfo  {journal} {Z. Phys. D},\ }\textbf {\bibinfo
  {volume} {26}},\ \bibinfo {pages} {301} (\bibinfo {year} {1993})}\BibitemShut
  {NoStop}%
\bibitem [{\citenamefont {Billas}\ \emph {et~al.}(1994)\citenamefont {Billas},
  \citenamefont {Ch\^atelain},\ and\ \citenamefont {de~Heer}}]{Billas94}%
  \BibitemOpen
  \bibfield  {author} {\bibinfo {author} {\bibfnamefont {I.~M.~L.}\
  \bibnamefont {Billas}}, \bibinfo {author} {\bibfnamefont {A.}~\bibnamefont
  {Ch\^atelain}}, \ and\ \bibinfo {author} {\bibfnamefont {W.~A.}\ \bibnamefont
  {de~Heer}},\ }\Doi {10.1126/science.265.5179.1682} {\bibfield  {journal}
  {\bibinfo  {journal} {Science},\ }\textbf {\bibinfo {volume} {265}},\
  \bibinfo {pages} {1682} (\bibinfo {year} {1994})}\BibitemShut {NoStop}%
\bibitem [{\citenamefont {Cox}\ \emph {et~al.}(1994)\citenamefont {Cox},
  \citenamefont {Louderback}, \citenamefont {Apsel},\ and\ \citenamefont
  {Bloomfield}}]{Cox94}%
  \BibitemOpen
  \bibfield  {author} {\bibinfo {author} {\bibfnamefont {A.~J.}\ \bibnamefont
  {Cox}}, \bibinfo {author} {\bibfnamefont {J.~G.}\ \bibnamefont {Louderback}},
  \bibinfo {author} {\bibfnamefont {S.~E.}\ \bibnamefont {Apsel}}, \ and\
  \bibinfo {author} {\bibfnamefont {L.~A.}\ \bibnamefont {Bloomfield}},\
  }\href@noop {} {\bibfield  {journal} {\bibinfo  {journal} {Phys. Rev. B},\
  }\textbf {\bibinfo {volume} {49}},\ \bibinfo {pages} {12295} (\bibinfo {year}
  {1994})}\BibitemShut {NoStop}%
\bibitem [{\citenamefont {Apsel}\ \emph {et~al.}(1996)\citenamefont {Apsel},
  \citenamefont {Emmert}, \citenamefont {Deng},\ and\ \citenamefont
  {Bloomfield}}]{Apsel96}%
  \BibitemOpen
  \bibfield  {author} {\bibinfo {author} {\bibfnamefont {S.~E.}\ \bibnamefont
  {Apsel}}, \bibinfo {author} {\bibfnamefont {J.~W.}\ \bibnamefont {Emmert}},
  \bibinfo {author} {\bibfnamefont {J.}~\bibnamefont {Deng}}, \ and\ \bibinfo
  {author} {\bibfnamefont {L.~A.}\ \bibnamefont {Bloomfield}},\ }\Doi
  {10.1103/PhysRevLett.76.1441} {\bibfield  {journal} {\bibinfo  {journal}
  {Phys. Rev. Lett.},\ }\textbf {\bibinfo {volume} {76}},\ \bibinfo {pages}
  {1441} (\bibinfo {year} {1996})}\BibitemShut {NoStop}%
\bibitem [{\citenamefont {Billas}\ \emph {et~al.}(1997)\citenamefont {Billas},
  \citenamefont {Ch\^atelain},\ and\ \citenamefont {de~Heer}}]{Billas97}%
  \BibitemOpen
  \bibfield  {author} {\bibinfo {author} {\bibfnamefont {I.~M.~L.}\
  \bibnamefont {Billas}}, \bibinfo {author} {\bibfnamefont {A.}~\bibnamefont
  {Ch\^atelain}}, \ and\ \bibinfo {author} {\bibfnamefont {W.~A.}\ \bibnamefont
  {de~Heer}},\ }\Doi {http://dx.doi.org/10.1016/S0304-8853(96)00694-4}
  {\bibfield  {journal} {\bibinfo  {journal} {J. Magn. Magn. Mater.},\ }\textbf
  {\bibinfo {volume} {168}},\ \bibinfo {pages} {64} (\bibinfo {year}
  {1997})}\BibitemShut {NoStop}%
\bibitem [{\citenamefont {Bean}\ and\ \citenamefont
  {Livingston}(1959)}]{Bean59}%
  \BibitemOpen
  \bibfield  {author} {\bibinfo {author} {\bibfnamefont {C.~P.}\ \bibnamefont
  {Bean}}\ and\ \bibinfo {author} {\bibfnamefont {J.~D.}\ \bibnamefont
  {Livingston}},\ }\Doi {10.1063/1.2185850} {\bibfield  {journal} {\bibinfo
  {journal} {J. Appl. Phys.},\ }\textbf {\bibinfo {volume} {30}},\ \bibinfo
  {pages} {S120} (\bibinfo {year} {1959})}\BibitemShut {NoStop}%
\bibitem [{\citenamefont {Cox}\ \emph {et~al.}(1993)\citenamefont {Cox},
  \citenamefont {Louderback},\ and\ \citenamefont {Bloomfield}}]{Cox93a}%
  \BibitemOpen
  \bibfield  {author} {\bibinfo {author} {\bibfnamefont {A.~J.}\ \bibnamefont
  {Cox}}, \bibinfo {author} {\bibfnamefont {J.~G.}\ \bibnamefont {Louderback}},
  \ and\ \bibinfo {author} {\bibfnamefont {L.~A.}\ \bibnamefont {Bloomfield}},\
  }\href@noop {} {\bibfield  {journal} {\bibinfo  {journal} {Phys. Rev.
  Lett.},\ }\textbf {\bibinfo {volume} {71}},\ \bibinfo {pages} {923} (\bibinfo
  {year} {1993})}\BibitemShut {NoStop}%
\bibitem [{\citenamefont {Knickelbein}(2001)}]{Knickelbein01}%
  \BibitemOpen
  \bibfield  {author} {\bibinfo {author} {\bibfnamefont {M.~B.}\ \bibnamefont
  {Knickelbein}},\ }\Doi {10.1103/PhysRevLett.86.5255} {\bibfield  {journal}
  {\bibinfo  {journal} {Phys. Rev. Lett.},\ }\textbf {\bibinfo {volume} {86}},\
  \bibinfo {pages} {5255} (\bibinfo {year} {2001})}\BibitemShut {NoStop}%
\bibitem [{\citenamefont {Knickelbein}(2004){\natexlab{a}}}]{Knickelbein04a}%
  \BibitemOpen
  \bibfield  {author} {\bibinfo {author} {\bibfnamefont {M.~B.}\ \bibnamefont
  {Knickelbein}},\ }\Doi {10.1103/PhysRevB.70.014424} {\bibfield  {journal}
  {\bibinfo  {journal} {Phys. Rev. B},\ }\textbf {\bibinfo {volume} {70}},\
  \bibinfo {eid} {014424} (\bibinfo {year} {2004}{\natexlab{a}})}\BibitemShut
  {NoStop}%
\bibitem [{\citenamefont {Payne}\ \emph {et~al.}(2006)\citenamefont {Payne},
  \citenamefont {Jiang},\ and\ \citenamefont {Bloomfield}}]{Payne06}%
  \BibitemOpen
  \bibfield  {author} {\bibinfo {author} {\bibfnamefont {F.~W.}\ \bibnamefont
  {Payne}}, \bibinfo {author} {\bibfnamefont {W.}~\bibnamefont {Jiang}}, \ and\
  \bibinfo {author} {\bibfnamefont {L.~A.}\ \bibnamefont {Bloomfield}},\ }\Doi
  {10.1103/PhysRevLett.97.193401} {\bibfield  {journal} {\bibinfo  {journal}
  {Phys. Rev. Lett.},\ }\textbf {\bibinfo {volume} {97}},\ \bibinfo {eid}
  {193401} (\bibinfo {year} {2006})}\BibitemShut {NoStop}%
\bibitem [{\citenamefont {Xu}\ \emph {et~al.}(2011)\citenamefont {Xu},
  \citenamefont {Yin}, \citenamefont {Moro}, \citenamefont {Liang},
  \citenamefont {Bowlan},\ and\ \citenamefont {de~Heer}}]{Xu11}%
  \BibitemOpen
  \bibfield  {author} {\bibinfo {author} {\bibfnamefont {X.}~\bibnamefont
  {Xu}}, \bibinfo {author} {\bibfnamefont {S.}~\bibnamefont {Yin}}, \bibinfo
  {author} {\bibfnamefont {R.}~\bibnamefont {Moro}}, \bibinfo {author}
  {\bibfnamefont {A.}~\bibnamefont {Liang}}, \bibinfo {author} {\bibfnamefont
  {J.}~\bibnamefont {Bowlan}}, \ and\ \bibinfo {author} {\bibfnamefont {W.~A.}\
  \bibnamefont {de~Heer}},\ }\Doi {10.1103/PhysRevLett.107.057203} {\bibfield
  {journal} {\bibinfo  {journal} {Phys. Rev. Lett.},\ }\textbf {\bibinfo
  {volume} {107}},\ \bibinfo {pages} {057203} (\bibinfo {year}
  {2011})}\BibitemShut {NoStop}%
\bibitem [{\citenamefont {Gambardella}\ \emph {et~al.}(2003)\citenamefont
  {Gambardella}, \citenamefont {Rusponi}, \citenamefont {Veronese},
  \citenamefont {Dhesi}, \citenamefont {Grazioli}, \citenamefont {Dallmeyer},
  \citenamefont {Cabria}, \citenamefont {Zeller}, \citenamefont {Dederichs},
  \citenamefont {Kern}, \citenamefont {Carbone},\ and\ \citenamefont
  {Brune}}]{Gambardella03}%
  \BibitemOpen
  \bibfield  {author} {\bibinfo {author} {\bibfnamefont {P.}~\bibnamefont
  {Gambardella}}, \bibinfo {author} {\bibfnamefont {S.}~\bibnamefont
  {Rusponi}}, \bibinfo {author} {\bibfnamefont {M.}~\bibnamefont {Veronese}},
  \bibinfo {author} {\bibfnamefont {S.~S.}\ \bibnamefont {Dhesi}}, \bibinfo
  {author} {\bibfnamefont {C.}~\bibnamefont {Grazioli}}, \bibinfo {author}
  {\bibfnamefont {A.}~\bibnamefont {Dallmeyer}}, \bibinfo {author}
  {\bibfnamefont {I.}~\bibnamefont {Cabria}}, \bibinfo {author} {\bibfnamefont
  {R.}~\bibnamefont {Zeller}}, \bibinfo {author} {\bibfnamefont {P.~H.}\
  \bibnamefont {Dederichs}}, \bibinfo {author} {\bibfnamefont {K.}~\bibnamefont
  {Kern}}, \bibinfo {author} {\bibfnamefont {C.}~\bibnamefont {Carbone}}, \
  and\ \bibinfo {author} {\bibfnamefont {H.}~\bibnamefont {Brune}},\ }\Doi
  {10.1126/science.1082857} {\bibfield  {journal} {\bibinfo  {journal}
  {Science},\ }\textbf {\bibinfo {volume} {300}},\ \bibinfo {pages} {1130}
  (\bibinfo {year} {2003})}\BibitemShut {NoStop}%
\bibitem [{\citenamefont {Meier}\ \emph {et~al.}(2008)\citenamefont {Meier},
  \citenamefont {Zhou}, \citenamefont {Wiebe},\ and\ \citenamefont
  {Wiesendanger}}]{Meier08}%
  \BibitemOpen
  \bibfield  {author} {\bibinfo {author} {\bibfnamefont {F.}~\bibnamefont
  {Meier}}, \bibinfo {author} {\bibfnamefont {L.}~\bibnamefont {Zhou}},
  \bibinfo {author} {\bibfnamefont {J.}~\bibnamefont {Wiebe}}, \ and\ \bibinfo
  {author} {\bibfnamefont {R.}~\bibnamefont {Wiesendanger}},\ }\Doi
  {10.1126/science.1154415} {\bibfield  {journal} {\bibinfo  {journal}
  {Science},\ }\textbf {\bibinfo {volume} {320}},\ \bibinfo {pages} {82}
  (\bibinfo {year} {2008})}\BibitemShut {NoStop}%
\bibitem [{\citenamefont {Brune}\ and\ \citenamefont
  {Gambardella}(2009)}]{Brune09}%
  \BibitemOpen
  \bibfield  {author} {\bibinfo {author} {\bibfnamefont {H.}~\bibnamefont
  {Brune}}\ and\ \bibinfo {author} {\bibfnamefont {P.}~\bibnamefont
  {Gambardella}},\ }\Doi {http://dx.doi.org/10.1016/j.susc.2008.11.055}
  {\bibfield  {journal} {\bibinfo  {journal} {Surf. Sci.},\ }\textbf {\bibinfo
  {volume} {603}},\ \bibinfo {pages} {1812 } (\bibinfo {year}
  {2009})}\BibitemShut {NoStop}%
\bibitem [{\citenamefont {Lau}\ \emph {et~al.}(2002){\natexlab{a}}\citenamefont
  {Lau}, \citenamefont {F\"ohlisch}, \citenamefont {Nietuby\`c}, \citenamefont
  {Reif},\ and\ \citenamefont {Wurth}}]{Lau02a}%
  \BibitemOpen
  \bibfield  {author} {\bibinfo {author} {\bibfnamefont {J.~T.}\ \bibnamefont
  {Lau}}, \bibinfo {author} {\bibfnamefont {A.}~\bibnamefont {F\"ohlisch}},
  \bibinfo {author} {\bibfnamefont {R.}~\bibnamefont {Nietuby\`c}}, \bibinfo
  {author} {\bibfnamefont {M.}~\bibnamefont {Reif}}, \ and\ \bibinfo {author}
  {\bibfnamefont {W.}~\bibnamefont {Wurth}},\ }\Doi
  {10.1103/PhysRevLett.89.057201} {\bibfield  {journal} {\bibinfo  {journal}
  {Phys. Rev. Lett.},\ }\textbf {\bibinfo {volume} {89}},\ \bibinfo {pages}
  {057201} (\bibinfo {year} {2002}{\natexlab{a}})}\BibitemShut {NoStop}%
\bibitem [{\citenamefont {Lau}\ \emph {et~al.}(2002){\natexlab{b}}\citenamefont
  {Lau}, \citenamefont {F\"ohlisch}, \citenamefont {Martins}, \citenamefont
  {Nietuby{\`c}}, \citenamefont {Reif},\ and\ \citenamefont {Wurth}}]{Lau02b}%
  \BibitemOpen
  \bibfield  {author} {\bibinfo {author} {\bibfnamefont {J.~T.}\ \bibnamefont
  {Lau}}, \bibinfo {author} {\bibfnamefont {A.}~\bibnamefont {F\"ohlisch}},
  \bibinfo {author} {\bibfnamefont {M.}~\bibnamefont {Martins}}, \bibinfo
  {author} {\bibfnamefont {R.}~\bibnamefont {Nietuby{\`c}}}, \bibinfo {author}
  {\bibfnamefont {M.}~\bibnamefont {Reif}}, \ and\ \bibinfo {author}
  {\bibfnamefont {W.}~\bibnamefont {Wurth}},\ }\Doi {10.1088/1367-2630/4/1/398}
  {\bibfield  {journal} {\bibinfo  {journal} {New J. Phys.},\ }\textbf
  {\bibinfo {volume} {4}},\ \bibinfo {pages} {98} (\bibinfo {year}
  {2002}{\natexlab{b}})}\BibitemShut {NoStop}%
\bibitem [{\citenamefont {Ballentine}\ \emph {et~al.}(2007)\citenamefont
  {Ballentine}, \citenamefont {He{\ss}ler}, \citenamefont {Kinza},\ and\
  \citenamefont {Fauth}}]{Ballentine07}%
  \BibitemOpen
  \bibfield  {author} {\bibinfo {author} {\bibfnamefont {G.}~\bibnamefont
  {Ballentine}}, \bibinfo {author} {\bibfnamefont {M.}~\bibnamefont
  {He{\ss}ler}}, \bibinfo {author} {\bibfnamefont {M.}~\bibnamefont {Kinza}}, \
  and\ \bibinfo {author} {\bibfnamefont {K.}~\bibnamefont {Fauth}},\ }\Doi
  {10.1140/epjd/e2007-00271-9} {\bibfield  {journal} {\bibinfo  {journal} {Eur.
  Phys. J. D},\ }\textbf {\bibinfo {volume} {45}},\ \bibinfo {pages} {535}
  (\bibinfo {year} {2007})}\BibitemShut {NoStop}%
\bibitem [{\citenamefont {Glaser}\ \emph {et~al.}(2012)\citenamefont {Glaser},
  \citenamefont {Chen}, \citenamefont {Fiedler}, \citenamefont {Wellh\"ofer},
  \citenamefont {Wurth},\ and\ \citenamefont {Martins}}]{Glaser12}%
  \BibitemOpen
  \bibfield  {author} {\bibinfo {author} {\bibfnamefont {L.}~\bibnamefont
  {Glaser}}, \bibinfo {author} {\bibfnamefont {K.}~\bibnamefont {Chen}},
  \bibinfo {author} {\bibfnamefont {S.}~\bibnamefont {Fiedler}}, \bibinfo
  {author} {\bibfnamefont {M.}~\bibnamefont {Wellh\"ofer}}, \bibinfo {author}
  {\bibfnamefont {W.}~\bibnamefont {Wurth}}, \ and\ \bibinfo {author}
  {\bibfnamefont {M.}~\bibnamefont {Martins}},\ }\Doi
  {10.1103/PhysRevB.86.075435} {\bibfield  {journal} {\bibinfo  {journal}
  {Phys. Rev. B},\ }\textbf {\bibinfo {volume} {86}},\ \bibinfo {pages}
  {075435} (\bibinfo {year} {2012})}\BibitemShut {NoStop}%
\bibitem [{\citenamefont {Edmonds}\ \emph {et~al.}(1999)\citenamefont
  {Edmonds}, \citenamefont {Binns}, \citenamefont {Baker}, \citenamefont
  {Thornton}, \citenamefont {Norris}, \citenamefont {Goedkoop}, \citenamefont
  {Finazzi},\ and\ \citenamefont {Brookes}}]{Edmonds99}%
  \BibitemOpen
  \bibfield  {author} {\bibinfo {author} {\bibfnamefont {K.~W.}\ \bibnamefont
  {Edmonds}}, \bibinfo {author} {\bibfnamefont {C.}~\bibnamefont {Binns}},
  \bibinfo {author} {\bibfnamefont {S.~H.}\ \bibnamefont {Baker}}, \bibinfo
  {author} {\bibfnamefont {S.~C.}\ \bibnamefont {Thornton}}, \bibinfo {author}
  {\bibfnamefont {C.}~\bibnamefont {Norris}}, \bibinfo {author} {\bibfnamefont
  {J.~B.}\ \bibnamefont {Goedkoop}}, \bibinfo {author} {\bibfnamefont
  {M.}~\bibnamefont {Finazzi}}, \ and\ \bibinfo {author} {\bibfnamefont
  {N.~B.}\ \bibnamefont {Brookes}},\ }\Doi {10.1103/PhysRevB.60.472} {\bibfield
   {journal} {\bibinfo  {journal} {Phys. Rev. B},\ }\textbf {\bibinfo {volume}
  {60}},\ \bibinfo {pages} {472} (\bibinfo {year} {1999})}\BibitemShut
  {NoStop}%
\bibitem [{\citenamefont {Pietzsch}\ \emph {et~al.}(2004)\citenamefont
  {Pietzsch}, \citenamefont {Kubetzka}, \citenamefont {Bode},\ and\
  \citenamefont {Wiesendanger}}]{Pietzsch04}%
  \BibitemOpen
  \bibfield  {author} {\bibinfo {author} {\bibfnamefont {O.}~\bibnamefont
  {Pietzsch}}, \bibinfo {author} {\bibfnamefont {A.}~\bibnamefont {Kubetzka}},
  \bibinfo {author} {\bibfnamefont {M.}~\bibnamefont {Bode}}, \ and\ \bibinfo
  {author} {\bibfnamefont {R.}~\bibnamefont {Wiesendanger}},\ }\Doi
  {10.1103/PhysRevLett.92.057202} {\bibfield  {journal} {\bibinfo  {journal}
  {Phys. Rev. Lett.},\ }\textbf {\bibinfo {volume} {92}},\ \bibinfo {pages}
  {057202} (\bibinfo {year} {2004})}\BibitemShut {NoStop}%
\bibitem [{\citenamefont {Kleibert}\ \emph {et~al.}(2009)\citenamefont
  {Kleibert}, \citenamefont {Meiwes-Broer},\ and\ \citenamefont
  {Bansmann}}]{Kleibert09}%
  \BibitemOpen
  \bibfield  {author} {\bibinfo {author} {\bibfnamefont {A.}~\bibnamefont
  {Kleibert}}, \bibinfo {author} {\bibfnamefont {K.-H.}\ \bibnamefont
  {Meiwes-Broer}}, \ and\ \bibinfo {author} {\bibfnamefont {J.}~\bibnamefont
  {Bansmann}},\ }\Doi {10.1103/PhysRevB.79.125423} {\bibfield  {journal}
  {\bibinfo  {journal} {Phys. Rev. B},\ }\textbf {\bibinfo {volume} {79}},\
  \bibinfo {pages} {125423} (\bibinfo {year} {2009})}\BibitemShut {NoStop}%
\bibitem [{\citenamefont {Bansmann}\ \emph {et~al.}(2010)\citenamefont
  {Bansmann}, \citenamefont {Kleibert}, \citenamefont {Getzlaff}, \citenamefont
  {Fraile~Rodr\'{\i}guez}, \citenamefont {Nolting}, \citenamefont {Boeglin},\
  and\ \citenamefont {Meiwes-Broer}}]{Bansmann10}%
  \BibitemOpen
  \bibfield  {author} {\bibinfo {author} {\bibfnamefont {J.}~\bibnamefont
  {Bansmann}}, \bibinfo {author} {\bibfnamefont {A.}~\bibnamefont {Kleibert}},
  \bibinfo {author} {\bibfnamefont {M.}~\bibnamefont {Getzlaff}}, \bibinfo
  {author} {\bibfnamefont {A.}~\bibnamefont {Fraile~Rodr\'{\i}guez}}, \bibinfo
  {author} {\bibfnamefont {F.}~\bibnamefont {Nolting}}, \bibinfo {author}
  {\bibfnamefont {C.}~\bibnamefont {Boeglin}}, \ and\ \bibinfo {author}
  {\bibfnamefont {K.-H.}\ \bibnamefont {Meiwes-Broer}},\ }\Doi
  {10.1002/pssb.200945516} {\bibfield  {journal} {\bibinfo  {journal} {Phys.
  Status Solidi B},\ }\textbf {\bibinfo {volume} {247}},\ \bibinfo {pages}
  {1152} (\bibinfo {year} {2010})}\BibitemShut {NoStop}%
\bibitem [{\citenamefont {Koide}\ \emph {et~al.}(2001)\citenamefont {Koide},
  \citenamefont {Miyauchi}, \citenamefont {Okamoto}, \citenamefont {Shidara},
  \citenamefont {Fujimori}, \citenamefont {Fukutani}, \citenamefont {Amemiya},
  \citenamefont {Takeshita}, \citenamefont {Yuasa}, \citenamefont {Katayama},\
  and\ \citenamefont {Suzuki}}]{Koide01}%
  \BibitemOpen
  \bibfield  {author} {\bibinfo {author} {\bibfnamefont {T.}~\bibnamefont
  {Koide}}, \bibinfo {author} {\bibfnamefont {H.}~\bibnamefont {Miyauchi}},
  \bibinfo {author} {\bibfnamefont {J.}~\bibnamefont {Okamoto}}, \bibinfo
  {author} {\bibfnamefont {T.}~\bibnamefont {Shidara}}, \bibinfo {author}
  {\bibfnamefont {A.}~\bibnamefont {Fujimori}}, \bibinfo {author}
  {\bibfnamefont {H.}~\bibnamefont {Fukutani}}, \bibinfo {author}
  {\bibfnamefont {K.}~\bibnamefont {Amemiya}}, \bibinfo {author} {\bibfnamefont
  {H.}~\bibnamefont {Takeshita}}, \bibinfo {author} {\bibfnamefont
  {S.}~\bibnamefont {Yuasa}}, \bibinfo {author} {\bibfnamefont
  {T.}~\bibnamefont {Katayama}}, \ and\ \bibinfo {author} {\bibfnamefont
  {Y.}~\bibnamefont {Suzuki}},\ }\Doi {10.1103/PhysRevLett.87.257201}
  {\bibfield  {journal} {\bibinfo  {journal} {Phys. Rev. Lett.},\ }\textbf
  {\bibinfo {volume} {87}},\ \bibinfo {pages} {257201} (\bibinfo {year}
  {2001})}\BibitemShut {NoStop}%
\bibitem [{\citenamefont {Fauth}\ \emph {et~al.}(2004)\citenamefont {Fauth},
  \citenamefont {Gold}, \citenamefont {He{\ss}ler}, \citenamefont {Schneider},\
  and\ \citenamefont {Sch\"utz}}]{Fauth04a}%
  \BibitemOpen
  \bibfield  {author} {\bibinfo {author} {\bibfnamefont {K.}~\bibnamefont
  {Fauth}}, \bibinfo {author} {\bibfnamefont {S.}~\bibnamefont {Gold}},
  \bibinfo {author} {\bibfnamefont {M.}~\bibnamefont {He{\ss}ler}}, \bibinfo
  {author} {\bibfnamefont {N.}~\bibnamefont {Schneider}}, \ and\ \bibinfo
  {author} {\bibfnamefont {G.}~\bibnamefont {Sch\"utz}},\ }\href@noop {}
  {\bibfield  {journal} {\bibinfo  {journal} {Chem. Phys. Lett.},\ }\textbf
  {\bibinfo {volume} {392}},\ \bibinfo {pages} {498} (\bibinfo {year}
  {2004})}\BibitemShut {NoStop}%
\bibitem [{\citenamefont {Boyen}\ \emph {et~al.}(2005)\citenamefont {Boyen},
  \citenamefont {Fauth}, \citenamefont {Stahl}, \citenamefont {Ziemann},
  \citenamefont {K\"{a}stle}, \citenamefont {Weigl}, \citenamefont {Banhart},
  \citenamefont {Hessler}, \citenamefont {Sch\"{u}tz}, \citenamefont
  {Gajbhiye}, \citenamefont {Ellrich}, \citenamefont {Hahn}, \citenamefont
  {B\"{u}ttner}, \citenamefont {Garnier},\ and\ \citenamefont
  {Oelhafen}}]{Boyen05a}%
  \BibitemOpen
  \bibfield  {author} {\bibinfo {author} {\bibfnamefont {H.-G.}\ \bibnamefont
  {Boyen}}, \bibinfo {author} {\bibfnamefont {K.}~\bibnamefont {Fauth}},
  \bibinfo {author} {\bibfnamefont {B.}~\bibnamefont {Stahl}}, \bibinfo
  {author} {\bibfnamefont {P.}~\bibnamefont {Ziemann}}, \bibinfo {author}
  {\bibfnamefont {G.}~\bibnamefont {K\"{a}stle}}, \bibinfo {author}
  {\bibfnamefont {F.}~\bibnamefont {Weigl}}, \bibinfo {author} {\bibfnamefont
  {F.}~\bibnamefont {Banhart}}, \bibinfo {author} {\bibfnamefont
  {M.}~\bibnamefont {Hessler}}, \bibinfo {author} {\bibfnamefont
  {G.}~\bibnamefont {Sch\"{u}tz}}, \bibinfo {author} {\bibfnamefont {N.~S.}\
  \bibnamefont {Gajbhiye}}, \bibinfo {author} {\bibfnamefont {J.}~\bibnamefont
  {Ellrich}}, \bibinfo {author} {\bibfnamefont {H.}~\bibnamefont {Hahn}},
  \bibinfo {author} {\bibfnamefont {M.}~\bibnamefont {B\"{u}ttner}}, \bibinfo
  {author} {\bibfnamefont {M.~G.}\ \bibnamefont {Garnier}}, \ and\ \bibinfo
  {author} {\bibfnamefont {P.}~\bibnamefont {Oelhafen}},\ }\Doi
  {10.1002/adma.200400748} {\bibfield  {journal} {\bibinfo  {journal} {Adv.
  Mater.},\ }\textbf {\bibinfo {volume} {17}},\ \bibinfo {pages} {574}
  (\bibinfo {year} {2005})}\BibitemShut {NoStop}%
\bibitem [{\citenamefont {Gambardella}\ \emph {et~al.}(2005)\citenamefont
  {Gambardella}, \citenamefont {Brune}, \citenamefont {Dhesi}, \citenamefont
  {Bencok}, \citenamefont {Krishnakumar}, \citenamefont {Gardonio},
  \citenamefont {Veronese}, \citenamefont {Grazioli},\ and\ \citenamefont
  {Carbone}}]{Gambardella05a}%
  \BibitemOpen
  \bibfield  {author} {\bibinfo {author} {\bibfnamefont {P.}~\bibnamefont
  {Gambardella}}, \bibinfo {author} {\bibfnamefont {H.}~\bibnamefont {Brune}},
  \bibinfo {author} {\bibfnamefont {S.~S.}\ \bibnamefont {Dhesi}}, \bibinfo
  {author} {\bibfnamefont {P.}~\bibnamefont {Bencok}}, \bibinfo {author}
  {\bibfnamefont {S.~R.}\ \bibnamefont {Krishnakumar}}, \bibinfo {author}
  {\bibfnamefont {S.}~\bibnamefont {Gardonio}}, \bibinfo {author}
  {\bibfnamefont {M.}~\bibnamefont {Veronese}}, \bibinfo {author}
  {\bibfnamefont {C.}~\bibnamefont {Grazioli}}, \ and\ \bibinfo {author}
  {\bibfnamefont {C.}~\bibnamefont {Carbone}},\ }\Doi
  {10.1103/PhysRevB.72.045337} {\bibfield  {journal} {\bibinfo  {journal}
  {Phys. Rev. B},\ }\textbf {\bibinfo {volume} {72}},\ \bibinfo {eid} {045337}
  (\bibinfo {year} {2005})}\BibitemShut {NoStop}%
\bibitem [{\citenamefont {Wiesendanger}\ \emph {et~al.}(1990)\citenamefont
  {Wiesendanger}, \citenamefont {G\"untherodt}, \citenamefont {G\"untherodt},
  \citenamefont {Gambino},\ and\ \citenamefont {Ruf}}]{Wiesendanger90}%
  \BibitemOpen
  \bibfield  {author} {\bibinfo {author} {\bibfnamefont {R.}~\bibnamefont
  {Wiesendanger}}, \bibinfo {author} {\bibfnamefont {H.-J.}\ \bibnamefont
  {G\"untherodt}}, \bibinfo {author} {\bibfnamefont {G.}~\bibnamefont
  {G\"untherodt}}, \bibinfo {author} {\bibfnamefont {R.~J.}\ \bibnamefont
  {Gambino}}, \ and\ \bibinfo {author} {\bibfnamefont {R.}~\bibnamefont
  {Ruf}},\ }\Doi {10.1103/PhysRevLett.65.247} {\bibfield  {journal} {\bibinfo
  {journal} {Phys. Rev. Lett.},\ }\textbf {\bibinfo {volume} {65}},\ \bibinfo
  {pages} {247} (\bibinfo {year} {1990})}\BibitemShut {NoStop}%
\bibitem [{\citenamefont {Wulfhekel}\ and\ \citenamefont
  {Kirschner}(2007)}]{Wulfhekel07}%
  \BibitemOpen
  \bibfield  {author} {\bibinfo {author} {\bibfnamefont {W.}~\bibnamefont
  {Wulfhekel}}\ and\ \bibinfo {author} {\bibfnamefont {J.}~\bibnamefont
  {Kirschner}},\ }\Doi {10.1146/annurev.matsci.37.052506.084342} {\bibfield
  {journal} {\bibinfo  {journal} {Annu. Rev. Mater. Res.},\ }\textbf {\bibinfo
  {volume} {37}},\ \bibinfo {pages} {69} (\bibinfo {year} {2007})}\BibitemShut
  {NoStop}%
\bibitem [{\citenamefont {Brune}(2006)}]{Brune06}%
  \BibitemOpen
  \bibfield  {author} {\bibinfo {author} {\bibfnamefont {H.}~\bibnamefont
  {Brune}},\ }\Doi {10.1126/science.1127387} {\bibfield  {journal} {\bibinfo
  {journal} {Science},\ }\textbf {\bibinfo {volume} {312}},\ \bibinfo {pages}
  {1005} (\bibinfo {year} {2006})}\BibitemShut {NoStop}%
\bibitem [{\citenamefont {Hirjibehedin}\ \emph {et~al.}(2007)\citenamefont
  {Hirjibehedin}, \citenamefont {Lin}, \citenamefont {Otte}, \citenamefont
  {Ternes}, \citenamefont {Lutz}, \citenamefont {Jones},\ and\ \citenamefont
  {Heinrich}}]{Hirjibehedin07}%
  \BibitemOpen
  \bibfield  {author} {\bibinfo {author} {\bibfnamefont {C.~F.}\ \bibnamefont
  {Hirjibehedin}}, \bibinfo {author} {\bibfnamefont {C.-Y.}\ \bibnamefont
  {Lin}}, \bibinfo {author} {\bibfnamefont {A.~F.}\ \bibnamefont {Otte}},
  \bibinfo {author} {\bibfnamefont {M.}~\bibnamefont {Ternes}}, \bibinfo
  {author} {\bibfnamefont {C.~P.}\ \bibnamefont {Lutz}}, \bibinfo {author}
  {\bibfnamefont {B.~A.}\ \bibnamefont {Jones}}, \ and\ \bibinfo {author}
  {\bibfnamefont {A.~J.}\ \bibnamefont {Heinrich}},\ }\Doi
  {10.1126/science.1146110} {\bibfield  {journal} {\bibinfo  {journal}
  {Science},\ }\textbf {\bibinfo {volume} {317}},\ \bibinfo {pages} {1199}
  (\bibinfo {year} {2007})}\BibitemShut {NoStop}%
\bibitem [{\citenamefont {Balashov}\ \emph {et~al.}(2009)\citenamefont
  {Balashov}, \citenamefont {Schuh}, \citenamefont {Tak\'acs}, \citenamefont
  {Ernst}, \citenamefont {Ostanin}, \citenamefont {Henk}, \citenamefont
  {Mertig}, \citenamefont {Bruno}, \citenamefont {Miyamachi}, \citenamefont
  {Suga},\ and\ \citenamefont {Wulfhekel}}]{Balashov09}%
  \BibitemOpen
  \bibfield  {author} {\bibinfo {author} {\bibfnamefont {T.}~\bibnamefont
  {Balashov}}, \bibinfo {author} {\bibfnamefont {T.}~\bibnamefont {Schuh}},
  \bibinfo {author} {\bibfnamefont {A.~F.}\ \bibnamefont {Tak\'acs}}, \bibinfo
  {author} {\bibfnamefont {A.}~\bibnamefont {Ernst}}, \bibinfo {author}
  {\bibfnamefont {S.}~\bibnamefont {Ostanin}}, \bibinfo {author} {\bibfnamefont
  {J.}~\bibnamefont {Henk}}, \bibinfo {author} {\bibfnamefont {I.}~\bibnamefont
  {Mertig}}, \bibinfo {author} {\bibfnamefont {P.}~\bibnamefont {Bruno}},
  \bibinfo {author} {\bibfnamefont {T.}~\bibnamefont {Miyamachi}}, \bibinfo
  {author} {\bibfnamefont {S.}~\bibnamefont {Suga}}, \ and\ \bibinfo {author}
  {\bibfnamefont {W.}~\bibnamefont {Wulfhekel}},\ }\Doi
  {10.1103/PhysRevLett.102.257203} {\bibfield  {journal} {\bibinfo  {journal}
  {Phys. Rev. Lett.},\ }\textbf {\bibinfo {volume} {102}},\ \bibinfo {pages}
  {257203} (\bibinfo {year} {2009})}\BibitemShut {NoStop}%
\bibitem [{\citenamefont {Donati}\ \emph {et~al.}(2013)\citenamefont {Donati},
  \citenamefont {Dubout}, \citenamefont {Aut\`es}, \citenamefont {Patthey},
  \citenamefont {Calleja}, \citenamefont {Gambardella}, \citenamefont
  {Yazyev},\ and\ \citenamefont {Brune}}]{Donati13}%
  \BibitemOpen
  \bibfield  {author} {\bibinfo {author} {\bibfnamefont {F.}~\bibnamefont
  {Donati}}, \bibinfo {author} {\bibfnamefont {Q.}~\bibnamefont {Dubout}},
  \bibinfo {author} {\bibfnamefont {G.}~\bibnamefont {Aut\`es}}, \bibinfo
  {author} {\bibfnamefont {F.}~\bibnamefont {Patthey}}, \bibinfo {author}
  {\bibfnamefont {F.}~\bibnamefont {Calleja}}, \bibinfo {author} {\bibfnamefont
  {P.}~\bibnamefont {Gambardella}}, \bibinfo {author} {\bibfnamefont {O.~V.}\
  \bibnamefont {Yazyev}}, \ and\ \bibinfo {author} {\bibfnamefont
  {H.}~\bibnamefont {Brune}},\ }\Doi {10.1103/PhysRevLett.111.236801}
  {\bibfield  {journal} {\bibinfo  {journal} {Phys. Rev. Lett.},\ }\textbf
  {\bibinfo {volume} {111}},\ \bibinfo {pages} {236801} (\bibinfo {year}
  {2013})}\BibitemShut {NoStop}%
\bibitem [{\citenamefont {Getzlaff}\ \emph {et~al.}(2006)\citenamefont
  {Getzlaff}, \citenamefont {Bansmann}, \citenamefont {Bulut}, \citenamefont
  {Gebhardt}, \citenamefont {Kleibert},\ and\ \citenamefont
  {Meiwes-Broer}}]{Getzlaff06}%
  \BibitemOpen
  \bibfield  {author} {\bibinfo {author} {\bibfnamefont {M.}~\bibnamefont
  {Getzlaff}}, \bibinfo {author} {\bibfnamefont {J.}~\bibnamefont {Bansmann}},
  \bibinfo {author} {\bibfnamefont {F.}~\bibnamefont {Bulut}}, \bibinfo
  {author} {\bibfnamefont {R.}~\bibnamefont {Gebhardt}}, \bibinfo {author}
  {\bibfnamefont {A.}~\bibnamefont {Kleibert}}, \ and\ \bibinfo {author}
  {\bibfnamefont {K.}~\bibnamefont {Meiwes-Broer}},\ }\Doi
  {10.1007/s00339-005-3347-5} {\bibfield  {journal} {\bibinfo  {journal} {Appl.
  Phys. A: Mater. Sci. Process.},\ }\textbf {\bibinfo {volume} {82}},\ \bibinfo
  {pages} {95} (\bibinfo {year} {2006})}\BibitemShut {NoStop}%
\bibitem [{\citenamefont {Cox}\ \emph {et~al.}(1985)\citenamefont {Cox},
  \citenamefont {Trevor}, \citenamefont {Whetten}, \citenamefont {Rohlfing},\
  and\ \citenamefont {Kaldor}}]{Cox85}%
  \BibitemOpen
  \bibfield  {author} {\bibinfo {author} {\bibfnamefont {D.~M.}\ \bibnamefont
  {Cox}}, \bibinfo {author} {\bibfnamefont {D.~J.}\ \bibnamefont {Trevor}},
  \bibinfo {author} {\bibfnamefont {R.~L.}\ \bibnamefont {Whetten}}, \bibinfo
  {author} {\bibfnamefont {E.~A.}\ \bibnamefont {Rohlfing}}, \ and\ \bibinfo
  {author} {\bibfnamefont {A.}~\bibnamefont {Kaldor}},\ }\Doi
  {10.1103/PhysRevB.32.7290} {\bibfield  {journal} {\bibinfo  {journal} {Phys.
  Rev. B},\ }\textbf {\bibinfo {volume} {32}},\ \bibinfo {pages} {7290}
  (\bibinfo {year} {1985})}\BibitemShut {NoStop}%
\bibitem [{\citenamefont {Knickelbein}(2006)}]{Knickelbein06}%
  \BibitemOpen
  \bibfield  {author} {\bibinfo {author} {\bibfnamefont {M.~B.}\ \bibnamefont
  {Knickelbein}},\ }\Doi {10.1063/1.2217951} {\bibfield  {journal} {\bibinfo
  {journal} {J. Chem. Phys.},\ }\textbf {\bibinfo {volume} {125}},\ \bibinfo
  {eid} {044308} (\bibinfo {year} {2006})}\BibitemShut {NoStop}%
\bibitem [{\citenamefont {Payne}\ \emph {et~al.}(2007)\citenamefont {Payne},
  \citenamefont {Jiang}, \citenamefont {Emmert}, \citenamefont {Deng},\ and\
  \citenamefont {Bloomfield}}]{Payne07}%
  \BibitemOpen
  \bibfield  {author} {\bibinfo {author} {\bibfnamefont {F.~W.}\ \bibnamefont
  {Payne}}, \bibinfo {author} {\bibfnamefont {W.}~\bibnamefont {Jiang}},
  \bibinfo {author} {\bibfnamefont {J.~W.}\ \bibnamefont {Emmert}}, \bibinfo
  {author} {\bibfnamefont {J.}~\bibnamefont {Deng}}, \ and\ \bibinfo {author}
  {\bibfnamefont {L.~A.}\ \bibnamefont {Bloomfield}},\ }\Doi
  {10.1103/PhysRevB.75.094431} {\bibfield  {journal} {\bibinfo  {journal}
  {Phys. Rev. B},\ }\textbf {\bibinfo {volume} {75}},\ \bibinfo {pages}
  {094431} (\bibinfo {year} {2007})}\BibitemShut {NoStop}%
\bibitem [{\citenamefont {Rohrmann}\ and\ \citenamefont
  {Sch\"afer}(2013)}]{Rohrmann13}%
  \BibitemOpen
  \bibfield  {author} {\bibinfo {author} {\bibfnamefont {U.}~\bibnamefont
  {Rohrmann}}\ and\ \bibinfo {author} {\bibfnamefont {R.}~\bibnamefont
  {Sch\"afer}},\ }\Doi {10.1103/PhysRevLett.111.133401} {\bibfield  {journal}
  {\bibinfo  {journal} {Phys. Rev. Lett.},\ }\textbf {\bibinfo {volume}
  {111}},\ \bibinfo {pages} {133401} (\bibinfo {year} {2013})}\BibitemShut
  {NoStop}%
\bibitem [{\citenamefont {Erskine}\ and\ \citenamefont
  {Stern}(1975)}]{Erskine75}%
  \BibitemOpen
  \bibfield  {author} {\bibinfo {author} {\bibfnamefont {J.~L.}\ \bibnamefont
  {Erskine}}\ and\ \bibinfo {author} {\bibfnamefont {E.~A.}\ \bibnamefont
  {Stern}},\ }\href@noop {} {\bibfield  {journal} {\bibinfo  {journal} {Phys.
  Rev. B},\ }\textbf {\bibinfo {volume} {12}},\ \bibinfo {pages} {5016}
  (\bibinfo {year} {1975})}\BibitemShut {NoStop}%
\bibitem [{\citenamefont {Sch\"utz}\ \emph {et~al.}(1987)\citenamefont
  {Sch\"utz}, \citenamefont {Wagner}, \citenamefont {Wilhelm}, \citenamefont
  {Kienle}, \citenamefont {Zeller}, \citenamefont {Frahm},\ and\ \citenamefont
  {Materlik}}]{Schuetz87}%
  \BibitemOpen
  \bibfield  {author} {\bibinfo {author} {\bibfnamefont {G.}~\bibnamefont
  {Sch\"utz}}, \bibinfo {author} {\bibfnamefont {W.}~\bibnamefont {Wagner}},
  \bibinfo {author} {\bibfnamefont {W.}~\bibnamefont {Wilhelm}}, \bibinfo
  {author} {\bibfnamefont {P.}~\bibnamefont {Kienle}}, \bibinfo {author}
  {\bibfnamefont {R.}~\bibnamefont {Zeller}}, \bibinfo {author} {\bibfnamefont
  {R.}~\bibnamefont {Frahm}}, \ and\ \bibinfo {author} {\bibfnamefont
  {G.}~\bibnamefont {Materlik}},\ }\Doi {10.1103/PhysRevLett.58.737} {\bibfield
   {journal} {\bibinfo  {journal} {Phys. Rev. Lett.},\ }\textbf {\bibinfo
  {volume} {58}},\ \bibinfo {pages} {737} (\bibinfo {year} {1987})}\BibitemShut
  {NoStop}%
\bibitem [{\citenamefont {Thole}\ \emph {et~al.}(1992)\citenamefont {Thole},
  \citenamefont {Carra}, \citenamefont {Sette},\ and\ \citenamefont {van~der
  Laan}}]{Thole92}%
  \BibitemOpen
  \bibfield  {author} {\bibinfo {author} {\bibfnamefont {B.~T.}\ \bibnamefont
  {Thole}}, \bibinfo {author} {\bibfnamefont {P.}~\bibnamefont {Carra}},
  \bibinfo {author} {\bibfnamefont {F.}~\bibnamefont {Sette}}, \ and\ \bibinfo
  {author} {\bibfnamefont {G.}~\bibnamefont {van~der Laan}},\ }\Doi
  {10.1103/PhysRevLett.68.1943} {\bibfield  {journal} {\bibinfo  {journal}
  {Phys. Rev. Lett.},\ }\textbf {\bibinfo {volume} {68}},\ \bibinfo {pages}
  {1943} (\bibinfo {year} {1992})}\BibitemShut {NoStop}%
\bibitem [{\citenamefont {Carra}\ \emph {et~al.}(1993)\citenamefont {Carra},
  \citenamefont {Thole}, \citenamefont {Altarelli},\ and\ \citenamefont
  {Wang}}]{Carra93}%
  \BibitemOpen
  \bibfield  {author} {\bibinfo {author} {\bibfnamefont {P.}~\bibnamefont
  {Carra}}, \bibinfo {author} {\bibfnamefont {B.~T.}\ \bibnamefont {Thole}},
  \bibinfo {author} {\bibfnamefont {M.}~\bibnamefont {Altarelli}}, \ and\
  \bibinfo {author} {\bibfnamefont {X.}~\bibnamefont {Wang}},\ }\Doi
  {10.1103/PhysRevLett.70.694} {\bibfield  {journal} {\bibinfo  {journal}
  {Phys. Rev. Lett.},\ }\textbf {\bibinfo {volume} {70}},\ \bibinfo {pages}
  {694} (\bibinfo {year} {1993})}\BibitemShut {NoStop}%
\bibitem [{\citenamefont {Chen}\ \emph {et~al.}(1995)\citenamefont {Chen},
  \citenamefont {Idzerda}, \citenamefont {Lin}, \citenamefont {Smith},
  \citenamefont {Meigs}, \citenamefont {Chaban}, \citenamefont {Ho},
  \citenamefont {Pellegrin},\ and\ \citenamefont {Sette}}]{Chen95}%
  \BibitemOpen
  \bibfield  {author} {\bibinfo {author} {\bibfnamefont {C.~T.}\ \bibnamefont
  {Chen}}, \bibinfo {author} {\bibfnamefont {Y.~U.}\ \bibnamefont {Idzerda}},
  \bibinfo {author} {\bibfnamefont {H.-J.}\ \bibnamefont {Lin}}, \bibinfo
  {author} {\bibfnamefont {N.~V.}\ \bibnamefont {Smith}}, \bibinfo {author}
  {\bibfnamefont {G.}~\bibnamefont {Meigs}}, \bibinfo {author} {\bibfnamefont
  {E.}~\bibnamefont {Chaban}}, \bibinfo {author} {\bibfnamefont {G.~H.}\
  \bibnamefont {Ho}}, \bibinfo {author} {\bibfnamefont {E.}~\bibnamefont
  {Pellegrin}}, \ and\ \bibinfo {author} {\bibfnamefont {F.}~\bibnamefont
  {Sette}},\ }\Doi {10.1103/PhysRevLett.75.152} {\bibfield  {journal} {\bibinfo
   {journal} {Phys. Rev. Lett.},\ }\textbf {\bibinfo {volume} {75}},\ \bibinfo
  {pages} {152} (\bibinfo {year} {1995})}\BibitemShut {NoStop}%
\bibitem [{\citenamefont {Peredkov}\ \emph {et~al.}(2011)\citenamefont
  {Peredkov}, \citenamefont {Neeb}, \citenamefont {Eberhardt}, \citenamefont
  {Meyer}, \citenamefont {Tombers}, \citenamefont {Kampschulte},\ and\
  \citenamefont {Niedner-Schatteburg}}]{Peredkov11b}%
  \BibitemOpen
  \bibfield  {author} {\bibinfo {author} {\bibfnamefont {S.}~\bibnamefont
  {Peredkov}}, \bibinfo {author} {\bibfnamefont {M.}~\bibnamefont {Neeb}},
  \bibinfo {author} {\bibfnamefont {W.}~\bibnamefont {Eberhardt}}, \bibinfo
  {author} {\bibfnamefont {J.}~\bibnamefont {Meyer}}, \bibinfo {author}
  {\bibfnamefont {M.}~\bibnamefont {Tombers}}, \bibinfo {author} {\bibfnamefont
  {H.}~\bibnamefont {Kampschulte}}, \ and\ \bibinfo {author} {\bibfnamefont
  {G.}~\bibnamefont {Niedner-Schatteburg}},\ }\Doi
  {10.1103/PhysRevLett.107.233401} {\bibfield  {journal} {\bibinfo  {journal}
  {Phys. Rev. Lett.},\ }\textbf {\bibinfo {volume} {107}},\ \bibinfo {pages}
  {233401} (\bibinfo {year} {2011})}\BibitemShut {NoStop}%
\bibitem [{\citenamefont {Niemeyer}\ \emph {et~al.}(2012)\citenamefont
  {Niemeyer}, \citenamefont {Hirsch}, \citenamefont {Zamudio-Bayer},
  \citenamefont {Langenberg}, \citenamefont {Vogel}, \citenamefont {Kossick},
  \citenamefont {Ebrecht}, \citenamefont {Egashira}, \citenamefont {Terasaki},
  \citenamefont {M\"oller}, \citenamefont {v.~Issendorff},\ and\ \citenamefont
  {Lau}}]{Niemeyer12}%
  \BibitemOpen
  \bibfield  {author} {\bibinfo {author} {\bibfnamefont {M.}~\bibnamefont
  {Niemeyer}}, \bibinfo {author} {\bibfnamefont {K.}~\bibnamefont {Hirsch}},
  \bibinfo {author} {\bibfnamefont {V.}~\bibnamefont {Zamudio-Bayer}}, \bibinfo
  {author} {\bibfnamefont {A.}~\bibnamefont {Langenberg}}, \bibinfo {author}
  {\bibfnamefont {M.}~\bibnamefont {Vogel}}, \bibinfo {author} {\bibfnamefont
  {M.}~\bibnamefont {Kossick}}, \bibinfo {author} {\bibfnamefont
  {C.}~\bibnamefont {Ebrecht}}, \bibinfo {author} {\bibfnamefont
  {K.}~\bibnamefont {Egashira}}, \bibinfo {author} {\bibfnamefont
  {A.}~\bibnamefont {Terasaki}}, \bibinfo {author} {\bibfnamefont
  {T.}~\bibnamefont {M\"oller}}, \bibinfo {author} {\bibfnamefont
  {B.}~\bibnamefont {v.~Issendorff}}, \ and\ \bibinfo {author} {\bibfnamefont
  {J.~T.}\ \bibnamefont {Lau}},\ }\Doi {10.1103/PhysRevLett.108.057201}
  {\bibfield  {journal} {\bibinfo  {journal} {Phys. Rev. Lett.},\ }\textbf
  {\bibinfo {volume} {108}},\ \bibinfo {pages} {057201} (\bibinfo {year}
  {2012})}\BibitemShut {NoStop}%
\bibitem [{\citenamefont {{Hirsch}}\ \emph {et~al.}(2013)\citenamefont
  {{Hirsch}}, \citenamefont {{Zamudio-Bayer}}, \citenamefont {{Langenberg}},
  \citenamefont {{Niemeyer}}, \citenamefont {{Langbehn}}, \citenamefont
  {{M{\"o}ller}}, \citenamefont {{Terasaki}}, \citenamefont {{Issendorff}},\
  and\ \citenamefont {{Lau}}}]{Hirsch13a}%
  \BibitemOpen
  \bibfield  {author} {\bibinfo {author} {\bibfnamefont {K.}~\bibnamefont
  {{Hirsch}}}, \bibinfo {author} {\bibfnamefont {V.}~\bibnamefont
  {{Zamudio-Bayer}}}, \bibinfo {author} {\bibfnamefont {A.}~\bibnamefont
  {{Langenberg}}}, \bibinfo {author} {\bibfnamefont {M.}~\bibnamefont
  {{Niemeyer}}}, \bibinfo {author} {\bibfnamefont {B.}~\bibnamefont
  {{Langbehn}}}, \bibinfo {author} {\bibfnamefont {T.}~\bibnamefont
  {{M{\"o}ller}}}, \bibinfo {author} {\bibfnamefont {A.}~\bibnamefont
  {{Terasaki}}}, \bibinfo {author} {\bibfnamefont {B.~v.}\ \bibnamefont
  {{Issendorff}}}, \ and\ \bibinfo {author} {\bibfnamefont {J.~T.}\
  \bibnamefont {{Lau}}},\ }\href@noop {} {\bibfield  {journal} {\bibinfo
  {journal} {arXiv e-prints}} (\bibinfo {year} {2013})},\ \Eprint
  {http://arxiv.org/abs/1304.7173} {arXiv:1304.7173} \BibitemShut {NoStop}%
\bibitem [{\citenamefont {Zamudio-Bayer}\ \emph {et~al.}(2013)\citenamefont
  {Zamudio-Bayer}, \citenamefont {Leppert}, \citenamefont {Hirsch},
  \citenamefont {Langenberg}, \citenamefont {Rittmann}, \citenamefont
  {Kossick}, \citenamefont {Vogel}, \citenamefont {Richter}, \citenamefont
  {Terasaki}, \citenamefont {M\"oller}, \citenamefont {v.~Issendorff},
  \citenamefont {K\"ummel},\ and\ \citenamefont {Lau}}]{ZamudioBayer13}%
  \BibitemOpen
  \bibfield  {author} {\bibinfo {author} {\bibfnamefont {V.}~\bibnamefont
  {Zamudio-Bayer}}, \bibinfo {author} {\bibfnamefont {L.}~\bibnamefont
  {Leppert}}, \bibinfo {author} {\bibfnamefont {K.}~\bibnamefont {Hirsch}},
  \bibinfo {author} {\bibfnamefont {A.}~\bibnamefont {Langenberg}}, \bibinfo
  {author} {\bibfnamefont {J.}~\bibnamefont {Rittmann}}, \bibinfo {author}
  {\bibfnamefont {M.}~\bibnamefont {Kossick}}, \bibinfo {author} {\bibfnamefont
  {M.}~\bibnamefont {Vogel}}, \bibinfo {author} {\bibfnamefont
  {R.}~\bibnamefont {Richter}}, \bibinfo {author} {\bibfnamefont
  {A.}~\bibnamefont {Terasaki}}, \bibinfo {author} {\bibfnamefont
  {T.}~\bibnamefont {M\"oller}}, \bibinfo {author} {\bibfnamefont
  {B.}~\bibnamefont {v.~Issendorff}}, \bibinfo {author} {\bibfnamefont
  {S.}~\bibnamefont {K\"ummel}}, \ and\ \bibinfo {author} {\bibfnamefont
  {J.~T.}\ \bibnamefont {Lau}},\ }\Doi {10.1103/PhysRevB.88.115425} {\bibfield
  {journal} {\bibinfo  {journal} {Phys. Rev. B},\ }\textbf {\bibinfo {volume}
  {88}},\ \bibinfo {pages} {115425} (\bibinfo {year} {2013})}\BibitemShut
  {NoStop}%
\bibitem [{\citenamefont {Bancroft}\ \emph {et~al.}(1956)\citenamefont
  {Bancroft}, \citenamefont {Peterson},\ and\ \citenamefont
  {Minshall}}]{Bancroft56}%
  \BibitemOpen
  \bibfield  {author} {\bibinfo {author} {\bibfnamefont {D.}~\bibnamefont
  {Bancroft}}, \bibinfo {author} {\bibfnamefont {E.~L.}\ \bibnamefont
  {Peterson}}, \ and\ \bibinfo {author} {\bibfnamefont {S.}~\bibnamefont
  {Minshall}},\ }\Doi {10.1063/1.1722359} {\bibfield  {journal} {\bibinfo
  {journal} {J. Appl. Phys.},\ }\textbf {\bibinfo {volume} {27}},\ \bibinfo
  {pages} {291} (\bibinfo {year} {1956})}\BibitemShut {NoStop}%
\bibitem [{\citenamefont {Jamieson}\ and\ \citenamefont
  {Lawson}(1962)}]{Jamieson62}%
  \BibitemOpen
  \bibfield  {author} {\bibinfo {author} {\bibfnamefont {J.~C.}\ \bibnamefont
  {Jamieson}}\ and\ \bibinfo {author} {\bibfnamefont {A.~W.}\ \bibnamefont
  {Lawson}},\ }\Doi {10.1063/1.1777167} {\bibfield  {journal} {\bibinfo
  {journal} {J. Appl. Phys.},\ }\textbf {\bibinfo {volume} {33}},\ \bibinfo
  {pages} {776} (\bibinfo {year} {1962})}\BibitemShut {NoStop}%
\bibitem [{\citenamefont {Taylor}\ \emph {et~al.}(1991)\citenamefont {Taylor},
  \citenamefont {Pasternak},\ and\ \citenamefont {Jeanloz}}]{Taylor91}%
  \BibitemOpen
  \bibfield  {author} {\bibinfo {author} {\bibfnamefont {R.~D.}\ \bibnamefont
  {Taylor}}, \bibinfo {author} {\bibfnamefont {M.~P.}\ \bibnamefont
  {Pasternak}}, \ and\ \bibinfo {author} {\bibfnamefont {R.}~\bibnamefont
  {Jeanloz}},\ }\Doi {10.1063/1.348779} {\bibfield  {journal} {\bibinfo
  {journal} {J. Appl. Phys.},\ }\textbf {\bibinfo {volume} {69}},\ \bibinfo
  {pages} {6126} (\bibinfo {year} {1991})}\BibitemShut {NoStop}%
\bibitem [{\citenamefont {Wang}\ and\ \citenamefont {Ingalls}(1998)}]{Wang98b}%
  \BibitemOpen
  \bibfield  {author} {\bibinfo {author} {\bibfnamefont {F.~M.}\ \bibnamefont
  {Wang}}\ and\ \bibinfo {author} {\bibfnamefont {R.}~\bibnamefont {Ingalls}},\
  }\Doi {10.1103/PhysRevB.57.5647} {\bibfield  {journal} {\bibinfo  {journal}
  {Phys. Rev. B},\ }\textbf {\bibinfo {volume} {57}},\ \bibinfo {pages} {5647}
  (\bibinfo {year} {1998})}\BibitemShut {NoStop}%
\bibitem [{\citenamefont {Rueff}\ \emph {et~al.}(1999)\citenamefont {Rueff},
  \citenamefont {Krisch}, \citenamefont {Cai}, \citenamefont {Kaprolat},
  \citenamefont {Hanfland}, \citenamefont {Lorenzen}, \citenamefont
  {Masciovecchio}, \citenamefont {Verbeni},\ and\ \citenamefont
  {Sette}}]{Rueff99}%
  \BibitemOpen
  \bibfield  {author} {\bibinfo {author} {\bibfnamefont {J.~P.}\ \bibnamefont
  {Rueff}}, \bibinfo {author} {\bibfnamefont {M.}~\bibnamefont {Krisch}},
  \bibinfo {author} {\bibfnamefont {Y.~Q.}\ \bibnamefont {Cai}}, \bibinfo
  {author} {\bibfnamefont {A.}~\bibnamefont {Kaprolat}}, \bibinfo {author}
  {\bibfnamefont {M.}~\bibnamefont {Hanfland}}, \bibinfo {author}
  {\bibfnamefont {M.}~\bibnamefont {Lorenzen}}, \bibinfo {author}
  {\bibfnamefont {C.}~\bibnamefont {Masciovecchio}}, \bibinfo {author}
  {\bibfnamefont {R.}~\bibnamefont {Verbeni}}, \ and\ \bibinfo {author}
  {\bibfnamefont {F.}~\bibnamefont {Sette}},\ }\Doi {10.1103/PhysRevB.60.14510}
  {\bibfield  {journal} {\bibinfo  {journal} {Phys. Rev. B},\ }\textbf
  {\bibinfo {volume} {60}},\ \bibinfo {pages} {14510} (\bibinfo {year}
  {1999})}\BibitemShut {NoStop}%
\bibitem [{\citenamefont {Mathon}\ \emph {et~al.}(2004)\citenamefont {Mathon},
  \citenamefont {Baudelet}, \citenamefont {Iti\'e}, \citenamefont {Polian},
  \citenamefont {d'Astuto}, \citenamefont {Chervin},\ and\ \citenamefont
  {Pascarelli}}]{Mathon04}%
  \BibitemOpen
  \bibfield  {author} {\bibinfo {author} {\bibfnamefont {O.}~\bibnamefont
  {Mathon}}, \bibinfo {author} {\bibfnamefont {F.}~\bibnamefont {Baudelet}},
  \bibinfo {author} {\bibfnamefont {J.~P.}\ \bibnamefont {Iti\'e}}, \bibinfo
  {author} {\bibfnamefont {A.}~\bibnamefont {Polian}}, \bibinfo {author}
  {\bibfnamefont {M.}~\bibnamefont {d'Astuto}}, \bibinfo {author}
  {\bibfnamefont {J.~C.}\ \bibnamefont {Chervin}}, \ and\ \bibinfo {author}
  {\bibfnamefont {S.}~\bibnamefont {Pascarelli}},\ }\Doi
  {10.1103/PhysRevLett.93.255503} {\bibfield  {journal} {\bibinfo  {journal}
  {Phys. Rev. Lett.},\ }\textbf {\bibinfo {volume} {93}},\ \bibinfo {pages}
  {255503} (\bibinfo {year} {2004})}\BibitemShut {NoStop}%
\bibitem [{\citenamefont {Ishimatsu}\ \emph {et~al.}(2007)\citenamefont
  {Ishimatsu}, \citenamefont {Maruyama}, \citenamefont {Kawamura},
  \citenamefont {Suzuki}, \citenamefont {Ohishi},\ and\ \citenamefont
  {Shimomura}}]{Ishimatsu07}%
  \BibitemOpen
  \bibfield  {author} {\bibinfo {author} {\bibfnamefont {N.}~\bibnamefont
  {Ishimatsu}}, \bibinfo {author} {\bibfnamefont {H.}~\bibnamefont {Maruyama}},
  \bibinfo {author} {\bibfnamefont {N.}~\bibnamefont {Kawamura}}, \bibinfo
  {author} {\bibfnamefont {M.}~\bibnamefont {Suzuki}}, \bibinfo {author}
  {\bibfnamefont {Y.}~\bibnamefont {Ohishi}}, \ and\ \bibinfo {author}
  {\bibfnamefont {O.}~\bibnamefont {Shimomura}},\ }\Doi
  {10.1143/JPSJ.76.064703} {\bibfield  {journal} {\bibinfo  {journal} {J. Phys.
  Soc. Jpn.},\ }\textbf {\bibinfo {volume} {76}},\ \bibinfo {pages} {064703}
  (\bibinfo {year} {2007})}\BibitemShut {NoStop}%
\bibitem [{\citenamefont {Lau}\ \emph {et~al.}(2008)\citenamefont {Lau},
  \citenamefont {Rittmann}, \citenamefont {Zamudio-Bayer}, \citenamefont
  {Vogel}, \citenamefont {Hirsch}, \citenamefont {Klar}, \citenamefont
  {Lofink}, \citenamefont {M\"{o}ller},\ and\ \citenamefont
  {v.~Issendorff}}]{Lau08}%
  \BibitemOpen
  \bibfield  {author} {\bibinfo {author} {\bibfnamefont {J.~T.}\ \bibnamefont
  {Lau}}, \bibinfo {author} {\bibfnamefont {J.}~\bibnamefont {Rittmann}},
  \bibinfo {author} {\bibfnamefont {V.}~\bibnamefont {Zamudio-Bayer}}, \bibinfo
  {author} {\bibfnamefont {M.}~\bibnamefont {Vogel}}, \bibinfo {author}
  {\bibfnamefont {K.}~\bibnamefont {Hirsch}}, \bibinfo {author} {\bibfnamefont
  {P.}~\bibnamefont {Klar}}, \bibinfo {author} {\bibfnamefont {F.}~\bibnamefont
  {Lofink}}, \bibinfo {author} {\bibfnamefont {T.}~\bibnamefont {M\"{o}ller}},
  \ and\ \bibinfo {author} {\bibfnamefont {B.}~\bibnamefont {v.~Issendorff}},\
  }\Doi {10.1103/PhysRevLett.101.153401} {\bibfield  {journal} {\bibinfo
  {journal} {Phys. Rev. Lett.},\ }\textbf {\bibinfo {volume} {101}},\ \bibinfo
  {eid} {153401} (\bibinfo {year} {2008})}\BibitemShut {NoStop}%
\bibitem [{\citenamefont {Lau}\ \emph {et~al.}(2009){\natexlab{a}}\citenamefont
  {Lau}, \citenamefont {Hirsch}, \citenamefont {Klar}, \citenamefont
  {Langenberg}, \citenamefont {Lofink}, \citenamefont {Richter}, \citenamefont
  {Rittmann}, \citenamefont {Vogel}, \citenamefont {Zamudio-Bayer},
  \citenamefont {M\"oller},\ and\ \citenamefont {v.~Issendorff}}]{Lau09a}%
  \BibitemOpen
  \bibfield  {author} {\bibinfo {author} {\bibfnamefont {J.~T.}\ \bibnamefont
  {Lau}}, \bibinfo {author} {\bibfnamefont {K.}~\bibnamefont {Hirsch}},
  \bibinfo {author} {\bibfnamefont {P.}~\bibnamefont {Klar}}, \bibinfo {author}
  {\bibfnamefont {A.}~\bibnamefont {Langenberg}}, \bibinfo {author}
  {\bibfnamefont {F.}~\bibnamefont {Lofink}}, \bibinfo {author} {\bibfnamefont
  {R.}~\bibnamefont {Richter}}, \bibinfo {author} {\bibfnamefont
  {J.}~\bibnamefont {Rittmann}}, \bibinfo {author} {\bibfnamefont
  {M.}~\bibnamefont {Vogel}}, \bibinfo {author} {\bibfnamefont
  {V.}~\bibnamefont {Zamudio-Bayer}}, \bibinfo {author} {\bibfnamefont
  {T.}~\bibnamefont {M\"oller}}, \ and\ \bibinfo {author} {\bibfnamefont
  {B.}~\bibnamefont {v.~Issendorff}},\ }\Doi {10.1103/PhysRevA.79.053201}
  {\bibfield  {journal} {\bibinfo  {journal} {Phys. Rev. A},\ }\textbf
  {\bibinfo {volume} {79}},\ \bibinfo {eid} {053201} (\bibinfo {year}
  {2009}{\natexlab{a}})}\BibitemShut {NoStop}%
\bibitem [{\citenamefont {Lau}\ \emph {et~al.}(2009){\natexlab{b}}\citenamefont
  {Lau}, \citenamefont {Hirsch}, \citenamefont {Langenberg}, \citenamefont
  {Probst}, \citenamefont {Richter}, \citenamefont {Rittmann}, \citenamefont
  {Vogel}, \citenamefont {Zamudio-Bayer}, \citenamefont {M\"oller},\ and\
  \citenamefont {von Issendorff}}]{Lau09b}%
  \BibitemOpen
  \bibfield  {author} {\bibinfo {author} {\bibfnamefont {J.~T.}\ \bibnamefont
  {Lau}}, \bibinfo {author} {\bibfnamefont {K.}~\bibnamefont {Hirsch}},
  \bibinfo {author} {\bibfnamefont {A.}~\bibnamefont {Langenberg}}, \bibinfo
  {author} {\bibfnamefont {J.}~\bibnamefont {Probst}}, \bibinfo {author}
  {\bibfnamefont {R.}~\bibnamefont {Richter}}, \bibinfo {author} {\bibfnamefont
  {J.}~\bibnamefont {Rittmann}}, \bibinfo {author} {\bibfnamefont
  {M.}~\bibnamefont {Vogel}}, \bibinfo {author} {\bibfnamefont
  {V.}~\bibnamefont {Zamudio-Bayer}}, \bibinfo {author} {\bibfnamefont
  {T.}~\bibnamefont {M\"oller}}, \ and\ \bibinfo {author} {\bibfnamefont
  {B.}~\bibnamefont {von Issendorff}},\ }\Doi {10.1103/PhysRevB.79.241102}
  {\bibfield  {journal} {\bibinfo  {journal} {Phys. Rev. B},\ }\textbf
  {\bibinfo {volume} {79}},\ \bibinfo {eid} {241102} (\bibinfo {year}
  {2009}{\natexlab{b}})}\BibitemShut {NoStop}%
\bibitem [{\citenamefont {Hirsch}\ \emph {et~al.}(2009)\citenamefont {Hirsch},
  \citenamefont {Lau}, \citenamefont {Klar}, \citenamefont {Langenberg},
  \citenamefont {Probst}, \citenamefont {Rittmann}, \citenamefont {Vogel},
  \citenamefont {Zamudio-Bayer}, \citenamefont {M\"oller},\ and\ \citenamefont
  {von Issendorff}}]{Hirsch09}%
  \BibitemOpen
  \bibfield  {author} {\bibinfo {author} {\bibfnamefont {K.}~\bibnamefont
  {Hirsch}}, \bibinfo {author} {\bibfnamefont {J.~T.}\ \bibnamefont {Lau}},
  \bibinfo {author} {\bibfnamefont {P.}~\bibnamefont {Klar}}, \bibinfo {author}
  {\bibfnamefont {A.}~\bibnamefont {Langenberg}}, \bibinfo {author}
  {\bibfnamefont {J.}~\bibnamefont {Probst}}, \bibinfo {author} {\bibfnamefont
  {J.}~\bibnamefont {Rittmann}}, \bibinfo {author} {\bibfnamefont
  {M.}~\bibnamefont {Vogel}}, \bibinfo {author} {\bibfnamefont
  {V.}~\bibnamefont {Zamudio-Bayer}}, \bibinfo {author} {\bibfnamefont
  {T.}~\bibnamefont {M\"oller}}, \ and\ \bibinfo {author} {\bibfnamefont
  {B.}~\bibnamefont {von Issendorff}},\ }\Doi
  {http://dx.doi.org/10.1088/0953-4075/42/15/154029} {\bibfield  {journal}
  {\bibinfo  {journal} {J. Phys. B: At. Mol. Opt. Phys.},\ }\textbf {\bibinfo
  {volume} {42}},\ \bibinfo {pages} {154029} (\bibinfo {year}
  {2009})}\BibitemShut {NoStop}%
\bibitem [{\citenamefont {Lau}\ \emph {et~al.}(2011)\citenamefont {Lau},
  \citenamefont {Vogel}, \citenamefont {Langenberg}, \citenamefont {Hirsch},
  \citenamefont {Rittmann}, \citenamefont {Zamudio-Bayer}, \citenamefont
  {M\"oller},\ and\ \citenamefont {von Issendorff}}]{Lau11}%
  \BibitemOpen
  \bibfield  {author} {\bibinfo {author} {\bibfnamefont {J.~T.}\ \bibnamefont
  {Lau}}, \bibinfo {author} {\bibfnamefont {M.}~\bibnamefont {Vogel}}, \bibinfo
  {author} {\bibfnamefont {A.}~\bibnamefont {Langenberg}}, \bibinfo {author}
  {\bibfnamefont {K.}~\bibnamefont {Hirsch}}, \bibinfo {author} {\bibfnamefont
  {J.}~\bibnamefont {Rittmann}}, \bibinfo {author} {\bibfnamefont
  {V.}~\bibnamefont {Zamudio-Bayer}}, \bibinfo {author} {\bibfnamefont
  {T.}~\bibnamefont {M\"oller}}, \ and\ \bibinfo {author} {\bibfnamefont
  {B.}~\bibnamefont {von Issendorff}},\ }\Doi {10.1063/1.3547699} {\bibfield
  {journal} {\bibinfo  {journal} {J. Chem. Phys.},\ }\textbf {\bibinfo {volume}
  {134}},\ \bibinfo {eid} {041102} (\bibinfo {year} {2011})}\BibitemShut
  {NoStop}%
\bibitem [{\citenamefont {Vogel}\ \emph {et~al.}(2012)\citenamefont {Vogel},
  \citenamefont {Kasigkeit}, \citenamefont {Hirsch}, \citenamefont
  {Langenberg}, \citenamefont {Rittmann}, \citenamefont {Zamudio-Bayer},
  \citenamefont {Kulesza}, \citenamefont {Mitri\'{c}}, \citenamefont
  {M\"oller}, \citenamefont {v.~Issendorff},\ and\ \citenamefont
  {Lau}}]{Vogel12}%
  \BibitemOpen
  \bibfield  {author} {\bibinfo {author} {\bibfnamefont {M.}~\bibnamefont
  {Vogel}}, \bibinfo {author} {\bibfnamefont {C.}~\bibnamefont {Kasigkeit}},
  \bibinfo {author} {\bibfnamefont {K.}~\bibnamefont {Hirsch}}, \bibinfo
  {author} {\bibfnamefont {A.}~\bibnamefont {Langenberg}}, \bibinfo {author}
  {\bibfnamefont {J.}~\bibnamefont {Rittmann}}, \bibinfo {author}
  {\bibfnamefont {V.}~\bibnamefont {Zamudio-Bayer}}, \bibinfo {author}
  {\bibfnamefont {A.}~\bibnamefont {Kulesza}}, \bibinfo {author} {\bibfnamefont
  {R.}~\bibnamefont {Mitri\'{c}}}, \bibinfo {author} {\bibfnamefont
  {T.}~\bibnamefont {M\"oller}}, \bibinfo {author} {\bibfnamefont
  {B.}~\bibnamefont {v.~Issendorff}}, \ and\ \bibinfo {author} {\bibfnamefont
  {J.~T.}\ \bibnamefont {Lau}},\ }\Doi {10.1103/PhysRevB.85.195454} {\bibfield
  {journal} {\bibinfo  {journal} {Phys. Rev. B},\ }\textbf {\bibinfo {volume}
  {85}},\ \bibinfo {pages} {195454} (\bibinfo {year} {2012})}\BibitemShut
  {NoStop}%
\bibitem [{\citenamefont {Hirsch}\ \emph
  {et~al.}(2012){\natexlab{a}}\citenamefont {Hirsch}, \citenamefont
  {Zamudio-Bayer}, \citenamefont {Rittmann}, \citenamefont {Langenberg},
  \citenamefont {Vogel}, \citenamefont {M\"oller}, \citenamefont
  {v.~Issendorff},\ and\ \citenamefont {Lau}}]{Hirsch12b}%
  \BibitemOpen
  \bibfield  {author} {\bibinfo {author} {\bibfnamefont {K.}~\bibnamefont
  {Hirsch}}, \bibinfo {author} {\bibfnamefont {V.}~\bibnamefont
  {Zamudio-Bayer}}, \bibinfo {author} {\bibfnamefont {J.}~\bibnamefont
  {Rittmann}}, \bibinfo {author} {\bibfnamefont {A.}~\bibnamefont
  {Langenberg}}, \bibinfo {author} {\bibfnamefont {M.}~\bibnamefont {Vogel}},
  \bibinfo {author} {\bibfnamefont {T.}~\bibnamefont {M\"oller}}, \bibinfo
  {author} {\bibfnamefont {B.}~\bibnamefont {v.~Issendorff}}, \ and\ \bibinfo
  {author} {\bibfnamefont {J.~T.}\ \bibnamefont {Lau}},\ }\Doi
  {10.1103/PhysRevB.86.165402} {\bibfield  {journal} {\bibinfo  {journal}
  {Phys. Rev. B},\ }\textbf {\bibinfo {volume} {86}},\ \bibinfo {pages}
  {165402} (\bibinfo {year} {2012}{\natexlab{a}})}\BibitemShut {NoStop}%
\bibitem [{\citenamefont {van~der Laan}\ and\ \citenamefont
  {Thole}(1991)}]{vanderLaan91}%
  \BibitemOpen
  \bibfield  {author} {\bibinfo {author} {\bibfnamefont {G.}~\bibnamefont
  {van~der Laan}}\ and\ \bibinfo {author} {\bibfnamefont {B.~T.}\ \bibnamefont
  {Thole}},\ }\Doi {10.1103/PhysRevB.43.13401} {\bibfield  {journal} {\bibinfo
  {journal} {Phys. Rev. B},\ }\textbf {\bibinfo {volume} {43}},\ \bibinfo
  {pages} {13401} (\bibinfo {year} {1991})}\BibitemShut {NoStop}%
\bibitem [{\citenamefont {Hirsch}\ \emph
  {et~al.}(2012){\natexlab{b}}\citenamefont {Hirsch}, \citenamefont
  {Zamudio-Bayer}, \citenamefont {Ameseder}, \citenamefont {Langenberg},
  \citenamefont {Rittmann}, \citenamefont {Vogel}, \citenamefont {M\"oller},
  \citenamefont {v.~Issendorff},\ and\ \citenamefont {Lau}}]{Hirsch12a}%
  \BibitemOpen
  \bibfield  {author} {\bibinfo {author} {\bibfnamefont {K.}~\bibnamefont
  {Hirsch}}, \bibinfo {author} {\bibfnamefont {V.}~\bibnamefont
  {Zamudio-Bayer}}, \bibinfo {author} {\bibfnamefont {F.}~\bibnamefont
  {Ameseder}}, \bibinfo {author} {\bibfnamefont {A.}~\bibnamefont
  {Langenberg}}, \bibinfo {author} {\bibfnamefont {J.}~\bibnamefont
  {Rittmann}}, \bibinfo {author} {\bibfnamefont {M.}~\bibnamefont {Vogel}},
  \bibinfo {author} {\bibfnamefont {T.}~\bibnamefont {M\"oller}}, \bibinfo
  {author} {\bibfnamefont {B.}~\bibnamefont {v.~Issendorff}}, \ and\ \bibinfo
  {author} {\bibfnamefont {J.~T.}\ \bibnamefont {Lau}},\ }\Doi
  {10.1103/PhysRevA.85.062501} {\bibfield  {journal} {\bibinfo  {journal}
  {Phys. Rev. A},\ }\textbf {\bibinfo {volume} {85}},\ \bibinfo {pages}
  {062501} (\bibinfo {year} {2012}{\natexlab{b}})}\BibitemShut {NoStop}%
\bibitem [{\citenamefont {Aguilera-Granja}\ \emph {et~al.}(1998)\citenamefont
  {Aguilera-Granja}, \citenamefont {Bouarab}, \citenamefont {L\'opez},
  \citenamefont {Vega}, \citenamefont {Montejano-Carrizales}, \citenamefont
  {I\~niguez},\ and\ \citenamefont {Alonso}}]{AguileraGranja98}%
  \BibitemOpen
  \bibfield  {author} {\bibinfo {author} {\bibfnamefont {F.}~\bibnamefont
  {Aguilera-Granja}}, \bibinfo {author} {\bibfnamefont {S.}~\bibnamefont
  {Bouarab}}, \bibinfo {author} {\bibfnamefont {M.~J.}\ \bibnamefont
  {L\'opez}}, \bibinfo {author} {\bibfnamefont {A.}~\bibnamefont {Vega}},
  \bibinfo {author} {\bibfnamefont {J.~M.}\ \bibnamefont
  {Montejano-Carrizales}}, \bibinfo {author} {\bibfnamefont {M.~P.}\
  \bibnamefont {I\~niguez}}, \ and\ \bibinfo {author} {\bibfnamefont {J.~A.}\
  \bibnamefont {Alonso}},\ }\Doi {10.1103/PhysRevB.57.12469} {\bibfield
  {journal} {\bibinfo  {journal} {Phys. Rev. B},\ }\textbf {\bibinfo {volume}
  {57}},\ \bibinfo {pages} {12469} (\bibinfo {year} {1998})}\BibitemShut
  {NoStop}%
\bibitem [{\citenamefont {S\"oderlind}\ \emph {et~al.}(1992)\citenamefont
  {S\"oderlind}, \citenamefont {Eriksson}, \citenamefont {Johansson},
  \citenamefont {Albers},\ and\ \citenamefont {Boring}}]{Soederlind92}%
  \BibitemOpen
  \bibfield  {author} {\bibinfo {author} {\bibfnamefont {P.}~\bibnamefont
  {S\"oderlind}}, \bibinfo {author} {\bibfnamefont {O.}~\bibnamefont
  {Eriksson}}, \bibinfo {author} {\bibfnamefont {B.}~\bibnamefont {Johansson}},
  \bibinfo {author} {\bibfnamefont {R.~C.}\ \bibnamefont {Albers}}, \ and\
  \bibinfo {author} {\bibfnamefont {A.~M.}\ \bibnamefont {Boring}},\ }\Doi
  {10.1103/PhysRevB.45.12911} {\bibfield  {journal} {\bibinfo  {journal} {Phys.
  Rev. B},\ }\textbf {\bibinfo {volume} {45}},\ \bibinfo {pages} {12911}
  (\bibinfo {year} {1992})}\BibitemShut {NoStop}%
\bibitem [{\citenamefont {Alvarado-Leyva}\ \emph {et~al.}(2013)\citenamefont
  {Alvarado-Leyva}, \citenamefont {Aguilera-Granja}, \citenamefont
  {Balb\'{a}s},\ and\ \citenamefont {Vega}}]{AlvaradoLeyva13}%
  \BibitemOpen
  \bibfield  {author} {\bibinfo {author} {\bibfnamefont {P.~G.}\ \bibnamefont
  {Alvarado-Leyva}}, \bibinfo {author} {\bibfnamefont {F.}~\bibnamefont
  {Aguilera-Granja}}, \bibinfo {author} {\bibfnamefont {L.~C.}\ \bibnamefont
  {Balb\'{a}s}}, \ and\ \bibinfo {author} {\bibfnamefont {A.}~\bibnamefont
  {Vega}},\ }\Doi {10.1039/C3CP51377G} {\bibfield  {journal} {\bibinfo
  {journal} {Phys. Chem. Chem. Phys.},\ }\textbf {\bibinfo {volume} {15}},\
  \bibinfo {pages} {14458} (\bibinfo {year} {2013})}\BibitemShut {NoStop}%
\bibitem [{\citenamefont {Wu}\ \emph {et~al.}(2012)\citenamefont {Wu},
  \citenamefont {Kandalam}, \citenamefont {Gutsev},\ and\ \citenamefont
  {Jena}}]{Wu12b}%
  \BibitemOpen
  \bibfield  {author} {\bibinfo {author} {\bibfnamefont {M.}~\bibnamefont
  {Wu}}, \bibinfo {author} {\bibfnamefont {A.~K.}\ \bibnamefont {Kandalam}},
  \bibinfo {author} {\bibfnamefont {G.~L.}\ \bibnamefont {Gutsev}}, \ and\
  \bibinfo {author} {\bibfnamefont {P.}~\bibnamefont {Jena}},\ }\Doi
  {10.1103/PhysRevB.86.174410} {\bibfield  {journal} {\bibinfo  {journal}
  {Phys. Rev. B},\ }\textbf {\bibinfo {volume} {86}},\ \bibinfo {pages}
  {174410} (\bibinfo {year} {2012})}\BibitemShut {NoStop}%
\bibitem [{\citenamefont {Yuan}\ \emph {et~al.}(2013)\citenamefont {Yuan},
  \citenamefont {Chen}, \citenamefont {Kuang}, \citenamefont {Tian},\ and\
  \citenamefont {Wang}}]{Yuan13}%
  \BibitemOpen
  \bibfield  {author} {\bibinfo {author} {\bibfnamefont {H.~K.}\ \bibnamefont
  {Yuan}}, \bibinfo {author} {\bibfnamefont {H.}~\bibnamefont {Chen}}, \bibinfo
  {author} {\bibfnamefont {A.~L.}\ \bibnamefont {Kuang}}, \bibinfo {author}
  {\bibfnamefont {C.~L.}\ \bibnamefont {Tian}}, \ and\ \bibinfo {author}
  {\bibfnamefont {J.~Z.}\ \bibnamefont {Wang}},\ }\Doi {10.1063/1.4813611}
  {\bibfield  {journal} {\bibinfo  {journal} {J. Chem. Phys.},\ }\textbf
  {\bibinfo {volume} {139}},\ \bibinfo {eid} {034314} (\bibinfo {year}
  {2013})}\BibitemShut {NoStop}%
\bibitem [{\citenamefont {Yeh}\ and\ \citenamefont {Lindau}(1985)}]{Yeh85}%
  \BibitemOpen
  \bibfield  {author} {\bibinfo {author} {\bibfnamefont {J.~J.}\ \bibnamefont
  {Yeh}}\ and\ \bibinfo {author} {\bibfnamefont {I.}~\bibnamefont {Lindau}},\
  }\href
  {http://www.sciencedirect.com/science/article/B6WBB-4DBJ6HV-54/2/f1ad08f23a7%
d57f5adab72d3ece17af5} {\bibfield  {journal} {\bibinfo  {journal} {Atom. Data
  Nucl. Data},\ }\textbf {\bibinfo {volume} {32}},\ \bibinfo {pages} {1}
  (\bibinfo {year} {1985})}\BibitemShut {NoStop}%
\bibitem [{\citenamefont {Henke}\ \emph {et~al.}(1993)\citenamefont {Henke},
  \citenamefont {Gullikson},\ and\ \citenamefont {Davis}}]{Henke93}%
  \BibitemOpen
  \bibfield  {author} {\bibinfo {author} {\bibfnamefont {B.}~\bibnamefont
  {Henke}}, \bibinfo {author} {\bibfnamefont {E.}~\bibnamefont {Gullikson}}, \
  and\ \bibinfo {author} {\bibfnamefont {J.}~\bibnamefont {Davis}},\ }\Doi
  {http://dx.doi.org/10.1006/adnd.1993.1013} {\bibfield  {journal} {\bibinfo
  {journal} {Atom. Data Nucl. Data},\ }\textbf {\bibinfo {volume} {54}},\
  \bibinfo {pages} {181 } (\bibinfo {year} {1993})}\BibitemShut {NoStop}%
\bibitem [{\citenamefont {Chen}\ \emph {et~al.}(1991)\citenamefont {Chen},
  \citenamefont {Smith},\ and\ \citenamefont {Sette}}]{Chen91}%
  \BibitemOpen
  \bibfield  {author} {\bibinfo {author} {\bibfnamefont {C.~T.}\ \bibnamefont
  {Chen}}, \bibinfo {author} {\bibfnamefont {N.~V.}\ \bibnamefont {Smith}}, \
  and\ \bibinfo {author} {\bibfnamefont {F.}~\bibnamefont {Sette}},\ }\Doi
  {10.1103/PhysRevB.43.6785} {\bibfield  {journal} {\bibinfo  {journal} {Phys.
  Rev. B},\ }\textbf {\bibinfo {volume} {43}},\ \bibinfo {pages} {6785}
  (\bibinfo {year} {1991})}\BibitemShut {NoStop}%
\bibitem [{\citenamefont {Regan}\ \emph {et~al.}(2001)\citenamefont {Regan},
  \citenamefont {Ohldag}, \citenamefont {Stamm}, \citenamefont {Nolting},
  \citenamefont {L\"uning}, \citenamefont {St\"ohr},\ and\ \citenamefont
  {White}}]{Regan01}%
  \BibitemOpen
  \bibfield  {author} {\bibinfo {author} {\bibfnamefont {T.~J.}\ \bibnamefont
  {Regan}}, \bibinfo {author} {\bibfnamefont {H.}~\bibnamefont {Ohldag}},
  \bibinfo {author} {\bibfnamefont {C.}~\bibnamefont {Stamm}}, \bibinfo
  {author} {\bibfnamefont {F.}~\bibnamefont {Nolting}}, \bibinfo {author}
  {\bibfnamefont {J.}~\bibnamefont {L\"uning}}, \bibinfo {author}
  {\bibfnamefont {J.}~\bibnamefont {St\"ohr}}, \ and\ \bibinfo {author}
  {\bibfnamefont {R.~L.}\ \bibnamefont {White}},\ }\Doi
  {10.1103/PhysRevB.64.214422} {\bibfield  {journal} {\bibinfo  {journal}
  {Phys. Rev. B},\ }\textbf {\bibinfo {volume} {64}},\ \bibinfo {pages}
  {214422} (\bibinfo {year} {2001})}\BibitemShut {NoStop}%
\bibitem [{\citenamefont {Bahn}\ \emph {et~al.}(2012)\citenamefont {Bahn},
  \citenamefont {Oel{\ss}ner}, \citenamefont {K\"{o}ther}, \citenamefont
  {Braun}, \citenamefont {Senz}, \citenamefont {Palutke}, \citenamefont
  {Martins}, \citenamefont {R\"{u}hl}, \citenamefont {Gantef\"{o}r},
  \citenamefont {M\"{o}ller}, \citenamefont {von Issendorff}, \citenamefont
  {Bauer}, \citenamefont {Tiggesb\"{a}umker},\ and\ \citenamefont
  {Meiwes-Broer}}]{Bahn12}%
  \BibitemOpen
  \bibfield  {author} {\bibinfo {author} {\bibfnamefont {J.}~\bibnamefont
  {Bahn}}, \bibinfo {author} {\bibfnamefont {P.}~\bibnamefont {Oel{\ss}ner}},
  \bibinfo {author} {\bibfnamefont {M.}~\bibnamefont {K\"{o}ther}}, \bibinfo
  {author} {\bibfnamefont {C.}~\bibnamefont {Braun}}, \bibinfo {author}
  {\bibfnamefont {V.}~\bibnamefont {Senz}}, \bibinfo {author} {\bibfnamefont
  {S.}~\bibnamefont {Palutke}}, \bibinfo {author} {\bibfnamefont
  {M.}~\bibnamefont {Martins}}, \bibinfo {author} {\bibfnamefont
  {E.}~\bibnamefont {R\"{u}hl}}, \bibinfo {author} {\bibfnamefont
  {G.}~\bibnamefont {Gantef\"{o}r}}, \bibinfo {author} {\bibfnamefont
  {T.}~\bibnamefont {M\"{o}ller}}, \bibinfo {author} {\bibfnamefont
  {B.}~\bibnamefont {von Issendorff}}, \bibinfo {author} {\bibfnamefont
  {D.}~\bibnamefont {Bauer}}, \bibinfo {author} {\bibfnamefont
  {J.}~\bibnamefont {Tiggesb\"{a}umker}}, \ and\ \bibinfo {author}
  {\bibfnamefont {K.-H.}\ \bibnamefont {Meiwes-Broer}},\ }\Doi
  {10.1088/1367-2630/14/7/075008} {\bibfield  {journal} {\bibinfo  {journal}
  {New J. Phys.},\ }\textbf {\bibinfo {volume} {14}},\ \bibinfo {pages}
  {075008} (\bibinfo {year} {2012})}\BibitemShut {NoStop}%
\bibitem [{\citenamefont {Wu}\ and\ \citenamefont {Freeman}(1994)}]{Wu94}%
  \BibitemOpen
  \bibfield  {author} {\bibinfo {author} {\bibfnamefont {R.}~\bibnamefont
  {Wu}}\ and\ \bibinfo {author} {\bibfnamefont {A.~J.}\ \bibnamefont
  {Freeman}},\ }\Doi {10.1103/PhysRevLett.73.1994} {\bibfield  {journal}
  {\bibinfo  {journal} {Phys. Rev. Lett.},\ }\textbf {\bibinfo {volume} {73}},\
  \bibinfo {pages} {1994} (\bibinfo {year} {1994})}\BibitemShut {NoStop}%
\bibitem [{\citenamefont {St\"ohr}(1995)}]{Stoehr95a}%
  \BibitemOpen
  \bibfield  {author} {\bibinfo {author} {\bibfnamefont {J.}~\bibnamefont
  {St\"ohr}},\ }\href@noop {} {\bibfield  {journal} {\bibinfo  {journal} {J.
  Electron Spectrosc. Relat. Phenom.},\ }\textbf {\bibinfo {volume} {75}},\
  \bibinfo {pages} {253} (\bibinfo {year} {1995})}\BibitemShut {NoStop}%
\bibitem [{\citenamefont {Wu}\ \emph {et~al.}(1993)\citenamefont {Wu},
  \citenamefont {Wang},\ and\ \citenamefont {Freeman}}]{Wu93}%
  \BibitemOpen
  \bibfield  {author} {\bibinfo {author} {\bibfnamefont {R.}~\bibnamefont
  {Wu}}, \bibinfo {author} {\bibfnamefont {D.}~\bibnamefont {Wang}}, \ and\
  \bibinfo {author} {\bibfnamefont {A.~J.}\ \bibnamefont {Freeman}},\ }\Doi
  {10.1103/PhysRevLett.71.3581} {\bibfield  {journal} {\bibinfo  {journal}
  {Phys. Rev. Lett.},\ }\textbf {\bibinfo {volume} {71}},\ \bibinfo {pages}
  {3581} (\bibinfo {year} {1993})}\BibitemShut {NoStop}%
\bibitem [{\citenamefont {Guo}\ \emph {et~al.}(1994)\citenamefont {Guo},
  \citenamefont {Ebert}, \citenamefont {Temmerman},\ and\ \citenamefont
  {Durham}}]{Guo94}%
  \BibitemOpen
  \bibfield  {author} {\bibinfo {author} {\bibfnamefont {G.~Y.}\ \bibnamefont
  {Guo}}, \bibinfo {author} {\bibfnamefont {H.}~\bibnamefont {Ebert}}, \bibinfo
  {author} {\bibfnamefont {W.~M.}\ \bibnamefont {Temmerman}}, \ and\ \bibinfo
  {author} {\bibfnamefont {P.~J.}\ \bibnamefont {Durham}},\ }\Doi
  {10.1103/PhysRevB.50.3861} {\bibfield  {journal} {\bibinfo  {journal} {Phys.
  Rev. B},\ }\textbf {\bibinfo {volume} {50}},\ \bibinfo {pages} {3861}
  (\bibinfo {year} {1994})}\BibitemShut {NoStop}%
\bibitem [{\citenamefont {Basch}\ \emph {et~al.}(1980)\citenamefont {Basch},
  \citenamefont {Newton},\ and\ \citenamefont {Moskowitz}}]{Basch80}%
  \BibitemOpen
  \bibfield  {author} {\bibinfo {author} {\bibfnamefont {H.}~\bibnamefont
  {Basch}}, \bibinfo {author} {\bibfnamefont {M.~D.}\ \bibnamefont {Newton}}, \
  and\ \bibinfo {author} {\bibfnamefont {J.~W.}\ \bibnamefont {Moskowitz}},\
  }\Doi {10.1063/1.440687} {\bibfield  {journal} {\bibinfo  {journal} {J. Chem.
  Phys.},\ }\textbf {\bibinfo {volume} {73}},\ \bibinfo {pages} {4492}
  (\bibinfo {year} {1980})}\BibitemShut {NoStop}%
\bibitem [{\citenamefont {Pacchioni}\ and\ \citenamefont
  {Fantucci}(1987)}]{Pacchioni87}%
  \BibitemOpen
  \bibfield  {author} {\bibinfo {author} {\bibfnamefont {G.}~\bibnamefont
  {Pacchioni}}\ and\ \bibinfo {author} {\bibfnamefont {P.}~\bibnamefont
  {Fantucci}},\ }\Doi {http://dx.doi.org/10.1016/0009-2614(87)87163-4}
  {\bibfield  {journal} {\bibinfo  {journal} {Chem. Phys. Lett.},\ }\textbf
  {\bibinfo {volume} {134}},\ \bibinfo {pages} {407 } (\bibinfo {year}
  {1987})}\BibitemShut {NoStop}%
\bibitem [{\citenamefont {\v{S}ipr}\ and\ \citenamefont
  {Ebert}(2005)}]{Sipr05}%
  \BibitemOpen
  \bibfield  {author} {\bibinfo {author} {\bibfnamefont {O.}~\bibnamefont
  {\v{S}ipr}}\ and\ \bibinfo {author} {\bibfnamefont {H.}~\bibnamefont
  {Ebert}},\ }\Doi {10.1103/PhysRevB.72.134406} {\bibfield  {journal} {\bibinfo
   {journal} {Phys. Rev. B},\ }\textbf {\bibinfo {volume} {72}},\ \bibinfo
  {eid} {134406} (\bibinfo {year} {2005})}\BibitemShut {NoStop}%
\bibitem [{\citenamefont {Min\'ar}\ \emph {et~al.}(2006)\citenamefont
  {Min\'ar}, \citenamefont {Bornemann}, \citenamefont {\v{S}ipr}, \citenamefont
  {Polesya},\ and\ \citenamefont {Ebert}}]{Minar06}%
  \BibitemOpen
  \bibfield  {author} {\bibinfo {author} {\bibfnamefont {J.}~\bibnamefont
  {Min\'ar}}, \bibinfo {author} {\bibfnamefont {S.}~\bibnamefont {Bornemann}},
  \bibinfo {author} {\bibfnamefont {O.}~\bibnamefont {\v{S}ipr}}, \bibinfo
  {author} {\bibfnamefont {S.}~\bibnamefont {Polesya}}, \ and\ \bibinfo
  {author} {\bibfnamefont {H.}~\bibnamefont {Ebert}},\ }\Doi
  {10.1007/s00339-005-3359-1} {\bibfield  {journal} {\bibinfo  {journal} {Appl.
  Phys. A: Mater. Sci. Process.},\ }\textbf {\bibinfo {volume} {82}},\ \bibinfo
  {pages} {139} (\bibinfo {year} {2006})}\BibitemShut {NoStop}%
\bibitem [{\citenamefont {Gutsev}\ \emph {et~al.}(2012)\citenamefont {Gutsev},
  \citenamefont {Weatherford}, \citenamefont {Jena}, \citenamefont {Johnson},\
  and\ \citenamefont {Ramachandran}}]{Gutsev12}%
  \BibitemOpen
  \bibfield  {author} {\bibinfo {author} {\bibfnamefont {G.~L.}\ \bibnamefont
  {Gutsev}}, \bibinfo {author} {\bibfnamefont {C.~A.}\ \bibnamefont
  {Weatherford}}, \bibinfo {author} {\bibfnamefont {P.}~\bibnamefont {Jena}},
  \bibinfo {author} {\bibfnamefont {E.}~\bibnamefont {Johnson}}, \ and\
  \bibinfo {author} {\bibfnamefont {B.~R.}\ \bibnamefont {Ramachandran}},\
  }\Doi {10.1021/jp307284v} {\bibfield  {journal} {\bibinfo  {journal} {J.
  Phys. Chem. A},\ }\textbf {\bibinfo {volume} {116}},\ \bibinfo {pages}
  {10218} (\bibinfo {year} {2012})}\BibitemShut {NoStop}%
\bibitem [{\citenamefont {Gutsev}\ and\ \citenamefont {{Bauschlicher,
  Jr.}}(2003)}]{Gutsev03a}%
  \BibitemOpen
  \bibfield  {author} {\bibinfo {author} {\bibfnamefont {G.~L.}\ \bibnamefont
  {Gutsev}}\ and\ \bibinfo {author} {\bibfnamefont {C.~W.}\ \bibnamefont
  {{Bauschlicher, Jr.}}},\ }\Doi {10.1021/jp030146v} {\bibfield  {journal}
  {\bibinfo  {journal} {J. Phys. Chem. A},\ }\textbf {\bibinfo {volume}
  {107}},\ \bibinfo {pages} {4755} (\bibinfo {year} {2003})}\BibitemShut
  {NoStop}%
\bibitem [{\citenamefont {Daalderop}\ \emph {et~al.}(1994)\citenamefont
  {Daalderop}, \citenamefont {Kelly},\ and\ \citenamefont
  {Schuurmans}}]{Daalderop94}%
  \BibitemOpen
  \bibfield  {author} {\bibinfo {author} {\bibfnamefont {G.~H.~O.}\
  \bibnamefont {Daalderop}}, \bibinfo {author} {\bibfnamefont {P.~J.}\
  \bibnamefont {Kelly}}, \ and\ \bibinfo {author} {\bibfnamefont {M.~F.~H.}\
  \bibnamefont {Schuurmans}},\ }\Doi {10.1103/PhysRevB.50.9989} {\bibfield
  {journal} {\bibinfo  {journal} {Phys. Rev. B},\ }\textbf {\bibinfo {volume}
  {50}},\ \bibinfo {pages} {9989} (\bibinfo {year} {1994})}\BibitemShut
  {NoStop}%
\bibitem [{\citenamefont {Srivastava}\ \emph {et~al.}(1997)\citenamefont
  {Srivastava}, \citenamefont {Haack}, \citenamefont {Wende}, \citenamefont
  {Chauvistr\'e},\ and\ \citenamefont {Baberschke}}]{Srivastava97}%
  \BibitemOpen
  \bibfield  {author} {\bibinfo {author} {\bibfnamefont {P.}~\bibnamefont
  {Srivastava}}, \bibinfo {author} {\bibfnamefont {N.}~\bibnamefont {Haack}},
  \bibinfo {author} {\bibfnamefont {H.}~\bibnamefont {Wende}}, \bibinfo
  {author} {\bibfnamefont {R.}~\bibnamefont {Chauvistr\'e}}, \ and\ \bibinfo
  {author} {\bibfnamefont {K.}~\bibnamefont {Baberschke}},\ }\Doi
  {10.1103/PhysRevB.56.R4398} {\bibfield  {journal} {\bibinfo  {journal} {Phys.
  Rev. B},\ }\textbf {\bibinfo {volume} {56}},\ \bibinfo {pages} {R4398}
  (\bibinfo {year} {1997})}\BibitemShut {NoStop}%
\bibitem [{\citenamefont {Dhesi}\ \emph {et~al.}(2000)\citenamefont {Dhesi},
  \citenamefont {Dudzik}, \citenamefont {D\"urr}, \citenamefont {Brookes},\
  and\ \citenamefont {van~der Laan}}]{Dhesi00}%
  \BibitemOpen
  \bibfield  {author} {\bibinfo {author} {\bibfnamefont {S.~S.}\ \bibnamefont
  {Dhesi}}, \bibinfo {author} {\bibfnamefont {E.}~\bibnamefont {Dudzik}},
  \bibinfo {author} {\bibfnamefont {H.~A.}\ \bibnamefont {D\"urr}}, \bibinfo
  {author} {\bibfnamefont {N.~B.}\ \bibnamefont {Brookes}}, \ and\ \bibinfo
  {author} {\bibfnamefont {G.}~\bibnamefont {van~der Laan}},\ }\href@noop {}
  {\bibfield  {journal} {\bibinfo  {journal} {Surf. Sci.},\ }\textbf {\bibinfo
  {volume} {454--456}},\ \bibinfo {pages} {930} (\bibinfo {year}
  {2000})}\BibitemShut {NoStop}%
\bibitem [{\citenamefont {Pearson}\ \emph {et~al.}(1993)\citenamefont
  {Pearson}, \citenamefont {Ahn},\ and\ \citenamefont {Fultz}}]{Pearson93}%
  \BibitemOpen
  \bibfield  {author} {\bibinfo {author} {\bibfnamefont {D.~H.}\ \bibnamefont
  {Pearson}}, \bibinfo {author} {\bibfnamefont {C.~C.}\ \bibnamefont {Ahn}}, \
  and\ \bibinfo {author} {\bibfnamefont {B.}~\bibnamefont {Fultz}},\ }\Doi
  {10.1103/PhysRevB.47.8471} {\bibfield  {journal} {\bibinfo  {journal} {Phys.
  Rev. B},\ }\textbf {\bibinfo {volume} {47}},\ \bibinfo {pages} {8471}
  (\bibinfo {year} {1993})}\BibitemShut {NoStop}%
\bibitem [{\citenamefont {Ankudinov}\ \emph {et~al.}(2001)\citenamefont
  {Ankudinov}, \citenamefont {Rehr}, \citenamefont {Low},\ and\ \citenamefont
  {Bare}}]{Ankudinov01}%
  \BibitemOpen
  \bibfield  {author} {\bibinfo {author} {\bibfnamefont {A.~L.}\ \bibnamefont
  {Ankudinov}}, \bibinfo {author} {\bibfnamefont {J.~J.}\ \bibnamefont {Rehr}},
  \bibinfo {author} {\bibfnamefont {J.~J.}\ \bibnamefont {Low}}, \ and\
  \bibinfo {author} {\bibfnamefont {S.~R.}\ \bibnamefont {Bare}},\ }\href@noop
  {} {\bibfield  {journal} {\bibinfo  {journal} {J. Synchrotron Rad.},\
  }\textbf {\bibinfo {volume} {8}},\ \bibinfo {pages} {578} (\bibinfo {year}
  {2001})}\BibitemShut {NoStop}%
\bibitem [{\citenamefont {Graetz}\ \emph {et~al.}(2004)\citenamefont {Graetz},
  \citenamefont {Ahn}, \citenamefont {Ouyang}, \citenamefont {Rez},\ and\
  \citenamefont {Fultz}}]{Graetz04}%
  \BibitemOpen
  \bibfield  {author} {\bibinfo {author} {\bibfnamefont {J.}~\bibnamefont
  {Graetz}}, \bibinfo {author} {\bibfnamefont {C.~C.}\ \bibnamefont {Ahn}},
  \bibinfo {author} {\bibfnamefont {H.}~\bibnamefont {Ouyang}}, \bibinfo
  {author} {\bibfnamefont {P.}~\bibnamefont {Rez}}, \ and\ \bibinfo {author}
  {\bibfnamefont {B.}~\bibnamefont {Fultz}},\ }\Doi
  {10.1103/PhysRevB.69.235103} {\bibfield  {journal} {\bibinfo  {journal}
  {Phys. Rev. B},\ }\textbf {\bibinfo {volume} {69}},\ \bibinfo {eid} {235103}
  (\bibinfo {year} {2004})}\BibitemShut {NoStop}%
\bibitem [{\citenamefont {St\"ohr}\ and\ \citenamefont
  {K\"onig}(1995)}]{Stoehr95b}%
  \BibitemOpen
  \bibfield  {author} {\bibinfo {author} {\bibfnamefont {J.}~\bibnamefont
  {St\"ohr}}\ and\ \bibinfo {author} {\bibfnamefont {H.}~\bibnamefont
  {K\"onig}},\ }\Doi {10.1103/PhysRevLett.75.3748} {\bibfield  {journal}
  {\bibinfo  {journal} {Phys. Rev. Lett.},\ }\textbf {\bibinfo {volume} {75}},\
  \bibinfo {pages} {3748} (\bibinfo {year} {1995})}\BibitemShut {NoStop}%
\bibitem [{\citenamefont {Gerlich}(1992)}]{Gerlich92}%
  \BibitemOpen
  \bibfield  {author} {\bibinfo {author} {\bibfnamefont {D.}~\bibnamefont
  {Gerlich}},\ }\enquote {\bibinfo {title} {Inhomogeneous rf fields: A
  versatile tool for the study of processes with slow ions},}\ in\ \Doi
  {10.1002/9780470141397.ch1} {\emph {\bibinfo {booktitle} {Advances in
  Chemical Physics}}},\ Vol.\ \bibinfo {volume} {LXXXII},\ \bibinfo {editor}
  {edited by\ \bibinfo {editor} {\bibfnamefont {C.-Y.}\ \bibnamefont {Ng}}\
  and\ \bibinfo {editor} {\bibfnamefont {M.}~\bibnamefont {Baer}}}\ (\bibinfo
  {publisher} {John Wiley \& Sons, Inc.},\ \bibinfo {address} {New York},\
  \bibinfo {year} {1992})\ pp.\ \bibinfo {pages} {1--176}\BibitemShut {NoStop}%
\bibitem [{\citenamefont {Moriwaki}\ \emph {et~al.}(1992)\citenamefont
  {Moriwaki}, \citenamefont {Tachikawa}, \citenamefont {Maeno},\ and\
  \citenamefont {Shimizu}}]{Moriwaki92}%
  \BibitemOpen
  \bibfield  {author} {\bibinfo {author} {\bibfnamefont {Y.}~\bibnamefont
  {Moriwaki}}, \bibinfo {author} {\bibfnamefont {M.}~\bibnamefont {Tachikawa}},
  \bibinfo {author} {\bibfnamefont {Y.}~\bibnamefont {Maeno}}, \ and\ \bibinfo
  {author} {\bibfnamefont {T.}~\bibnamefont {Shimizu}},\ }\Doi
  {10.1143/JJAP.31.L1640} {\bibfield  {journal} {\bibinfo  {journal} {Jpn. J.
  Appl. Phys.},\ }\textbf {\bibinfo {volume} {31}},\ \bibinfo {pages} {L1640}
  (\bibinfo {year} {1992})}\BibitemShut {NoStop}%
\bibitem [{\citenamefont {Gronert}(1998)}]{Gronert98}%
  \BibitemOpen
  \bibfield  {author} {\bibinfo {author} {\bibfnamefont {S.}~\bibnamefont
  {Gronert}},\ }\Doi {10.1016/S1044-0305(98)00055-5} {\bibfield  {journal}
  {\bibinfo  {journal} {J. Am. Soc. Mass Spectrom.},\ }\textbf {\bibinfo
  {volume} {9}},\ \bibinfo {pages} {845} (\bibinfo {year} {1998})}\BibitemShut
  {NoStop}%
\bibitem [{\citenamefont {Gerlich}\ and\ \citenamefont
  {Borodi}(2009)}]{Gerlich09}%
  \BibitemOpen
  \bibfield  {author} {\bibinfo {author} {\bibfnamefont {D.}~\bibnamefont
  {Gerlich}}\ and\ \bibinfo {author} {\bibfnamefont {G.}~\bibnamefont
  {Borodi}},\ }\Doi {10.1039/B820977D} {\bibfield  {journal} {\bibinfo
  {journal} {Faraday Discuss.},\ }\textbf {\bibinfo {volume} {142}},\ \bibinfo
  {pages} {57} (\bibinfo {year} {2009})}\BibitemShut {NoStop}%
\bibitem [{\citenamefont {Asvany}\ and\ \citenamefont
  {Schlemmer}(2009)}]{Asvany09}%
  \BibitemOpen
  \bibfield  {author} {\bibinfo {author} {\bibfnamefont {O.}~\bibnamefont
  {Asvany}}\ and\ \bibinfo {author} {\bibfnamefont {S.}~\bibnamefont
  {Schlemmer}},\ }\Doi {http://dx.doi.org/10.1016/j.ijms.2008.10.022}
  {\bibfield  {journal} {\bibinfo  {journal} {Int. J. Mass Spectrom.},\
  }\textbf {\bibinfo {volume} {279}},\ \bibinfo {pages} {147 } (\bibinfo {year}
  {2009})}\BibitemShut {NoStop}%
\bibitem [{\citenamefont {Otto}\ \emph {et~al.}(2013)\citenamefont {Otto},
  \citenamefont {von Zastrow}, \citenamefont {Best},\ and\ \citenamefont
  {Wester}}]{Otto13}%
  \BibitemOpen
  \bibfield  {author} {\bibinfo {author} {\bibfnamefont {R.}~\bibnamefont
  {Otto}}, \bibinfo {author} {\bibfnamefont {A.}~\bibnamefont {von Zastrow}},
  \bibinfo {author} {\bibfnamefont {T.}~\bibnamefont {Best}}, \ and\ \bibinfo
  {author} {\bibfnamefont {R.}~\bibnamefont {Wester}},\ }\Doi
  {10.1039/C2CP43186F} {\bibfield  {journal} {\bibinfo  {journal} {Phys. Chem.
  Chem. Phys.},\ }\textbf {\bibinfo {volume} {15}},\ \bibinfo {pages} {612}
  (\bibinfo {year} {2013})}\BibitemShut {NoStop}%
\bibitem [{\citenamefont {Gutsev}\ \emph {et~al.}(2013)\citenamefont {Gutsev},
  \citenamefont {Weatherford}, \citenamefont {Belay}, \citenamefont
  {Ramachandran},\ and\ \citenamefont {Jena}}]{Gutsev13}%
  \BibitemOpen
  \bibfield  {author} {\bibinfo {author} {\bibfnamefont {G.~L.}\ \bibnamefont
  {Gutsev}}, \bibinfo {author} {\bibfnamefont {C.~W.}\ \bibnamefont
  {Weatherford}}, \bibinfo {author} {\bibfnamefont {K.~G.}\ \bibnamefont
  {Belay}}, \bibinfo {author} {\bibfnamefont {B.~R.}\ \bibnamefont
  {Ramachandran}}, \ and\ \bibinfo {author} {\bibfnamefont {P.}~\bibnamefont
  {Jena}},\ }\Doi {10.1063/1.4799917} {\bibfield  {journal} {\bibinfo
  {journal} {J. Chem. Phys.},\ }\textbf {\bibinfo {volume} {138}},\ \bibinfo
  {eid} {164303} (\bibinfo {year} {2013})}\BibitemShut {NoStop}%
\bibitem [{\citenamefont {Guirado-L\'opez}\ \emph {et~al.}(2003)\citenamefont
  {Guirado-L\'opez}, \citenamefont {Dorantes-D\'avila},\ and\ \citenamefont
  {Pastor}}]{GuiradoLopez03}%
  \BibitemOpen
  \bibfield  {author} {\bibinfo {author} {\bibfnamefont {R.~A.}\ \bibnamefont
  {Guirado-L\'opez}}, \bibinfo {author} {\bibfnamefont {J.}~\bibnamefont
  {Dorantes-D\'avila}}, \ and\ \bibinfo {author} {\bibfnamefont {G.~M.}\
  \bibnamefont {Pastor}},\ }\Doi {10.1103/PhysRevLett.90.226402} {\bibfield
  {journal} {\bibinfo  {journal} {Phys. Rev. Lett.},\ }\textbf {\bibinfo
  {volume} {90}},\ \bibinfo {pages} {226402} (\bibinfo {year}
  {2003})}\BibitemShut {NoStop}%
\bibitem [{\citenamefont {Strandberg}\ \emph {et~al.}(2007)\citenamefont
  {Strandberg}, \citenamefont {Canali},\ and\ \citenamefont
  {MacDonald}}]{Strandberg07}%
  \BibitemOpen
  \bibfield  {author} {\bibinfo {author} {\bibfnamefont {T.~O.}\ \bibnamefont
  {Strandberg}}, \bibinfo {author} {\bibfnamefont {C.~M.}\ \bibnamefont
  {Canali}}, \ and\ \bibinfo {author} {\bibfnamefont {A.~H.}\ \bibnamefont
  {MacDonald}},\ }\Doi {10.1038/nmat1968} {\bibfield  {journal} {\bibinfo
  {journal} {Nat. Mater.},\ }\textbf {\bibinfo {volume} {6}},\ \bibinfo {pages}
  {648} (\bibinfo {year} {2007})}\BibitemShut {NoStop}%
\bibitem [{\citenamefont {Fritsch}\ \emph {et~al.}(2008)\citenamefont
  {Fritsch}, \citenamefont {Koepernik}, \citenamefont {Richter},\ and\
  \citenamefont {Eschrig}}]{Fritsch08}%
  \BibitemOpen
  \bibfield  {author} {\bibinfo {author} {\bibfnamefont {D.}~\bibnamefont
  {Fritsch}}, \bibinfo {author} {\bibfnamefont {K.}~\bibnamefont {Koepernik}},
  \bibinfo {author} {\bibfnamefont {M.}~\bibnamefont {Richter}}, \ and\
  \bibinfo {author} {\bibfnamefont {H.}~\bibnamefont {Eschrig}},\ }\Doi
  {10.1002/jcc.21012} {\bibfield  {journal} {\bibinfo  {journal} {J. Comput.
  Chem.},\ }\textbf {\bibinfo {volume} {29}},\ \bibinfo {pages} {2210}
  (\bibinfo {year} {2008})}\BibitemShut {NoStop}%
\bibitem [{\citenamefont {Jamet}\ \emph {et~al.}(2004)\citenamefont {Jamet},
  \citenamefont {Wernsdorfer}, \citenamefont {Thirion}, \citenamefont {Dupuis},
  \citenamefont {M\'elinon}, \citenamefont {P\'erez},\ and\ \citenamefont
  {Mailly}}]{Jamet04}%
  \BibitemOpen
  \bibfield  {author} {\bibinfo {author} {\bibfnamefont {M.}~\bibnamefont
  {Jamet}}, \bibinfo {author} {\bibfnamefont {W.}~\bibnamefont {Wernsdorfer}},
  \bibinfo {author} {\bibfnamefont {C.}~\bibnamefont {Thirion}}, \bibinfo
  {author} {\bibfnamefont {V.}~\bibnamefont {Dupuis}}, \bibinfo {author}
  {\bibfnamefont {P.}~\bibnamefont {M\'elinon}}, \bibinfo {author}
  {\bibfnamefont {A.}~\bibnamefont {P\'erez}}, \ and\ \bibinfo {author}
  {\bibfnamefont {D.}~\bibnamefont {Mailly}},\ }\Doi
  {10.1103/PhysRevB.69.024401} {\bibfield  {journal} {\bibinfo  {journal}
  {Phys. Rev. B},\ }\textbf {\bibinfo {volume} {69}},\ \bibinfo {pages}
  {024401} (\bibinfo {year} {2004})}\BibitemShut {NoStop}%
\bibitem [{\citenamefont {Stiles}\ \emph {et~al.}(2001)\citenamefont {Stiles},
  \citenamefont {Halilov}, \citenamefont {Hyman},\ and\ \citenamefont
  {Zangwill}}]{Stiles01}%
  \BibitemOpen
  \bibfield  {author} {\bibinfo {author} {\bibfnamefont {M.~D.}\ \bibnamefont
  {Stiles}}, \bibinfo {author} {\bibfnamefont {S.~V.}\ \bibnamefont {Halilov}},
  \bibinfo {author} {\bibfnamefont {R.~A.}\ \bibnamefont {Hyman}}, \ and\
  \bibinfo {author} {\bibfnamefont {A.}~\bibnamefont {Zangwill}},\ }\Doi
  {10.1103/PhysRevB.64.104430} {\bibfield  {journal} {\bibinfo  {journal}
  {Phys. Rev. B},\ }\textbf {\bibinfo {volume} {64}},\ \bibinfo {pages}
  {104430} (\bibinfo {year} {2001})}\BibitemShut {NoStop}%
\bibitem [{\citenamefont {Yang}\ \emph {et~al.}(2001)\citenamefont {Yang},
  \citenamefont {Savrasov},\ and\ \citenamefont {Kotliar}}]{Yang01}%
  \BibitemOpen
  \bibfield  {author} {\bibinfo {author} {\bibfnamefont {I.}~\bibnamefont
  {Yang}}, \bibinfo {author} {\bibfnamefont {S.~Y.}\ \bibnamefont {Savrasov}},
  \ and\ \bibinfo {author} {\bibfnamefont {G.}~\bibnamefont {Kotliar}},\ }\Doi
  {10.1103/PhysRevLett.87.216405} {\bibfield  {journal} {\bibinfo  {journal}
  {Phys. Rev. Lett.},\ }\textbf {\bibinfo {volume} {87}},\ \bibinfo {pages}
  {216405} (\bibinfo {year} {2001})}\BibitemShut {NoStop}%
\bibitem [{\citenamefont {Lehnert}\ \emph {et~al.}(2010)\citenamefont
  {Lehnert}, \citenamefont {Rusponi}, \citenamefont {Etzkorn}, \citenamefont
  {Ouazi}, \citenamefont {Thakur},\ and\ \citenamefont {Brune}}]{Lehnert10}%
  \BibitemOpen
  \bibfield  {author} {\bibinfo {author} {\bibfnamefont {A.}~\bibnamefont
  {Lehnert}}, \bibinfo {author} {\bibfnamefont {S.}~\bibnamefont {Rusponi}},
  \bibinfo {author} {\bibfnamefont {M.}~\bibnamefont {Etzkorn}}, \bibinfo
  {author} {\bibfnamefont {S.}~\bibnamefont {Ouazi}}, \bibinfo {author}
  {\bibfnamefont {P.}~\bibnamefont {Thakur}}, \ and\ \bibinfo {author}
  {\bibfnamefont {H.}~\bibnamefont {Brune}},\ }\Doi
  {10.1103/PhysRevB.81.104430} {\bibfield  {journal} {\bibinfo  {journal}
  {Phys. Rev. B},\ }\textbf {\bibinfo {volume} {81}},\ \bibinfo {pages}
  {104430} (\bibinfo {year} {2010})}\BibitemShut {NoStop}%
\bibitem [{\citenamefont {Eriksson}\ \emph {et~al.}(1992)\citenamefont
  {Eriksson}, \citenamefont {Boring}, \citenamefont {Albers}, \citenamefont
  {Fernando},\ and\ \citenamefont {Cooper}}]{Eriksson92}%
  \BibitemOpen
  \bibfield  {author} {\bibinfo {author} {\bibfnamefont {O.}~\bibnamefont
  {Eriksson}}, \bibinfo {author} {\bibfnamefont {A.~M.}\ \bibnamefont
  {Boring}}, \bibinfo {author} {\bibfnamefont {R.~C.}\ \bibnamefont {Albers}},
  \bibinfo {author} {\bibfnamefont {G.~W.}\ \bibnamefont {Fernando}}, \ and\
  \bibinfo {author} {\bibfnamefont {B.~R.}\ \bibnamefont {Cooper}},\ }\Doi
  {10.1103/PhysRevB.45.2868} {\bibfield  {journal} {\bibinfo  {journal} {Phys.
  Rev. B},\ }\textbf {\bibinfo {volume} {45}},\ \bibinfo {pages} {2868}
  (\bibinfo {year} {1992})}\BibitemShut {NoStop}%
\bibitem [{\citenamefont {Liu}\ \emph {et~al.}(1989)\citenamefont {Liu},
  \citenamefont {Press}, \citenamefont {Khanna},\ and\ \citenamefont
  {Jena}}]{Liu89}%
  \BibitemOpen
  \bibfield  {author} {\bibinfo {author} {\bibfnamefont {F.}~\bibnamefont
  {Liu}}, \bibinfo {author} {\bibfnamefont {M.~R.}\ \bibnamefont {Press}},
  \bibinfo {author} {\bibfnamefont {S.~N.}\ \bibnamefont {Khanna}}, \ and\
  \bibinfo {author} {\bibfnamefont {P.}~\bibnamefont {Jena}},\ }\Doi
  {10.1103/PhysRevB.39.6914} {\bibfield  {journal} {\bibinfo  {journal} {Phys.
  Rev. B},\ }\textbf {\bibinfo {volume} {39}},\ \bibinfo {pages} {6914}
  (\bibinfo {year} {1989})}\BibitemShut {NoStop}%
\bibitem [{\citenamefont {Di\'eguez}\ \emph {et~al.}(2001)\citenamefont
  {Di\'eguez}, \citenamefont {Alemany}, \citenamefont {Rey}, \citenamefont
  {Ordej\'on},\ and\ \citenamefont {Gallego}}]{Dieguez01}%
  \BibitemOpen
  \bibfield  {author} {\bibinfo {author} {\bibfnamefont {O.}~\bibnamefont
  {Di\'eguez}}, \bibinfo {author} {\bibfnamefont {M.~M.~G.}\ \bibnamefont
  {Alemany}}, \bibinfo {author} {\bibfnamefont {C.}~\bibnamefont {Rey}},
  \bibinfo {author} {\bibfnamefont {P.}~\bibnamefont {Ordej\'on}}, \ and\
  \bibinfo {author} {\bibfnamefont {L.~J.}\ \bibnamefont {Gallego}},\ }\href
  {http://link.aps.org/abstract/PRB/v63/e205407} {\bibfield  {journal}
  {\bibinfo  {journal} {Phys. Rev. B},\ }\textbf {\bibinfo {volume} {63}},\
  \bibinfo {eid} {205407} (\bibinfo {year} {2001})}\BibitemShut {NoStop}%
\bibitem [{\citenamefont {Bobadova-Parvanova}\ \emph
  {et~al.}(2002)\citenamefont {Bobadova-Parvanova}, \citenamefont {Jackson},
  \citenamefont {Srinivas},\ and\ \citenamefont
  {Horoi}}]{BobadovaParvanova02a}%
  \BibitemOpen
  \bibfield  {author} {\bibinfo {author} {\bibfnamefont {P.}~\bibnamefont
  {Bobadova-Parvanova}}, \bibinfo {author} {\bibfnamefont {K.~A.}\ \bibnamefont
  {Jackson}}, \bibinfo {author} {\bibfnamefont {S.}~\bibnamefont {Srinivas}}, \
  and\ \bibinfo {author} {\bibfnamefont {M.}~\bibnamefont {Horoi}},\ }\Doi
  {10.1103/PhysRevB.66.195402} {\bibfield  {journal} {\bibinfo  {journal}
  {Phys. Rev. B},\ }\textbf {\bibinfo {volume} {66}},\ \bibinfo {pages}
  {195402} (\bibinfo {year} {2002})}\BibitemShut {NoStop}%
\bibitem [{\citenamefont {Singh}\ and\ \citenamefont {Kroll}(2008)}]{Singh08}%
  \BibitemOpen
  \bibfield  {author} {\bibinfo {author} {\bibfnamefont {R.}~\bibnamefont
  {Singh}}\ and\ \bibinfo {author} {\bibfnamefont {P.}~\bibnamefont {Kroll}},\
  }\Doi {10.1103/PhysRevB.78.245404} {\bibfield  {journal} {\bibinfo  {journal}
  {Phys. Rev. B},\ }\textbf {\bibinfo {volume} {78}},\ \bibinfo {pages}
  {245404} (\bibinfo {year} {2008})}\BibitemShut {NoStop}%
\bibitem [{\citenamefont {Rollmann}\ \emph {et~al.}(2006)\citenamefont
  {Rollmann}, \citenamefont {Entel},\ and\ \citenamefont {Sahoo}}]{Rollmann06}%
  \BibitemOpen
  \bibfield  {author} {\bibinfo {author} {\bibfnamefont {G.}~\bibnamefont
  {Rollmann}}, \bibinfo {author} {\bibfnamefont {P.}~\bibnamefont {Entel}}, \
  and\ \bibinfo {author} {\bibfnamefont {S.}~\bibnamefont {Sahoo}},\ }\Doi
  {10.1016/j.commatsci.2004.09.059} {\bibfield  {journal} {\bibinfo  {journal}
  {Comput. Mater. Sci.},\ }\textbf {\bibinfo {volume} {35}},\ \bibinfo {pages}
  {275 } (\bibinfo {year} {2006})}\BibitemShut {NoStop}%
\bibitem [{\citenamefont {Dunlap}(1990)}]{Dunlap90}%
  \BibitemOpen
  \bibfield  {author} {\bibinfo {author} {\bibfnamefont {B.~I.}\ \bibnamefont
  {Dunlap}},\ }\Doi {10.1103/PhysRevA.41.5691} {\bibfield  {journal} {\bibinfo
  {journal} {Phys. Rev. A},\ }\textbf {\bibinfo {volume} {41}},\ \bibinfo
  {pages} {5691} (\bibinfo {year} {1990})}\BibitemShut {NoStop}%
\bibitem [{\citenamefont {Wilburn}\ and\ \citenamefont
  {Bassett}(1978)}]{Wilburn78}%
  \BibitemOpen
  \bibfield  {author} {\bibinfo {author} {\bibfnamefont {D.~R.}\ \bibnamefont
  {Wilburn}}\ and\ \bibinfo {author} {\bibfnamefont {W.~A.}\ \bibnamefont
  {Bassett}},\ }\href
  {http://ammin.geoscienceworld.org/content/63/5-6/591.short} {\bibfield
  {journal} {\bibinfo  {journal} {Am. Mineral.},\ }\textbf {\bibinfo {volume}
  {63}},\ \bibinfo {pages} {591} (\bibinfo {year} {1978})}\BibitemShut
  {NoStop}%
\bibitem [{\citenamefont {Iota}\ \emph {et~al.}(2007)\citenamefont {Iota},
  \citenamefont {Klepeis}, \citenamefont {Yoo}, \citenamefont {Lang},
  \citenamefont {Haskel},\ and\ \citenamefont {Srajer}}]{Iota07}%
  \BibitemOpen
  \bibfield  {author} {\bibinfo {author} {\bibfnamefont {V.}~\bibnamefont
  {Iota}}, \bibinfo {author} {\bibfnamefont {J.-H.~P.}\ \bibnamefont
  {Klepeis}}, \bibinfo {author} {\bibfnamefont {C.-S.}\ \bibnamefont {Yoo}},
  \bibinfo {author} {\bibfnamefont {J.}~\bibnamefont {Lang}}, \bibinfo {author}
  {\bibfnamefont {D.}~\bibnamefont {Haskel}}, \ and\ \bibinfo {author}
  {\bibfnamefont {G.}~\bibnamefont {Srajer}},\ }\Doi {10.1063/1.2434184}
  {\bibfield  {journal} {\bibinfo  {journal} {Appl. Phys. Lett.},\ }\textbf
  {\bibinfo {volume} {90}},\ \bibinfo {eid} {042505} (\bibinfo {year}
  {2007})}\BibitemShut {NoStop}%
\bibitem [{\citenamefont {Dewaele}\ \emph {et~al.}(2008)\citenamefont
  {Dewaele}, \citenamefont {Torrent}, \citenamefont {Loubeyre},\ and\
  \citenamefont {Mezouar}}]{Dewaele08}%
  \BibitemOpen
  \bibfield  {author} {\bibinfo {author} {\bibfnamefont {A.}~\bibnamefont
  {Dewaele}}, \bibinfo {author} {\bibfnamefont {M.}~\bibnamefont {Torrent}},
  \bibinfo {author} {\bibfnamefont {P.}~\bibnamefont {Loubeyre}}, \ and\
  \bibinfo {author} {\bibfnamefont {M.}~\bibnamefont {Mezouar}},\ }\Doi
  {10.1103/PhysRevB.78.104102} {\bibfield  {journal} {\bibinfo  {journal}
  {Phys. Rev. B},\ }\textbf {\bibinfo {volume} {78}},\ \bibinfo {pages}
  {104102} (\bibinfo {year} {2008})}\BibitemShut {NoStop}%
\bibitem [{\citenamefont {Abrahams}\ \emph {et~al.}(1962)\citenamefont
  {Abrahams}, \citenamefont {Guttman},\ and\ \citenamefont
  {Kasper}}]{Abrahams62}%
  \BibitemOpen
  \bibfield  {author} {\bibinfo {author} {\bibfnamefont {S.~C.}\ \bibnamefont
  {Abrahams}}, \bibinfo {author} {\bibfnamefont {L.}~\bibnamefont {Guttman}}, \
  and\ \bibinfo {author} {\bibfnamefont {J.~S.}\ \bibnamefont {Kasper}},\ }\Doi
  {10.1103/PhysRev.127.2052} {\bibfield  {journal} {\bibinfo  {journal} {Phys.
  Rev.},\ }\textbf {\bibinfo {volume} {127}},\ \bibinfo {pages} {2052}
  (\bibinfo {year} {1962})}\BibitemShut {NoStop}%
\bibitem [{\citenamefont {Gonser}\ \emph {et~al.}(1963)\citenamefont {Gonser},
  \citenamefont {Meechan}, \citenamefont {Muir},\ and\ \citenamefont
  {Wiedersich}}]{Gonser63}%
  \BibitemOpen
  \bibfield  {author} {\bibinfo {author} {\bibfnamefont {U.}~\bibnamefont
  {Gonser}}, \bibinfo {author} {\bibfnamefont {C.~J.}\ \bibnamefont {Meechan}},
  \bibinfo {author} {\bibfnamefont {A.~H.}\ \bibnamefont {Muir}}, \ and\
  \bibinfo {author} {\bibfnamefont {H.}~\bibnamefont {Wiedersich}},\ }\Doi
  {10.1063/1.1702749} {\bibfield  {journal} {\bibinfo  {journal} {J. Appl.
  Phys.},\ }\textbf {\bibinfo {volume} {34}},\ \bibinfo {pages} {2373}
  (\bibinfo {year} {1963})}\BibitemShut {NoStop}%
\bibitem [{\citenamefont {Johanson}\ \emph {et~al.}(1970)\citenamefont
  {Johanson}, \citenamefont {McGirr},\ and\ \citenamefont
  {Wheeler}}]{Johanson70}%
  \BibitemOpen
  \bibfield  {author} {\bibinfo {author} {\bibfnamefont {G.~J.}\ \bibnamefont
  {Johanson}}, \bibinfo {author} {\bibfnamefont {M.~B.}\ \bibnamefont
  {McGirr}}, \ and\ \bibinfo {author} {\bibfnamefont {D.~A.}\ \bibnamefont
  {Wheeler}},\ }\Doi {10.1103/PhysRevB.1.3208} {\bibfield  {journal} {\bibinfo
  {journal} {Phys. Rev. B},\ }\textbf {\bibinfo {volume} {1}},\ \bibinfo
  {pages} {3208} (\bibinfo {year} {1970})}\BibitemShut {NoStop}%
\bibitem [{\citenamefont {Onodera}\ \emph {et~al.}(1994)\citenamefont
  {Onodera}, \citenamefont {Tsunoda}, \citenamefont {Kunitomi}, \citenamefont
  {Pringle}, \citenamefont {Nicklow},\ and\ \citenamefont {Moon}}]{Onodera94}%
  \BibitemOpen
  \bibfield  {author} {\bibinfo {author} {\bibfnamefont {A.}~\bibnamefont
  {Onodera}}, \bibinfo {author} {\bibfnamefont {Y.}~\bibnamefont {Tsunoda}},
  \bibinfo {author} {\bibfnamefont {N.}~\bibnamefont {Kunitomi}}, \bibinfo
  {author} {\bibfnamefont {O.~A.}\ \bibnamefont {Pringle}}, \bibinfo {author}
  {\bibfnamefont {R.~M.}\ \bibnamefont {Nicklow}}, \ and\ \bibinfo {author}
  {\bibfnamefont {R.~M.}\ \bibnamefont {Moon}},\ }\Doi
  {10.1103/PhysRevB.50.3532} {\bibfield  {journal} {\bibinfo  {journal} {Phys.
  Rev. B},\ }\textbf {\bibinfo {volume} {50}},\ \bibinfo {pages} {3532}
  (\bibinfo {year} {1994})}\BibitemShut {NoStop}%
\bibitem [{\citenamefont {Keune}\ \emph {et~al.}(1977)\citenamefont {Keune},
  \citenamefont {Halbauer}, \citenamefont {Gonser}, \citenamefont {Lauer},\
  and\ \citenamefont {Williamson}}]{Keune77}%
  \BibitemOpen
  \bibfield  {author} {\bibinfo {author} {\bibfnamefont {W.}~\bibnamefont
  {Keune}}, \bibinfo {author} {\bibfnamefont {R.}~\bibnamefont {Halbauer}},
  \bibinfo {author} {\bibfnamefont {U.}~\bibnamefont {Gonser}}, \bibinfo
  {author} {\bibfnamefont {J.}~\bibnamefont {Lauer}}, \ and\ \bibinfo {author}
  {\bibfnamefont {D.~L.}\ \bibnamefont {Williamson}},\ }\Doi {10.1063/1.324113}
  {\bibfield  {journal} {\bibinfo  {journal} {J. Appl. Phys.},\ }\textbf
  {\bibinfo {volume} {48}},\ \bibinfo {pages} {2976} (\bibinfo {year}
  {1977})}\BibitemShut {NoStop}%
\bibitem [{\citenamefont {Himpsel}(1991)}]{Himpsel91a}%
  \BibitemOpen
  \bibfield  {author} {\bibinfo {author} {\bibfnamefont {F.~J.}\ \bibnamefont
  {Himpsel}},\ }\Doi {10.1103/PhysRevLett.67.2363} {\bibfield  {journal}
  {\bibinfo  {journal} {Phys. Rev. Lett.},\ }\textbf {\bibinfo {volume} {67}},\
  \bibinfo {pages} {2363} (\bibinfo {year} {1991})}\BibitemShut {NoStop}%
\bibitem [{\citenamefont {Li}\ \emph {et~al.}(1994)\citenamefont {Li},
  \citenamefont {Freitag}, \citenamefont {Pearson}, \citenamefont {Qiu},\ and\
  \citenamefont {Bader}}]{Li94}%
  \BibitemOpen
  \bibfield  {author} {\bibinfo {author} {\bibfnamefont {D.}~\bibnamefont
  {Li}}, \bibinfo {author} {\bibfnamefont {M.}~\bibnamefont {Freitag}},
  \bibinfo {author} {\bibfnamefont {J.}~\bibnamefont {Pearson}}, \bibinfo
  {author} {\bibfnamefont {Z.~Q.}\ \bibnamefont {Qiu}}, \ and\ \bibinfo
  {author} {\bibfnamefont {S.~D.}\ \bibnamefont {Bader}},\ }\Doi
  {10.1103/PhysRevLett.72.3112} {\bibfield  {journal} {\bibinfo  {journal}
  {Phys. Rev. Lett.},\ }\textbf {\bibinfo {volume} {72}},\ \bibinfo {pages}
  {3112} (\bibinfo {year} {1994})}\BibitemShut {NoStop}%
\bibitem [{\citenamefont {Meyerheim}\ \emph {et~al.}(2009)\citenamefont
  {Meyerheim}, \citenamefont {Tonnerre}, \citenamefont {Sandratskii},
  \citenamefont {Tolentino}, \citenamefont {Przybylski}, \citenamefont {Gabi},
  \citenamefont {Yildiz}, \citenamefont {Fu}, \citenamefont {Bontempi},
  \citenamefont {Grenier},\ and\ \citenamefont {Kirschner}}]{Meyerheim09}%
  \BibitemOpen
  \bibfield  {author} {\bibinfo {author} {\bibfnamefont {H.~L.}\ \bibnamefont
  {Meyerheim}}, \bibinfo {author} {\bibfnamefont {J.-M.}\ \bibnamefont
  {Tonnerre}}, \bibinfo {author} {\bibfnamefont {L.}~\bibnamefont
  {Sandratskii}}, \bibinfo {author} {\bibfnamefont {H.~C.~N.}\ \bibnamefont
  {Tolentino}}, \bibinfo {author} {\bibfnamefont {M.}~\bibnamefont
  {Przybylski}}, \bibinfo {author} {\bibfnamefont {Y.}~\bibnamefont {Gabi}},
  \bibinfo {author} {\bibfnamefont {F.}~\bibnamefont {Yildiz}}, \bibinfo
  {author} {\bibfnamefont {X.~L.}\ \bibnamefont {Fu}}, \bibinfo {author}
  {\bibfnamefont {E.}~\bibnamefont {Bontempi}}, \bibinfo {author}
  {\bibfnamefont {S.}~\bibnamefont {Grenier}}, \ and\ \bibinfo {author}
  {\bibfnamefont {J.}~\bibnamefont {Kirschner}},\ }\Doi
  {10.1103/PhysRevLett.103.267202} {\bibfield  {journal} {\bibinfo  {journal}
  {Phys. Rev. Lett.},\ }\textbf {\bibinfo {volume} {103}},\ \bibinfo {pages}
  {267202} (\bibinfo {year} {2009})}\BibitemShut {NoStop}%
\bibitem [{\citenamefont {Yoo}\ \emph {et~al.}(1998)\citenamefont {Yoo},
  \citenamefont {S\"{o}derlind},\ and\ \citenamefont {Cynn}}]{Yoo98}%
  \BibitemOpen
  \bibfield  {author} {\bibinfo {author} {\bibfnamefont {C.-S.}\ \bibnamefont
  {Yoo}}, \bibinfo {author} {\bibfnamefont {P.}~\bibnamefont {S\"{o}derlind}},
  \ and\ \bibinfo {author} {\bibfnamefont {H.}~\bibnamefont {Cynn}},\ }\Doi
  {10.1088/0953-8984/10/20/001} {\bibfield  {journal} {\bibinfo  {journal} {J.
  Phys.: Condens. Matter},\ }\textbf {\bibinfo {volume} {10}},\ \bibinfo
  {pages} {L311} (\bibinfo {year} {1998})}\BibitemShut {NoStop}%
\bibitem [{\citenamefont {Yoo}\ \emph {et~al.}(2000)\citenamefont {Yoo},
  \citenamefont {Cynn}, \citenamefont {S\"oderlind},\ and\ \citenamefont
  {Iota}}]{Yoo00}%
  \BibitemOpen
  \bibfield  {author} {\bibinfo {author} {\bibfnamefont {C.~S.}\ \bibnamefont
  {Yoo}}, \bibinfo {author} {\bibfnamefont {H.}~\bibnamefont {Cynn}}, \bibinfo
  {author} {\bibfnamefont {P.}~\bibnamefont {S\"oderlind}}, \ and\ \bibinfo
  {author} {\bibfnamefont {V.}~\bibnamefont {Iota}},\ }\Doi
  {10.1103/PhysRevLett.84.4132} {\bibfield  {journal} {\bibinfo  {journal}
  {Phys. Rev. Lett.},\ }\textbf {\bibinfo {volume} {84}},\ \bibinfo {pages}
  {4132} (\bibinfo {year} {2000})}\BibitemShut {NoStop}%
\bibitem [{\citenamefont {Goncharov}\ \emph {et~al.}(2004)\citenamefont
  {Goncharov}, \citenamefont {Crowhurst},\ and\ \citenamefont
  {Zaug}}]{Goncharov04}%
  \BibitemOpen
  \bibfield  {author} {\bibinfo {author} {\bibfnamefont {A.~F.}\ \bibnamefont
  {Goncharov}}, \bibinfo {author} {\bibfnamefont {J.}~\bibnamefont
  {Crowhurst}}, \ and\ \bibinfo {author} {\bibfnamefont {J.~M.}\ \bibnamefont
  {Zaug}},\ }\Doi {10.1103/PhysRevLett.92.115502} {\bibfield  {journal}
  {\bibinfo  {journal} {Phys. Rev. Lett.},\ }\textbf {\bibinfo {volume} {92}},\
  \bibinfo {pages} {115502} (\bibinfo {year} {2004})}\BibitemShut {NoStop}%
\bibitem [{\citenamefont {Antonangeli}\ \emph {et~al.}(2008)\citenamefont
  {Antonangeli}, \citenamefont {Benedetti}, \citenamefont {Farber},
  \citenamefont {Steinle-Neumann}, \citenamefont {Auzende}, \citenamefont
  {Badro}, \citenamefont {Hanfland},\ and\ \citenamefont
  {Krisch}}]{Antonangeli08}%
  \BibitemOpen
  \bibfield  {author} {\bibinfo {author} {\bibfnamefont {D.}~\bibnamefont
  {Antonangeli}}, \bibinfo {author} {\bibfnamefont {L.~R.}\ \bibnamefont
  {Benedetti}}, \bibinfo {author} {\bibfnamefont {D.~L.}\ \bibnamefont
  {Farber}}, \bibinfo {author} {\bibfnamefont {G.}~\bibnamefont
  {Steinle-Neumann}}, \bibinfo {author} {\bibfnamefont {A.-L.}\ \bibnamefont
  {Auzende}}, \bibinfo {author} {\bibfnamefont {J.}~\bibnamefont {Badro}},
  \bibinfo {author} {\bibfnamefont {M.}~\bibnamefont {Hanfland}}, \ and\
  \bibinfo {author} {\bibfnamefont {M.}~\bibnamefont {Krisch}},\ }\Doi
  {10.1063/1.2897038} {\bibfield  {journal} {\bibinfo  {journal} {Appl. Phys.
  Lett.},\ }\textbf {\bibinfo {volume} {92}},\ \bibinfo {eid} {111911}
  (\bibinfo {year} {2008})}\BibitemShut {NoStop}%
\bibitem [{\citenamefont {Ishimatsu}\ \emph {et~al.}(2011)\citenamefont
  {Ishimatsu}, \citenamefont {Kawamura}, \citenamefont {Maruyama},
  \citenamefont {Mizumaki}, \citenamefont {Matsuoka}, \citenamefont {Yumoto},
  \citenamefont {Ohashi},\ and\ \citenamefont {Suzuki}}]{Ishimatsu11}%
  \BibitemOpen
  \bibfield  {author} {\bibinfo {author} {\bibfnamefont {N.}~\bibnamefont
  {Ishimatsu}}, \bibinfo {author} {\bibfnamefont {N.}~\bibnamefont {Kawamura}},
  \bibinfo {author} {\bibfnamefont {H.}~\bibnamefont {Maruyama}}, \bibinfo
  {author} {\bibfnamefont {M.}~\bibnamefont {Mizumaki}}, \bibinfo {author}
  {\bibfnamefont {T.}~\bibnamefont {Matsuoka}}, \bibinfo {author}
  {\bibfnamefont {H.}~\bibnamefont {Yumoto}}, \bibinfo {author} {\bibfnamefont
  {H.}~\bibnamefont {Ohashi}}, \ and\ \bibinfo {author} {\bibfnamefont
  {M.}~\bibnamefont {Suzuki}},\ }\Doi {10.1103/PhysRevB.83.180409} {\bibfield
  {journal} {\bibinfo  {journal} {Phys. Rev. B},\ }\textbf {\bibinfo {volume}
  {83}},\ \bibinfo {pages} {180409} (\bibinfo {year} {2011})}\BibitemShut
  {NoStop}%
\bibitem [{\citenamefont {Torchio}\ \emph
  {et~al.}(2011){\natexlab{a}}\citenamefont {Torchio}, \citenamefont {Monza},
  \citenamefont {Baudelet}, \citenamefont {Pascarelli}, \citenamefont {Mathon},
  \citenamefont {Pugh}, \citenamefont {Antonangeli},\ and\ \citenamefont
  {Iti\'e}}]{Torchio11a}%
  \BibitemOpen
  \bibfield  {author} {\bibinfo {author} {\bibfnamefont {R.}~\bibnamefont
  {Torchio}}, \bibinfo {author} {\bibfnamefont {A.}~\bibnamefont {Monza}},
  \bibinfo {author} {\bibfnamefont {F.}~\bibnamefont {Baudelet}}, \bibinfo
  {author} {\bibfnamefont {S.}~\bibnamefont {Pascarelli}}, \bibinfo {author}
  {\bibfnamefont {O.}~\bibnamefont {Mathon}}, \bibinfo {author} {\bibfnamefont
  {E.}~\bibnamefont {Pugh}}, \bibinfo {author} {\bibfnamefont {D.}~\bibnamefont
  {Antonangeli}}, \ and\ \bibinfo {author} {\bibfnamefont {J.~P.}\ \bibnamefont
  {Iti\'e}},\ }\Doi {10.1103/PhysRevB.84.060403} {\bibfield  {journal}
  {\bibinfo  {journal} {Phys. Rev. B},\ }\textbf {\bibinfo {volume} {84}},\
  \bibinfo {pages} {060403} (\bibinfo {year} {2011}{\natexlab{a}})}\BibitemShut
  {NoStop}%
\bibitem [{\citenamefont {Torchio}\ \emph
  {et~al.}(2011){\natexlab{b}}\citenamefont {Torchio}, \citenamefont
  {Kvashnin}, \citenamefont {Pascarelli}, \citenamefont {Mathon}, \citenamefont
  {Marini}, \citenamefont {Genovese}, \citenamefont {Bruno}, \citenamefont
  {Garbarino}, \citenamefont {Dewaele}, \citenamefont {Occelli},\ and\
  \citenamefont {Loubeyre}}]{Torchio11b}%
  \BibitemOpen
  \bibfield  {author} {\bibinfo {author} {\bibfnamefont {R.}~\bibnamefont
  {Torchio}}, \bibinfo {author} {\bibfnamefont {Y.~O.}\ \bibnamefont
  {Kvashnin}}, \bibinfo {author} {\bibfnamefont {S.}~\bibnamefont
  {Pascarelli}}, \bibinfo {author} {\bibfnamefont {O.}~\bibnamefont {Mathon}},
  \bibinfo {author} {\bibfnamefont {C.}~\bibnamefont {Marini}}, \bibinfo
  {author} {\bibfnamefont {L.}~\bibnamefont {Genovese}}, \bibinfo {author}
  {\bibfnamefont {P.}~\bibnamefont {Bruno}}, \bibinfo {author} {\bibfnamefont
  {G.}~\bibnamefont {Garbarino}}, \bibinfo {author} {\bibfnamefont
  {A.}~\bibnamefont {Dewaele}}, \bibinfo {author} {\bibfnamefont
  {F.}~\bibnamefont {Occelli}}, \ and\ \bibinfo {author} {\bibfnamefont
  {P.}~\bibnamefont {Loubeyre}},\ }\Doi {10.1103/PhysRevLett.107.237202}
  {\bibfield  {journal} {\bibinfo  {journal} {Phys. Rev. Lett.},\ }\textbf
  {\bibinfo {volume} {107}},\ \bibinfo {pages} {237202} (\bibinfo {year}
  {2011}{\natexlab{b}})}\BibitemShut {NoStop}%
\bibitem [{\citenamefont {Rodr\'{\i}guez-L\'opez}\ \emph
  {et~al.}(2003)\citenamefont {Rodr\'{\i}guez-L\'opez}, \citenamefont
  {Aguilera-Granja}, \citenamefont {Michaelian},\ and\ \citenamefont
  {Vega}}]{RodriguezLopez03}%
  \BibitemOpen
  \bibfield  {author} {\bibinfo {author} {\bibfnamefont {J.~L.}\ \bibnamefont
  {Rodr\'{\i}guez-L\'opez}}, \bibinfo {author} {\bibfnamefont {F.}~\bibnamefont
  {Aguilera-Granja}}, \bibinfo {author} {\bibfnamefont {K.}~\bibnamefont
  {Michaelian}}, \ and\ \bibinfo {author} {\bibfnamefont {A.}~\bibnamefont
  {Vega}},\ }\href {http://link.aps.org/abstract/PRB/v67/e174413} {\bibfield
  {journal} {\bibinfo  {journal} {Phys. Rev. B},\ }\textbf {\bibinfo {volume}
  {67}},\ \bibinfo {eid} {174413} (\bibinfo {year} {2003})}\BibitemShut
  {NoStop}%
\bibitem [{\citenamefont {Datta}\ \emph {et~al.}(2007)\citenamefont {Datta},
  \citenamefont {Kabir}, \citenamefont {Ganguly}, \citenamefont {Sanyal},
  \citenamefont {Saha-Dasgupta},\ and\ \citenamefont {Mookerjee}}]{Datta07}%
  \BibitemOpen
  \bibfield  {author} {\bibinfo {author} {\bibfnamefont {S.}~\bibnamefont
  {Datta}}, \bibinfo {author} {\bibfnamefont {M.}~\bibnamefont {Kabir}},
  \bibinfo {author} {\bibfnamefont {S.}~\bibnamefont {Ganguly}}, \bibinfo
  {author} {\bibfnamefont {B.}~\bibnamefont {Sanyal}}, \bibinfo {author}
  {\bibfnamefont {T.}~\bibnamefont {Saha-Dasgupta}}, \ and\ \bibinfo {author}
  {\bibfnamefont {A.}~\bibnamefont {Mookerjee}},\ }\Doi
  {10.1103/PhysRevB.76.014429} {\bibfield  {journal} {\bibinfo  {journal}
  {Phys. Rev. B},\ }\textbf {\bibinfo {volume} {76}},\ \bibinfo {pages}
  {014429} (\bibinfo {year} {2007})}\BibitemShut {NoStop}%
\bibitem [{\citenamefont {Ma}\ \emph {et~al.}(2006)\citenamefont {Ma},
  \citenamefont {Xie}, \citenamefont {Wang}, \citenamefont {Liu},\ and\
  \citenamefont {Li}}]{Ma06a}%
  \BibitemOpen
  \bibfield  {author} {\bibinfo {author} {\bibfnamefont {Q.-M.}\ \bibnamefont
  {Ma}}, \bibinfo {author} {\bibfnamefont {Z.}~\bibnamefont {Xie}}, \bibinfo
  {author} {\bibfnamefont {J.}~\bibnamefont {Wang}}, \bibinfo {author}
  {\bibfnamefont {Y.}~\bibnamefont {Liu}}, \ and\ \bibinfo {author}
  {\bibfnamefont {Y.-C.}\ \bibnamefont {Li}},\ }\Doi
  {http://dx.doi.org/10.1016/j.physleta.2006.05.033} {\bibfield  {journal}
  {\bibinfo  {journal} {Phys. Lett. A},\ }\textbf {\bibinfo {volume} {358}},\
  \bibinfo {pages} {289 } (\bibinfo {year} {2006})}\BibitemShut {NoStop}%
\bibitem [{\citenamefont {Guevara}\ \emph {et~al.}(1999)\citenamefont
  {Guevara}, \citenamefont {Llois}, \citenamefont {Aguilera-Granja},\ and\
  \citenamefont {Montejano-Carrizales}}]{Guevara99}%
  \BibitemOpen
  \bibfield  {author} {\bibinfo {author} {\bibfnamefont {J.}~\bibnamefont
  {Guevara}}, \bibinfo {author} {\bibfnamefont {A.}~\bibnamefont {Llois}},
  \bibinfo {author} {\bibfnamefont {F.}~\bibnamefont {Aguilera-Granja}}, \ and\
  \bibinfo {author} {\bibfnamefont {J.}~\bibnamefont {Montejano-Carrizales}},\
  }\Doi {http://dx.doi.org/10.1016/S0038-1098(99)00189-1} {\bibfield  {journal}
  {\bibinfo  {journal} {Solid State Commun.},\ }\textbf {\bibinfo {volume}
  {111}},\ \bibinfo {pages} {335 } (\bibinfo {year} {1999})}\BibitemShut
  {NoStop}%
\bibitem [{\citenamefont {Fan}\ \emph {et~al.}(1997)\citenamefont {Fan},
  \citenamefont {Liu},\ and\ \citenamefont {Liao}}]{Fan97}%
  \BibitemOpen
  \bibfield  {author} {\bibinfo {author} {\bibfnamefont {H.-J.}\ \bibnamefont
  {Fan}}, \bibinfo {author} {\bibfnamefont {C.-W.}\ \bibnamefont {Liu}}, \ and\
  \bibinfo {author} {\bibfnamefont {M.-S.}\ \bibnamefont {Liao}},\ }\Doi
  {http://dx.doi.org/10.1016/S0009-2614(97)00534-4} {\bibfield  {journal}
  {\bibinfo  {journal} {Chem. Phys. Lett.},\ }\textbf {\bibinfo {volume}
  {273}},\ \bibinfo {pages} {353 } (\bibinfo {year} {1997})}\BibitemShut
  {NoStop}%
\bibitem [{\citenamefont {Reddy}\ \emph {et~al.}(1998)\citenamefont {Reddy},
  \citenamefont {Nayak}, \citenamefont {Khanna}, \citenamefont {Rao},\ and\
  \citenamefont {Jena}}]{Reddy98}%
  \BibitemOpen
  \bibfield  {author} {\bibinfo {author} {\bibfnamefont {B.~V.}\ \bibnamefont
  {Reddy}}, \bibinfo {author} {\bibfnamefont {S.~K.}\ \bibnamefont {Nayak}},
  \bibinfo {author} {\bibfnamefont {S.~N.}\ \bibnamefont {Khanna}}, \bibinfo
  {author} {\bibfnamefont {B.~K.}\ \bibnamefont {Rao}}, \ and\ \bibinfo
  {author} {\bibfnamefont {P.}~\bibnamefont {Jena}},\ }\Doi {10.1021/jp980262b}
  {\bibfield  {journal} {\bibinfo  {journal} {J. Phys. Chem. A},\ }\textbf
  {\bibinfo {volume} {102}},\ \bibinfo {pages} {1748} (\bibinfo {year}
  {1998})}\BibitemShut {NoStop}%
\bibitem [{\citenamefont {Andriotis}\ and\ \citenamefont
  {Menon}(1998)}]{Andriotis98}%
  \BibitemOpen
  \bibfield  {author} {\bibinfo {author} {\bibfnamefont {A.~N.}\ \bibnamefont
  {Andriotis}}\ and\ \bibinfo {author} {\bibfnamefont {M.}~\bibnamefont
  {Menon}},\ }\Doi {10.1103/PhysRevB.57.10069} {\bibfield  {journal} {\bibinfo
  {journal} {Phys. Rev. B},\ }\textbf {\bibinfo {volume} {57}},\ \bibinfo
  {pages} {10069} (\bibinfo {year} {1998})}\BibitemShut {NoStop}%
\bibitem [{\citenamefont {Parks}\ \emph {et~al.}(1988)\citenamefont {Parks},
  \citenamefont {Weiller}, \citenamefont {Bechthold}, \citenamefont {Hoffman},
  \citenamefont {Nieman}, \citenamefont {Pobo},\ and\ \citenamefont
  {Riley}}]{Parks88}%
  \BibitemOpen
  \bibfield  {author} {\bibinfo {author} {\bibfnamefont {E.~K.}\ \bibnamefont
  {Parks}}, \bibinfo {author} {\bibfnamefont {B.~H.}\ \bibnamefont {Weiller}},
  \bibinfo {author} {\bibfnamefont {P.~S.}\ \bibnamefont {Bechthold}}, \bibinfo
  {author} {\bibfnamefont {W.~F.}\ \bibnamefont {Hoffman}}, \bibinfo {author}
  {\bibfnamefont {G.~C.}\ \bibnamefont {Nieman}}, \bibinfo {author}
  {\bibfnamefont {L.~G.}\ \bibnamefont {Pobo}}, \ and\ \bibinfo {author}
  {\bibfnamefont {S.~J.}\ \bibnamefont {Riley}},\ }\href
  {http://link.aip.org/link/?JCP/88/1622/1} {\bibfield  {journal} {\bibinfo
  {journal} {J. Chem. Phys.},\ }\textbf {\bibinfo {volume} {88}},\ \bibinfo
  {pages} {1622} (\bibinfo {year} {1988})}\BibitemShut {NoStop}%
\bibitem [{\citenamefont {Parks}\ \emph {et~al.}(1992)\citenamefont {Parks},
  \citenamefont {Winter}, \citenamefont {Klots},\ and\ \citenamefont
  {Riley}}]{Parks92}%
  \BibitemOpen
  \bibfield  {author} {\bibinfo {author} {\bibfnamefont {E.~K.}\ \bibnamefont
  {Parks}}, \bibinfo {author} {\bibfnamefont {B.~J.}\ \bibnamefont {Winter}},
  \bibinfo {author} {\bibfnamefont {T.~D.}\ \bibnamefont {Klots}}, \ and\
  \bibinfo {author} {\bibfnamefont {S.~J.}\ \bibnamefont {Riley}},\ }\Doi
  {http://dx.doi.org/10.1063/1.462330} {\bibfield  {journal} {\bibinfo
  {journal} {J. Chem. Phys.},\ }\textbf {\bibinfo {volume} {96}},\ \bibinfo
  {pages} {8267} (\bibinfo {year} {1992})}\BibitemShut {NoStop}%
\bibitem [{\citenamefont {Parks}\ \emph {et~al.}(1993)\citenamefont {Parks},
  \citenamefont {Zhu}, \citenamefont {Ho},\ and\ \citenamefont
  {Riley}}]{Parks93a}%
  \BibitemOpen
  \bibfield  {author} {\bibinfo {author} {\bibfnamefont {E.}~\bibnamefont
  {Parks}}, \bibinfo {author} {\bibfnamefont {L.}~\bibnamefont {Zhu}}, \bibinfo
  {author} {\bibfnamefont {J.}~\bibnamefont {Ho}}, \ and\ \bibinfo {author}
  {\bibfnamefont {S.}~\bibnamefont {Riley}},\ }\Doi {10.1007/BF01429103}
  {\bibfield  {journal} {\bibinfo  {journal} {Z. Phys. D},\ }\textbf {\bibinfo
  {volume} {26}},\ \bibinfo {pages} {41} (\bibinfo {year} {1993})}\BibitemShut
  {NoStop}%
\bibitem [{\citenamefont {Lian}\ \emph {et~al.}(1992)\citenamefont {Lian},
  \citenamefont {Su},\ and\ \citenamefont {Armentrout}}]{Lian92a}%
  \BibitemOpen
  \bibfield  {author} {\bibinfo {author} {\bibfnamefont {L.}~\bibnamefont
  {Lian}}, \bibinfo {author} {\bibfnamefont {C.-X.}\ \bibnamefont {Su}}, \ and\
  \bibinfo {author} {\bibfnamefont {P.~B.}\ \bibnamefont {Armentrout}},\ }\Doi
  {10.1063/1.463912} {\bibfield  {journal} {\bibinfo  {journal} {J. Chem.
  Phys.},\ }\textbf {\bibinfo {volume} {97}},\ \bibinfo {pages} {4072}
  (\bibinfo {year} {1992})}\BibitemShut {NoStop}%
\bibitem [{\citenamefont {Sakurai}\ \emph {et~al.}(1999)\citenamefont
  {Sakurai}, \citenamefont {Watanabe}, \citenamefont {Sumiyama},\ and\
  \citenamefont {Suzuki}}]{Sakurai99}%
  \BibitemOpen
  \bibfield  {author} {\bibinfo {author} {\bibfnamefont {M.}~\bibnamefont
  {Sakurai}}, \bibinfo {author} {\bibfnamefont {K.}~\bibnamefont {Watanabe}},
  \bibinfo {author} {\bibfnamefont {K.}~\bibnamefont {Sumiyama}}, \ and\
  \bibinfo {author} {\bibfnamefont {K.}~\bibnamefont {Suzuki}},\ }\Doi
  {10.1063/1.479268} {\bibfield  {journal} {\bibinfo  {journal} {J. Chem.
  Phys.},\ }\textbf {\bibinfo {volume} {111}},\ \bibinfo {pages} {235}
  (\bibinfo {year} {1999})}\BibitemShut {NoStop}%
\bibitem [{\citenamefont {Rohlfing}\ \emph {et~al.}(1984)\citenamefont
  {Rohlfing}, \citenamefont {Cox}, \citenamefont {Kaldor},\ and\ \citenamefont
  {Johnson}}]{Rohlfing84}%
  \BibitemOpen
  \bibfield  {author} {\bibinfo {author} {\bibfnamefont {E.~A.}\ \bibnamefont
  {Rohlfing}}, \bibinfo {author} {\bibfnamefont {D.~M.}\ \bibnamefont {Cox}},
  \bibinfo {author} {\bibfnamefont {A.}~\bibnamefont {Kaldor}}, \ and\ \bibinfo
  {author} {\bibfnamefont {K.~H.}\ \bibnamefont {Johnson}},\ }\Doi
  {http://dx.doi.org/10.1063/1.448168} {\bibfield  {journal} {\bibinfo
  {journal} {J. Chem. Phys.},\ }\textbf {\bibinfo {volume} {81}},\ \bibinfo
  {pages} {3846} (\bibinfo {year} {1984})}\BibitemShut {NoStop}%
\bibitem [{\citenamefont {Yang}\ and\ \citenamefont
  {Knickelbein}(1990)}]{Yang90}%
  \BibitemOpen
  \bibfield  {author} {\bibinfo {author} {\bibfnamefont {S.}~\bibnamefont
  {Yang}}\ and\ \bibinfo {author} {\bibfnamefont {M.~B.}\ \bibnamefont
  {Knickelbein}},\ }\Doi {http://dx.doi.org/10.1063/1.459131} {\bibfield
  {journal} {\bibinfo  {journal} {J. Chem. Phys.},\ }\textbf {\bibinfo {volume}
  {93}},\ \bibinfo {pages} {1533} (\bibinfo {year} {1990})}\BibitemShut
  {NoStop}%
\bibitem [{\citenamefont {Parks}\ \emph {et~al.}(1990)\citenamefont {Parks},
  \citenamefont {Klots},\ and\ \citenamefont {Riley}}]{Parks90}%
  \BibitemOpen
  \bibfield  {author} {\bibinfo {author} {\bibfnamefont {E.~K.}\ \bibnamefont
  {Parks}}, \bibinfo {author} {\bibfnamefont {T.~D.}\ \bibnamefont {Klots}}, \
  and\ \bibinfo {author} {\bibfnamefont {S.~J.}\ \bibnamefont {Riley}},\ }\Doi
  {http://dx.doi.org/10.1063/1.457839} {\bibfield  {journal} {\bibinfo
  {journal} {J. Chem. Phys.},\ }\textbf {\bibinfo {volume} {92}},\ \bibinfo
  {pages} {3813} (\bibinfo {year} {1990})}\BibitemShut {NoStop}%
\bibitem [{\citenamefont {Rapps}\ \emph {et~al.}(2013)\citenamefont {Rapps},
  \citenamefont {Ahlrichs}, \citenamefont {Waldt}, \citenamefont {Kappes},\
  and\ \citenamefont {Schooss}}]{Rapps13}%
  \BibitemOpen
  \bibfield  {author} {\bibinfo {author} {\bibfnamefont {T.}~\bibnamefont
  {Rapps}}, \bibinfo {author} {\bibfnamefont {R.}~\bibnamefont {Ahlrichs}},
  \bibinfo {author} {\bibfnamefont {E.}~\bibnamefont {Waldt}}, \bibinfo
  {author} {\bibfnamefont {M.~M.}\ \bibnamefont {Kappes}}, \ and\ \bibinfo
  {author} {\bibfnamefont {D.}~\bibnamefont {Schooss}},\ }\Doi
  {10.1002/anie.201302165} {\bibfield  {journal} {\bibinfo  {journal} {Angew.
  Chem. Int. Ed.},\ }\textbf {\bibinfo {volume} {52}},\ \bibinfo {pages} {6102}
  (\bibinfo {year} {2013})}\BibitemShut {NoStop}%
\bibitem [{\citenamefont {Gehrke}\ \emph {et~al.}(2009)\citenamefont {Gehrke},
  \citenamefont {Gruene}, \citenamefont {Fielicke}, \citenamefont {Meijer},\
  and\ \citenamefont {Reuter}}]{Gehrke09}%
  \BibitemOpen
  \bibfield  {author} {\bibinfo {author} {\bibfnamefont {R.}~\bibnamefont
  {Gehrke}}, \bibinfo {author} {\bibfnamefont {P.}~\bibnamefont {Gruene}},
  \bibinfo {author} {\bibfnamefont {A.}~\bibnamefont {Fielicke}}, \bibinfo
  {author} {\bibfnamefont {G.}~\bibnamefont {Meijer}}, \ and\ \bibinfo {author}
  {\bibfnamefont {K.}~\bibnamefont {Reuter}},\ }\Doi {10.1063/1.3058637}
  {\bibfield  {journal} {\bibinfo  {journal} {J. Chem. Phys.},\ }\textbf
  {\bibinfo {volume} {130}},\ \bibinfo {eid} {034306} (\bibinfo {year}
  {2009})}\BibitemShut {NoStop}%
\bibitem [{\citenamefont {Kakar}\ \emph {et~al.}(1997)\citenamefont {Kakar},
  \citenamefont {Bj{\o}rneholm}, \citenamefont {Weigelt}, \citenamefont
  {de~Castro}, \citenamefont {Tr\"oger}, \citenamefont {Frahm}, \citenamefont
  {M\"oller}, \citenamefont {Knop},\ and\ \citenamefont {R\"uhl}}]{Kakar97}%
  \BibitemOpen
  \bibfield  {author} {\bibinfo {author} {\bibfnamefont {S.}~\bibnamefont
  {Kakar}}, \bibinfo {author} {\bibfnamefont {O.}~\bibnamefont
  {Bj{\o}rneholm}}, \bibinfo {author} {\bibfnamefont {J.}~\bibnamefont
  {Weigelt}}, \bibinfo {author} {\bibfnamefont {A.~R.~B.}\ \bibnamefont
  {de~Castro}}, \bibinfo {author} {\bibfnamefont {L.}~\bibnamefont {Tr\"oger}},
  \bibinfo {author} {\bibfnamefont {R.}~\bibnamefont {Frahm}}, \bibinfo
  {author} {\bibfnamefont {T.}~\bibnamefont {M\"oller}}, \bibinfo {author}
  {\bibfnamefont {A.}~\bibnamefont {Knop}}, \ and\ \bibinfo {author}
  {\bibfnamefont {E.}~\bibnamefont {R\"uhl}},\ }\href
  {http://link.aps.org/abstract/PRL/v78/p1675} {\bibfield  {journal} {\bibinfo
  {journal} {Phys. Rev. Lett.},\ }\textbf {\bibinfo {volume} {78}},\ \bibinfo
  {pages} {1675} (\bibinfo {year} {1997})}\BibitemShut {NoStop}%
\bibitem [{\citenamefont {Reuse}\ and\ \citenamefont
  {Khanna}(1995)}]{Reuse95b}%
  \BibitemOpen
  \bibfield  {author} {\bibinfo {author} {\bibfnamefont {F.}~\bibnamefont
  {Reuse}}\ and\ \bibinfo {author} {\bibfnamefont {S.}~\bibnamefont {Khanna}},\
  }\Doi {http://dx.doi.org/10.1016/0009-2614(95)00012-S} {\bibfield  {journal}
  {\bibinfo  {journal} {Chem. Phys. Lett.},\ }\textbf {\bibinfo {volume}
  {234}},\ \bibinfo {pages} {77 } (\bibinfo {year} {1995})}\BibitemShut
  {NoStop}%
\bibitem [{\citenamefont {Shenston}(1970)}]{Shenston70}%
  \BibitemOpen
  \bibfield  {author} {\bibinfo {author} {\bibfnamefont {A.~G.}\ \bibnamefont
  {Shenston}},\ }\href@noop {} {\bibfield  {journal} {\bibinfo  {journal} {J.
  Res. Nat. Inst. Stand. Technol.},\ }\textbf {\bibinfo {volume} {74A}},\
  \bibinfo {pages} {801} (\bibinfo {year} {1970})}\BibitemShut {NoStop}%
\bibitem [{\citenamefont {Litzen}\ \emph {et~al.}(1993)\citenamefont {Litzen},
  \citenamefont {Brault},\ and\ \citenamefont {Thorne}}]{Litzen93}%
  \BibitemOpen
  \bibfield  {author} {\bibinfo {author} {\bibfnamefont {U.}~\bibnamefont
  {Litzen}}, \bibinfo {author} {\bibfnamefont {J.~W.}\ \bibnamefont {Brault}},
  \ and\ \bibinfo {author} {\bibfnamefont {A.~P.}\ \bibnamefont {Thorne}},\
  }\href@noop {} {\bibfield  {journal} {\bibinfo  {journal} {Phys. Scr.},\
  }\textbf {\bibinfo {volume} {47}},\ \bibinfo {pages} {628} (\bibinfo {year}
  {1993})}\BibitemShut {NoStop}%
\bibitem [{\citenamefont {Pickering}\ and\ \citenamefont
  {Thorne}(1996)}]{Pickering96}%
  \BibitemOpen
  \bibfield  {author} {\bibinfo {author} {\bibfnamefont {J.}~\bibnamefont
  {Pickering}}\ and\ \bibinfo {author} {\bibfnamefont {A.}~\bibnamefont
  {Thorne}},\ }\href@noop {} {\bibfield  {journal} {\bibinfo  {journal}
  {Astrophys. J. Suppl. Ser.},\ }\textbf {\bibinfo {volume} {107}},\ \bibinfo
  {pages} {761} (\bibinfo {year} {1996})}\BibitemShut {NoStop}%
\bibitem [{\citenamefont {Pickering}\ \emph {et~al.}(1998)\citenamefont
  {Pickering}, \citenamefont {Raassen}, \citenamefont {Uylings},\ and\
  \citenamefont {Johansson}}]{Pickering98}%
  \BibitemOpen
  \bibfield  {author} {\bibinfo {author} {\bibfnamefont {J.}~\bibnamefont
  {Pickering}}, \bibinfo {author} {\bibfnamefont {A.}~\bibnamefont {Raassen}},
  \bibinfo {author} {\bibfnamefont {P.}~\bibnamefont {Uylings}}, \ and\
  \bibinfo {author} {\bibfnamefont {S.}~\bibnamefont {Johansson}},\ }\href@noop
  {} {\bibfield  {journal} {\bibinfo  {journal} {Astrophys. J. Suppl. Ser.},\
  }\textbf {\bibinfo {volume} {117}},\ \bibinfo {pages} {261} (\bibinfo {year}
  {1998})}\BibitemShut {NoStop}%
\bibitem [{\citenamefont {Johansson}(1978)}]{Johansson78}%
  \BibitemOpen
  \bibfield  {author} {\bibinfo {author} {\bibfnamefont {S.}~\bibnamefont
  {Johansson}},\ }\href@noop {} {\bibfield  {journal} {\bibinfo  {journal}
  {Phys. Scr.},\ }\textbf {\bibinfo {volume} {18}},\ \bibinfo {pages} {217}
  (\bibinfo {year} {1978})}\BibitemShut {NoStop}%
\bibitem [{\citenamefont {Nave}\ \emph {et~al.}(1994)\citenamefont {Nave},
  \citenamefont {Johansson}, \citenamefont {Learner}, \citenamefont {Thorne},\
  and\ \citenamefont {Brault}}]{Nave94}%
  \BibitemOpen
  \bibfield  {author} {\bibinfo {author} {\bibfnamefont {G.}~\bibnamefont
  {Nave}}, \bibinfo {author} {\bibfnamefont {S.}~\bibnamefont {Johansson}},
  \bibinfo {author} {\bibfnamefont {R.~C.~M.}\ \bibnamefont {Learner}},
  \bibinfo {author} {\bibfnamefont {A.~P.}\ \bibnamefont {Thorne}}, \ and\
  \bibinfo {author} {\bibfnamefont {J.~W.}\ \bibnamefont {Brault}},\
  }\href@noop {} {\bibfield  {journal} {\bibinfo  {journal} {Astrophys. J.
  Suppl. Ser.},\ }\textbf {\bibinfo {volume} {94}},\ \bibinfo {pages} {221}
  (\bibinfo {year} {1994})}\BibitemShut {NoStop}%
\bibitem [{\citenamefont {Eriksson}\ \emph {et~al.}(1990)\citenamefont
  {Eriksson}, \citenamefont {Johansson}, \citenamefont {Albers}, \citenamefont
  {Boring},\ and\ \citenamefont {Brooks}}]{Eriksson90}%
  \BibitemOpen
  \bibfield  {author} {\bibinfo {author} {\bibfnamefont {O.}~\bibnamefont
  {Eriksson}}, \bibinfo {author} {\bibfnamefont {B.}~\bibnamefont {Johansson}},
  \bibinfo {author} {\bibfnamefont {R.~C.}\ \bibnamefont {Albers}}, \bibinfo
  {author} {\bibfnamefont {A.~M.}\ \bibnamefont {Boring}}, \ and\ \bibinfo
  {author} {\bibfnamefont {M.~S.~S.}\ \bibnamefont {Brooks}},\ }\Doi
  {10.1103/PhysRevB.42.2707} {\bibfield  {journal} {\bibinfo  {journal} {Phys.
  Rev. B},\ }\textbf {\bibinfo {volume} {42}},\ \bibinfo {pages} {2707}
  (\bibinfo {year} {1990})}\BibitemShut {NoStop}%
\bibitem [{\citenamefont {Ma}\ \emph {et~al.}(2007)\citenamefont {Ma},
  \citenamefont {Xie}, \citenamefont {Wang}, \citenamefont {Liu},\ and\
  \citenamefont {Li}}]{Ma07a}%
  \BibitemOpen
  \bibfield  {author} {\bibinfo {author} {\bibfnamefont {Q.-M.}\ \bibnamefont
  {Ma}}, \bibinfo {author} {\bibfnamefont {Z.}~\bibnamefont {Xie}}, \bibinfo
  {author} {\bibfnamefont {J.}~\bibnamefont {Wang}}, \bibinfo {author}
  {\bibfnamefont {Y.}~\bibnamefont {Liu}}, \ and\ \bibinfo {author}
  {\bibfnamefont {Y.-C.}\ \bibnamefont {Li}},\ }\Doi {DOI:
  10.1016/j.ssc.2006.12.023} {\bibfield  {journal} {\bibinfo  {journal} {Solid
  State Commun.},\ }\textbf {\bibinfo {volume} {142}},\ \bibinfo {pages} {114 }
  (\bibinfo {year} {2007})}\BibitemShut {NoStop}%
\bibitem [{\citenamefont {Popescu}\ \emph {et~al.}(2001)\citenamefont
  {Popescu}, \citenamefont {Ebert}, \citenamefont {Nonas},\ and\ \citenamefont
  {Dederichs}}]{Popescu01}%
  \BibitemOpen
  \bibfield  {author} {\bibinfo {author} {\bibfnamefont {V.}~\bibnamefont
  {Popescu}}, \bibinfo {author} {\bibfnamefont {H.}~\bibnamefont {Ebert}},
  \bibinfo {author} {\bibfnamefont {B.}~\bibnamefont {Nonas}}, \ and\ \bibinfo
  {author} {\bibfnamefont {P.~H.}\ \bibnamefont {Dederichs}},\ }\Doi
  {10.1103/PhysRevB.64.184407} {\bibfield  {journal} {\bibinfo  {journal}
  {Phys. Rev. B},\ }\textbf {\bibinfo {volume} {64}},\ \bibinfo {pages}
  {184407} (\bibinfo {year} {2001})}\BibitemShut {NoStop}%
\bibitem [{\citenamefont {S\'anchez-Barriga}\ \emph {et~al.}(2010)\citenamefont
  {S\'anchez-Barriga}, \citenamefont {Min\'ar}, \citenamefont {Braun},
  \citenamefont {Varykhalov}, \citenamefont {Boni}, \citenamefont {Di~Marco},
  \citenamefont {Rader}, \citenamefont {Bellini}, \citenamefont {Manghi},
  \citenamefont {Ebert}, \citenamefont {Katsnelson}, \citenamefont
  {Lichtenstein}, \citenamefont {Eriksson}, \citenamefont {Eberhardt},
  \citenamefont {D\"urr},\ and\ \citenamefont {Fink}}]{SanchezBarriga10}%
  \BibitemOpen
  \bibfield  {author} {\bibinfo {author} {\bibfnamefont {J.}~\bibnamefont
  {S\'anchez-Barriga}}, \bibinfo {author} {\bibfnamefont {J.}~\bibnamefont
  {Min\'ar}}, \bibinfo {author} {\bibfnamefont {J.}~\bibnamefont {Braun}},
  \bibinfo {author} {\bibfnamefont {A.}~\bibnamefont {Varykhalov}}, \bibinfo
  {author} {\bibfnamefont {V.}~\bibnamefont {Boni}}, \bibinfo {author}
  {\bibfnamefont {I.}~\bibnamefont {Di~Marco}}, \bibinfo {author}
  {\bibfnamefont {O.}~\bibnamefont {Rader}}, \bibinfo {author} {\bibfnamefont
  {V.}~\bibnamefont {Bellini}}, \bibinfo {author} {\bibfnamefont
  {F.}~\bibnamefont {Manghi}}, \bibinfo {author} {\bibfnamefont
  {H.}~\bibnamefont {Ebert}}, \bibinfo {author} {\bibfnamefont {M.~I.}\
  \bibnamefont {Katsnelson}}, \bibinfo {author} {\bibfnamefont {A.~I.}\
  \bibnamefont {Lichtenstein}}, \bibinfo {author} {\bibfnamefont
  {O.}~\bibnamefont {Eriksson}}, \bibinfo {author} {\bibfnamefont
  {W.}~\bibnamefont {Eberhardt}}, \bibinfo {author} {\bibfnamefont {H.~A.}\
  \bibnamefont {D\"urr}}, \ and\ \bibinfo {author} {\bibfnamefont
  {J.}~\bibnamefont {Fink}},\ }\Doi {10.1103/PhysRevB.82.104414} {\bibfield
  {journal} {\bibinfo  {journal} {Phys. Rev. B},\ }\textbf {\bibinfo {volume}
  {82}},\ \bibinfo {pages} {104414} (\bibinfo {year} {2010})}\BibitemShut
  {NoStop}%
\bibitem [{\citenamefont {Knickelbein}(2002){\natexlab{a}}}]{Knickelbein02}%
  \BibitemOpen
  \bibfield  {author} {\bibinfo {author} {\bibfnamefont {M.~B.}\ \bibnamefont
  {Knickelbein}},\ }\Doi {10.1016/S0009-2614(02)00024-6} {\bibfield  {journal}
  {\bibinfo  {journal} {Chem. Phys. Lett.},\ }\textbf {\bibinfo {volume}
  {353}},\ \bibinfo {pages} {221} (\bibinfo {year}
  {2002}{\natexlab{a}})}\BibitemShut {NoStop}%
\bibitem [{\citenamefont {Douglass}\ \emph {et~al.}(1993)\citenamefont
  {Douglass}, \citenamefont {Cox}, \citenamefont {Bucher},\ and\ \citenamefont
  {Bloomfield}}]{Douglass93}%
  \BibitemOpen
  \bibfield  {author} {\bibinfo {author} {\bibfnamefont {D.~C.}\ \bibnamefont
  {Douglass}}, \bibinfo {author} {\bibfnamefont {A.~J.}\ \bibnamefont {Cox}},
  \bibinfo {author} {\bibfnamefont {J.~P.}\ \bibnamefont {Bucher}}, \ and\
  \bibinfo {author} {\bibfnamefont {L.~A.}\ \bibnamefont {Bloomfield}},\ }\Doi
  {10.1103/PhysRevB.47.12874} {\bibfield  {journal} {\bibinfo  {journal} {Phys.
  Rev. B},\ }\textbf {\bibinfo {volume} {47}},\ \bibinfo {pages} {12874}
  (\bibinfo {year} {1993})}\BibitemShut {NoStop}%
\bibitem [{\citenamefont {Xu}\ \emph {et~al.}(2005)\citenamefont {Xu},
  \citenamefont {Yin}, \citenamefont {Moro},\ and\ \citenamefont
  {de~Heer}}]{Xu05}%
  \BibitemOpen
  \bibfield  {author} {\bibinfo {author} {\bibfnamefont {X.}~\bibnamefont
  {Xu}}, \bibinfo {author} {\bibfnamefont {S.}~\bibnamefont {Yin}}, \bibinfo
  {author} {\bibfnamefont {R.}~\bibnamefont {Moro}}, \ and\ \bibinfo {author}
  {\bibfnamefont {W.~A.}\ \bibnamefont {de~Heer}},\ }\Doi
  {10.1103/PhysRevLett.95.237209} {\bibfield  {journal} {\bibinfo  {journal}
  {Phys. Rev. Lett.},\ }\textbf {\bibinfo {volume} {95}},\ \bibinfo {eid}
  {237209} (\bibinfo {year} {2005})}\BibitemShut {NoStop}%
\bibitem [{\citenamefont {Xu}\ \emph {et~al.}(2008)\citenamefont {Xu},
  \citenamefont {Yin}, \citenamefont {Moro},\ and\ \citenamefont
  {de~Heer}}]{Xu08}%
  \BibitemOpen
  \bibfield  {author} {\bibinfo {author} {\bibfnamefont {X.}~\bibnamefont
  {Xu}}, \bibinfo {author} {\bibfnamefont {S.}~\bibnamefont {Yin}}, \bibinfo
  {author} {\bibfnamefont {R.}~\bibnamefont {Moro}}, \ and\ \bibinfo {author}
  {\bibfnamefont {W.~A.}\ \bibnamefont {de~Heer}},\ }\Doi
  {10.1103/PhysRevB.78.054430} {\bibfield  {journal} {\bibinfo  {journal}
  {Phys. Rev. B},\ }\textbf {\bibinfo {volume} {78}},\ \bibinfo {eid} {054430}
  (\bibinfo {year} {2008})}\BibitemShut {NoStop}%
\bibitem [{\citenamefont {Knickelbein}(2002){\natexlab{b}}}]{Knickelbein02b}%
  \BibitemOpen
  \bibfield  {author} {\bibinfo {author} {\bibfnamefont {M.~B.}\ \bibnamefont
  {Knickelbein}},\ }\Doi {10.1063/1.1477175} {\bibfield  {journal} {\bibinfo
  {journal} {J. Chem. Phys.},\ }\textbf {\bibinfo {volume} {116}},\ \bibinfo
  {pages} {9703} (\bibinfo {year} {2002}{\natexlab{b}})}\BibitemShut {NoStop}%
\bibitem [{\citenamefont {Jensen}\ \emph {et~al.}(1991)\citenamefont {Jensen},
  \citenamefont {Mukherjee},\ and\ \citenamefont {Bennemann}}]{Jensen91}%
  \BibitemOpen
  \bibfield  {author} {\bibinfo {author} {\bibfnamefont {P.~J.}\ \bibnamefont
  {Jensen}}, \bibinfo {author} {\bibfnamefont {S.}~\bibnamefont {Mukherjee}}, \
  and\ \bibinfo {author} {\bibfnamefont {K.~H.}\ \bibnamefont {Bennemann}},\
  }\Doi {10.1007/BF01438408} {\bibfield  {journal} {\bibinfo  {journal} {Z.
  Phys. D},\ }\textbf {\bibinfo {volume} {21}},\ \bibinfo {pages} {349}
  (\bibinfo {year} {1991})}\BibitemShut {NoStop}%
\bibitem [{\citenamefont {Ballone}\ \emph {et~al.}(1991)\citenamefont
  {Ballone}, \citenamefont {Milani},\ and\ \citenamefont
  {de~Heer}}]{Ballone91}%
  \BibitemOpen
  \bibfield  {author} {\bibinfo {author} {\bibfnamefont {P.}~\bibnamefont
  {Ballone}}, \bibinfo {author} {\bibfnamefont {P.}~\bibnamefont {Milani}}, \
  and\ \bibinfo {author} {\bibfnamefont {W.~A.}\ \bibnamefont {de~Heer}},\
  }\Doi {10.1103/PhysRevB.44.10350} {\bibfield  {journal} {\bibinfo  {journal}
  {Phys. Rev. B},\ }\textbf {\bibinfo {volume} {44}},\ \bibinfo {pages} {10350}
  (\bibinfo {year} {1991})}\BibitemShut {NoStop}%
\bibitem [{\citenamefont {Maiti}\ and\ \citenamefont
  {Falicov}(1993)}]{Maiti93}%
  \BibitemOpen
  \bibfield  {author} {\bibinfo {author} {\bibfnamefont {A.}~\bibnamefont
  {Maiti}}\ and\ \bibinfo {author} {\bibfnamefont {L.~M.}\ \bibnamefont
  {Falicov}},\ }\Doi {10.1103/PhysRevB.48.13596} {\bibfield  {journal}
  {\bibinfo  {journal} {Phys. Rev. B},\ }\textbf {\bibinfo {volume} {48}},\
  \bibinfo {pages} {13596} (\bibinfo {year} {1993})}\BibitemShut {NoStop}%
\bibitem [{\citenamefont {Visuthikraisee}\ and\ \citenamefont
  {Bertsch}(1996)}]{Visuthikraisee96}%
  \BibitemOpen
  \bibfield  {author} {\bibinfo {author} {\bibfnamefont {V.}~\bibnamefont
  {Visuthikraisee}}\ and\ \bibinfo {author} {\bibfnamefont {G.~F.}\
  \bibnamefont {Bertsch}},\ }\Doi {10.1103/PhysRevA.54.5104} {\bibfield
  {journal} {\bibinfo  {journal} {Phys. Rev. A},\ }\textbf {\bibinfo {volume}
  {54}},\ \bibinfo {pages} {5104} (\bibinfo {year} {1996})}\BibitemShut
  {NoStop}%
\bibitem [{\citenamefont {Hamamoto}\ \emph {et~al.}(2000)\citenamefont
  {Hamamoto}, \citenamefont {Onishi},\ and\ \citenamefont
  {Bertsch}}]{Hamamoto00}%
  \BibitemOpen
  \bibfield  {author} {\bibinfo {author} {\bibfnamefont {N.}~\bibnamefont
  {Hamamoto}}, \bibinfo {author} {\bibfnamefont {N.}~\bibnamefont {Onishi}}, \
  and\ \bibinfo {author} {\bibfnamefont {G.}~\bibnamefont {Bertsch}},\ }\Doi
  {10.1103/PhysRevB.61.1336} {\bibfield  {journal} {\bibinfo  {journal} {Phys.
  Rev. B},\ }\textbf {\bibinfo {volume} {61}},\ \bibinfo {pages} {1336}
  (\bibinfo {year} {2000})}\BibitemShut {NoStop}%
\bibitem [{\citenamefont {Knickelbein}(2004){\natexlab{b}}}]{Knickelbein04b}%
  \BibitemOpen
  \bibfield  {author} {\bibinfo {author} {\bibfnamefont {M.~B.}\ \bibnamefont
  {Knickelbein}},\ }\Doi {10.1063/1.1781156} {\bibfield  {journal} {\bibinfo
  {journal} {J. Chem. Phys.},\ }\textbf {\bibinfo {volume} {121}},\ \bibinfo
  {pages} {5281} (\bibinfo {year} {2004}{\natexlab{b}})}\BibitemShut {NoStop}%
\bibitem [{\citenamefont {Das}\ \emph {et~al.}(2005)\citenamefont {Das},
  \citenamefont {Konar},\ and\ \citenamefont {Dattagupta}}]{Das05}%
  \BibitemOpen
  \bibfield  {author} {\bibinfo {author} {\bibfnamefont {R.~K.}\ \bibnamefont
  {Das}}, \bibinfo {author} {\bibfnamefont {A.}~\bibnamefont {Konar}}, \ and\
  \bibinfo {author} {\bibfnamefont {S.}~\bibnamefont {Dattagupta}},\ }\Doi
  {10.1103/PhysRevB.71.014442} {\bibfield  {journal} {\bibinfo  {journal}
  {Phys. Rev. B},\ }\textbf {\bibinfo {volume} {71}},\ \bibinfo {pages}
  {014442} (\bibinfo {year} {2005})}\BibitemShut {NoStop}%
\end{thebibliography}
\end{document}